# Multidimensional Economic Complexity and Inclusive Green Growth

Viktor Stojkoski[1,2], Philipp Koch[1,3], César A. Hidalgo[1,4,5,*]

[1]Center for Collective Learning, ANITI, TSE-R, IAST, IRIT, Université de Toulouse, Toulouse, France.

[2]Faculty of Economics – Skopje, University Ss. Cyril and Methodius in Skopje, Skopje, North Macedonia.

[3]EcoAustria – Institute for Economic Research, Vienna, Austria.

[4]Alliance Manchester Business School, University of Manchester, Manchester, United Kingdom.

[5]Center for Collective Learning, CIAS, Corvinus University, Budapest, Hungary.

**Abstract**

To achieve inclusive green growth, countries need to consider a multiplicity of economic, social, and environmental factors. These are often captured by metrics of economic complexity derived from the geography of trade, thus missing key information on innovative activities. To bridge this gap, we combine trade data with data on patent applications and research publications to build models that significantly and robustly improve the ability of economic complexity metrics to explain international variations in inclusive green growth. We show that measures of complexity built on trade and patent data combine to explain future economic growth and income inequality and that countries that score high in all three metrics tend to exhibit lower emission intensities. These findings illustrate how the geography of trade, technology, and research combine to explain inclusive green growth.

# Introduction

Sustainable development is often defined as the process of meeting human development goals while simultaneously sustaining the natural environment[1–4]. This approach implies that development and environment are interdependent, and that economic growth can be sustained only if it is inclusive and green[5,6].

To achieve sustainable development, countries need to consider a web of economic, social, and environmental factors[7–12]. This multiplicity of factors, however, can be hard to quantify and compare.

---

[*] Corresponding author, e-mail: cesar.hidalgo@univ-toulouse.fr

Economic complexity methods provide a solution to this problem[13,14] by leveraging data on the geographic distribution of economic activities to estimate the implicit presence of multiple economic factors. These estimates have been validated by their ability to explain international variations in economic growth[15–24], income inequality[25–27], and emissions[28–31]. The reason why complexity metrics work is because they capture information about productive structures that escapes simple aggregate metrics, such as GDP or market concentration indexes. Unlike these metrics, which aggregate values regardless of the activities involved, economic complexity metrics capture information about the sophistication of activities that is implicit in their geographic distribution. For instance, according to a market concentration index (such as the Herfindahl-Hirschman index or information entropy), a country that exports 80% bananas and 20% cars is the same as a country that exports 80% cars and 20% bananas. Economic complexity metrics break this symmetry by incorporating information about the sophistication of each activity that is implicit in patterns of specialization.

Today, the most commonly used metrics of complexity are based on trade data[23,30,32]. Trade data, however, can miss key information about innovative activities, such as patent applications and research publications that could be relevant for the geography of inclusive green growth. For example, research and technology can shape production processes, affecting the skills and compensation of workers and the emission intensity of industrial activities. Moreover, trade-based metrics of complexity can systematically underestimate the complexity of economies that are distant from global markets, which in turn might distort predictions about their inclusive green growth[33,34]. That is, the complexity of some economies that are rich in natural resource exports, but distant to markets, such as Australia, Chile, and New Zealand, might be better reflected in their ability to produce outputs such as scientific research and patentable innovations than sophisticated exports. The same may be true, but in reverse, for manufacturing heavy economies that are deeply integrated into their neighbors' value chains, such as Mexico or Czechia. These are countries with a complex tradeable product sector, but as we will show, with comparatively less sophisticated research and innovation sectors.

That is why the recent literature in economic complexity has begun using data on patents[35], employment[36,37], and research papers[38], to estimate the complexity of countries, cities, and regions. But these metrics are rarely combined in work using complexity methods to explain the geography of inclusive green growth[39,40].

To bridge this gap, we introduce a multidimensional approach to economic complexity that combines data on the geography of exports by product, patents by technology, and scientific publications by field of research. We use this approach to explain variations in economic growth, income inequality, and greenhouse emissions.

But why would the complexity of economies explain the geographic variation of inclusive green growth?

Consider the exports of X-rays and iron ore. The contribution of these exports to GDP is equal to their export value, but their contribution to economic complexity is quite different since X-rays are a high complexity product (pushing the complexity of an economy up) while iron ore is not. In fact, according to data on the Observatory of Economic Complexity[41], X-rays have a product complexity of 1.46 whereas iron ore has a product complexity of -1.84. Since complexity metrics are related to the sophistication of economic activities, a unit of GDP generated through the production of X-rays should be cleaner and more inclusive than a unit of GDP generated through iron ore mining.

This is an opportunity cost argument. Consider the economies of Switzerland, Singapore, or Sweden. These economies engage an important part of their population in relatively sophisticated activities (they are high complexity economies). While these activities have an associated level of emissions, an ability to contribute to economic growth, and affect the way in which income is distributed, complexity metrics do not capture their contribution to these outcomes in absolute terms. Instead, they capture their contribution relative to other activities. In simple terms, they capture the idea that, in the absence of X-ray equipment production, some of these engineers would be involved in mining.

Thus, we expect measures of economic complexity to help us explain variations in macroeconomic outcomes if they are effective at capturing information about economic structures. Also, we expect these methods to benefit from data about multiple activities (e.g. trade, patents, and research).

In fact, we find that the combination of trade, patent, and research publication data significantly and robustly improves the ability of economic complexity methods to explain inclusive green growth. In particular, metrics of trade and technology complexity—but not of research complexity—combine to explain international differences in economic growth and income inequality. In addition, countries that score high in all three metrics tend to have lower emission intensities. We also find that there is a negative interaction between trade and technology complexity when explaining growth, indicating that some of the information captured by these two metrics is redundant (and hence the metrics are partly substitutes). However, we find no negative interaction when explaining income inequality. Finally, when it comes to emissions, we find that interaction terms dominate the models, meaning that countries with lower emissions tend to score high in all complexity metrics. These results are robust to a variety of controls (total exports, number of patents, number of publications, GDP per capita, etc.) and are confirmed by an instrumental variable robustness check where the complexity of each country is replaced by the average of its most structurally similar non-neighbors.

These findings expand the knowledge about the role of economic complexity in inclusive green growth and help open a new avenue of research that explores the combination of multiple sources of data to create improved policies for achieving sustainable development.

**Results**

We use the Economic Complexity Index (ECI) method (see the Methods Section) to estimate three separate metrics of economic complexity: 1) trade complexity (*ECI (trade)*), using export data from the Observatory of Economic Complexity[41], 2) technology complexity (*ECI (technology)*), using patent applications data from World Intellectual Property Organization's International Patent System; and 3) research complexity (*ECI (research)*), using published documents data from SCImago Journal & Country Rank portal[42]. We investigate their individual and combined contribution to explaining international variations in economic growth, income inequality, and emissions intensity. The economic growth and emissions intensity of a country are estimated using GDP and emissions data from the World Development Indicators[43], whereas the income inequality data are taken from the Estimated Household Income Inequality[44,45]. See Supplementary Information (SI) Supplementary Note 1 for a detailed description of the data.

**International differences in Multidimensional Economic Complexity**

Figure 1a presents three binary specialization matrices ($M_{cp}$) for countries' exports by product, patents by technology, and publications by research area for the year 2014. Colored dots indicate that a country is specialized in an activity, i.e., that the share of its exports, patent applications, or number of papers are larger than the share of that activity in the world output ($M_{cp} = 1$).

Figs. 1b and c compare the three ECI rankings and Fig. 2 compares the ECI values. The figures show that, while the ECI metrics are correlated, they recover the qualitative behavior motivating this research: that trade-based measures of complexity tend to underestimate the complexity of some countries that are far from global markets (e.g., Australia and New Zealand) and overestimate the complexity of some manufacturing economies (e.g., Mexico and Czechia).

For example, consider Mexico (MEX), Czechia (CZE), Australia (AUS), and New Zealand (NZL). Mexico and Czechia rank high in trade complexity (MEX is #24 and CZE is #6) but lower in technology and research complexity. Mexico drops to #26 in the technology rankings and to #44 in the research rankings, whereas Czechia ranks #22 in technology and #34 in research. This could be explained in part by the fact that Mexico's and Czechia's exports do not serve global markets, but the value chains of their neighbors. In fact, over the last decade, 76% of Mexico's exports went to the United States (ranked #12 in trade complexity) and 31% of Czechia's exports went to Germany (ranked #3 in trade complexity)[41]. For comparison, the number one export destination of the median country represents 21% of its total exports, meaning that the United States and Germany are, respectively, heavily overrepresented in Mexico and Czechia's exports (see also SI Supplementary Note 1).

Australia and New Zealand show the opposite pattern. Both countries rank relatively low in trade complexity (AUS is #76 and NZL is #47) but are global leaders in technology and research rankings. Australia ranks #8 in technology complexity and #3 in research complexity, while New Zealand ranks respectively #12 and #10. This is explained in part by the fact that Australia and New Zealand are far from global markets and export commodities to China, a country that is over 7,000 kilometers away from their capitals. Thus, trade data miss key aspects of the complexity of these economies that is recovered using data on patents and research.

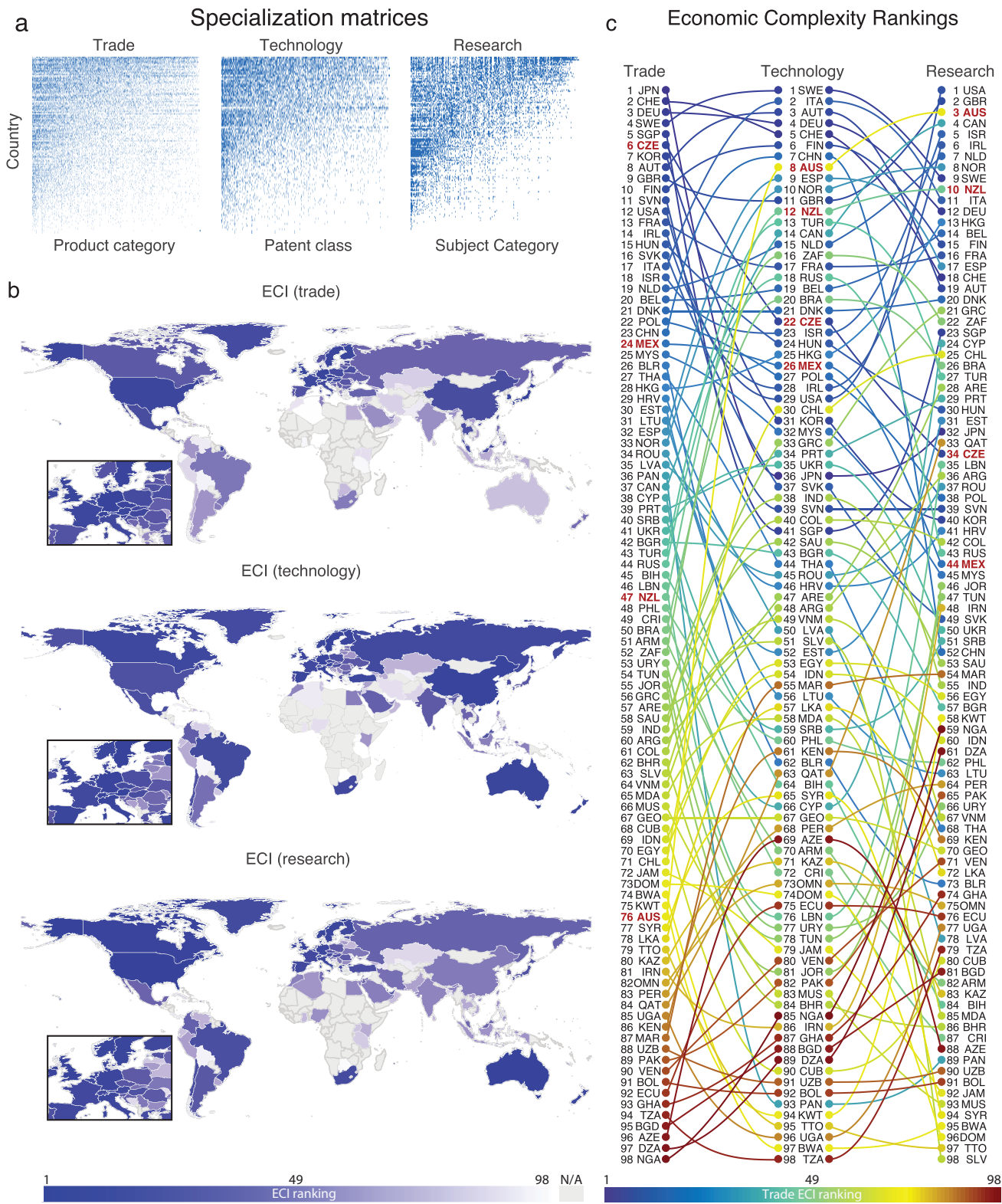

**Fig. 1. Multidimensional Economic Complexity. a** Specialization matrices of countries considering exports by product, patents by technology, and publications by subject category. **b** Maps showing the rankings of ECI (trade), ECI (technology), and ECI (research). **c** Comparison between the ECI rankings of countries based on ECI (trade), ECI (technology), and ECI (research). **a-c** All data is from the year 2014.

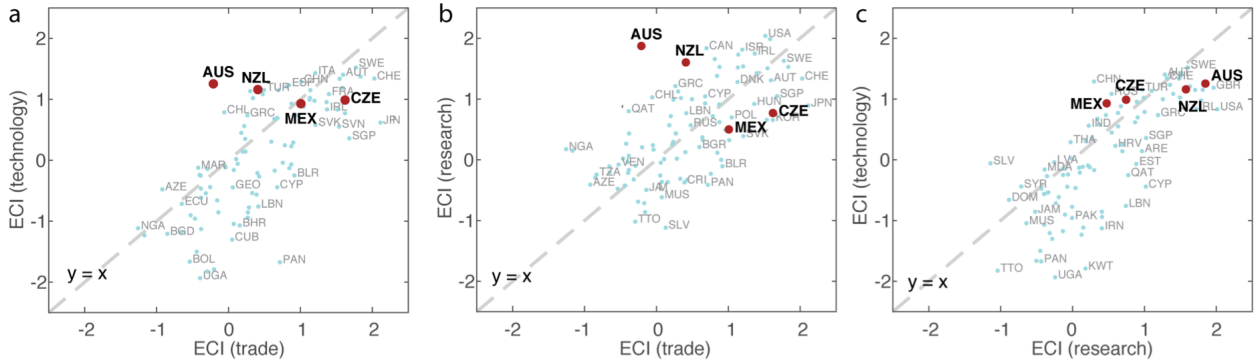

**Fig. 2. Comparison between trade, technology, and complexity ECI using 2014 data.** **a** Scatterplot for the relationships between ECI (trade) and ECI (technology) ($R^2 = 0.51$, $p$-value<$10^{-12}$), **b** ECI (trade) and ECI (research) ($R^2 = 0.44$, $p$-value<$10^{-12}$), and **c** ECI (research) and ECI (technology) ($R^2 = 0.54$, $p$-value < $10^{-12}$).

**Multidimensional Economic Complexity and Inclusive Green Growth**

Next, we explore how the information provided by technology and research complexity combines with trade complexity to explain international variations in future economic growth, income inequality, and greenhouse gas emissions. We investigate this question piecemeal, first by employing models that include each variable separately, then, by including variables together, and finally, by using interaction terms. In addition, we test for robustness by using an instrumental variable approach and several controls.

We follow the literature[15,30,32] and set up panel regressions of the form

$$y_{ct} = f(ECI_{ct}^d) + a^T X_{ct} + \mu_t + b_0 + e_{ct},$$

where $y_{ct}$ is the dependent variable for country $c$ in year $t$ (economic growth, income inequality, emission intensity), $f(ECI_{ct}^d)$ is a function of the three complexity indices ($d$ = trade, technology, or research), $X_{ct}$ is a vector of control variables that account for other key factors (e.g. population, GDP per capita, etc.), $\mu_t$ describes time-fixed effects to account for any unobserved period-specific factors, $b_0$ is the intercept, and $e_{ct}$ is the error term, (see Supplementary Information (SI) Supplementary Note 1 for more information about the data and SI Supplementary Note 2 about the regression specification).

We then validate and select a separate "multidimensional model" for growth, inequality, and emission intensity using the following criteria. First, the multidimensional model must lead to the largest significant increase in explanatory power over the baseline model (given by the coefficient of determination adjusted-$R^2$ and validated by a Wald F-test). The baseline models are defined in each respective section. Second, in the multidimensional model all included complexity coefficients (individual and interaction terms) must be statistically significant considering clustered standard errors. Finally, we require the model to pass two types of robustness checks.

First, we check for robustness by exploring whether the effects hold after including additional variables. These are measures of size (population), human capital (years of education), dependence on natural resources (natural resource exports per capita), and metrics of the intensity of each respective output (exports per capita, patent applications per capita, and number of research documents per capita). We also try alternative definitions of complexity[40,46,47] and check whether the results hold for non-complexity metrics of economic structure, such as measures of market concentration (Shannon information entropy and the Herfindahl–Hirschman index (HHI)) (see SI Supplementary Note 3). Unfortunately, because of limited time series data, our panels do not allow us to control for country-fixed effects in the growth and inequality model (e.g. the growth model consist of only two time periods). We do add country-fixed effects as a robustness check to the emission intensity model. We call the model with all significant and robust explanatory variables the "final model." This is the best model at explaining variations in economic growth, income inequality, and emission intensity.

Second, we also use an instrumental variable approach where complexity values are replaced by the average complexities of three similarly specialized non-neighboring countries. This is designed to address the possibility that the relationship between economic complexity and the studied macroeconomic outcomes may be endogenous when local conditions lead to both higher complexity and better outcomes. By replacing complexity estimates with the average of non-neighboring countries with similar specialization patterns, we decouple complexity estimates from other local conditions.

**Economic growth:** Economies with high levels of complexity relative to their GDP per capita are known to experience faster long-term economic growth[15–18,21,24,48–51]. The idea is that higher complexity economies can participate in sophisticated sectors that support higher wages. But while this relationship has been repeatedly validated using trade[15–17,51] and employment data[21,37], there is a lack of research exploring whether technology and research complexity play a similar role.

Here we test the effect of trade, technology, and research complexity on economic growth by looking at the 10-year annualized GDP per capita growth (in constant PPP dollars) using two periods 1999-2009 and 2009-2019. The baseline model includes the log of the initial GDP per capita (in constant PPP dollars) and time fixed-effects (see SI Supplementary Note 4). This captures Solow's idea of economic convergence[52] (baseline model is presented in column 1 of table 1, adjusted $R^2 = 0.25$).

Table 1 shows the effect of the three complexity metrics and their interactions. We find that trade complexity is a significant and positive predictor of economic growth (column 2, adjusted $R^2 = 0.34$) and that technological complexity has a similar explanatory power (column 3, adjusted $R^2 = 0.33$). Research complexity, however, is not significantly related to future economic growth (column 4, adjusted $R^2 = 0.24$). We also find that technological complexity significantly enhances the ability of trade complexity to explain future economic growth (columns 5-8 of table 1). This effect increases when we interact them (columns 9-12 of table 1), leading to our multidimensional model (column 9). The multidimensional model leads to an improvement in explanatory power over the trade complexity regression of 7 percentage points (adjusted $R^2 = 0.41$). In this regression, both trade and technology complexities have a positive impact on growth, but their interaction term is negative and significant, suggesting a strong substitute relationship. In general, countries with larger trade ECI than technology ECI experience higher growth, but countries that score poorly in both dimensions experience lower growth. Also, the F-statistics imply that the coefficients of the trade and technology ECI remain significant even when including the log of population and the log of human capital. In addition, the multidimensional model clearly outperforms similar models based on production intensity, measures of diversification, and other measures of complexity. Trade and technology ECIs also outperform measures of concentration (entropy and Herfindahl-Hirschman). The final model includes the multidimensional ECI (trade, technology, and their interaction), the Solow term (GDP per capita), the log of the human capital, and the log of natural resource exports per capita (see SI Supplementary Note 4).

# Table 1. Multidimensional Complexity and Economic Growth

| | (1) | (2) | (3) | (4) | (5) | (6) | (7) | (8) | (9) | (10) | (11) | (12) |
|---|---|---|---|---|---|---|---|---|---|---|---|---|
| | *Dependent variable:* Annualized GDP pc growth (1999-09, 2009-19) | | | | | | | | | | | |
| ECI (trade) | | 5.658*** | | | 4.006*** | 5.981*** | | 4.022*** | 12.255*** | 12.134*** | | 17.331 |
| | | (1.172) | | | (1.405) | (1.274) | | (1.469) | (2.955) | (3.863) | | (10.986) |
| ECI (technology) | | | 2.577*** | | 1.351 | | 3.323*** | 2.098** | 9.099*** | | 5.483** | 12.756 |
| | | | (0.745) | | (0.893) | | (0.765) | (0.928) | (2.497) | | (2.647) | (10.129) |
| ECI (research) | | | | 1.184 | | -0.890 | -2.541 | -2.563 | | 6.318 | 0.380 | -5.469 |
| | | | | (1.724) | | (1.568) | (1.617) | (1.607) | | (4.847) | (4.282) | (12.688) |
| ECI (trade) x ECI (technology) | | | | | | | | | -12.260*** | | | -22.692 |
| | | | | | | | | | (3.656) | | | (15.524) |
| ECI (trade) x ECI (research) | | | | | | | | | | -10.111* | | -3.392 |
| | | | | | | | | | | (5.831) | | (20.029) |
| ECI (research) x ECI (technology) | | | | | | | | | | | -3.856 | 0.435 |
| | | | | | | | | | | | (4.556) | (16.737) |
| ECI (trade) x ECI (research) x ECI (technology) | | | | | | | | | | | | 9.443 |
| | | | | | | | | | | | | (25.142) |
| Controls | ✓ | ✓ | ✓ | ✓ | ✓ | ✓ | ✓ | ✓ | ✓ | ✓ | ✓ | ✓ |
| Log of population F-Statistic | | 34.15*** | 22.29*** | 0.88 | | | | | 14.87*** | | | |
| Log of human capital F-Statistic | | 7.81*** | 5.13** | 0.50 | | | | | 5.06*** | | | |
| Log of natural resource exports per capita F-Statistic | | 32.51*** | 15.08*** | 1.46 | | | | | 12.52*** | | | |
| Log of production intensity F-Statistic | | 23.09*** | 3.49* | 0.49 | | | | | 27.50*** | | | |
| HHI F-Statistic | | 8.96*** | 5.05** | 0.15 | | | | | 13.70*** | | | |
| Entropy F-Statistic | | 8.61*** | 4.91** | 0.48 | | | | | 13.60*** | | | |
| Log of Fitness F-Statistic | | 6.79** | 0.34 | 2.31 | | | | | 9.10** | | | |
| i-ECI F-Statistic | | 8.94*** | 4.26** | 0.01 | | | | | 21.80*** | | | |
| Instrumental variables model F-Statistic | | 19.50*** | 8.2*** | 0.48 | | | | | 20.00*** | | | |
| Observations | 152 | 152 | 152 | 152 | 152 | 152 | 152 | 152 | 152 | 152 | 152 | 152 |
| R² | | 0.256 | 0.358 | 0.341 | 0.260 | 0.373 | 0.361 | 0.355 | 0.388 | 0.427 | 0.377 | 0.361 | 0.452 |
| Adjusted R² | | 0.246 | 0.345 | 0.327 | 0.245 | 0.356 | 0.343 | 0.338 | 0.367 | 0.407 | 0.356 | 0.339 | 0.417 |

Notes: Each regression includes period fixed effects. Clustered standard errors in brackets. *p<0.1, **p<0.05, ***p<0.01. The F-statistics for the models in columns 1-3 were estimated using models given in Supplementary Tables 1-3. The F-statistics for the model in column 9 were estimated using models estimated in Supplementary Tables 4-9. The vertical cell borders highlight the multidimensional model.

**Income inequality:** Economies with less complex trade structures are also known to exhibit higher levels of income inequality[25–27]. The idea is that firms operating in knowledge intense activities promote inclusive institutions because of their need to attract and retain talent. Firms in less complex activities, do not face this constraint, and benefit from a more extractive institutional environment. Thus, we should expect higher levels of economic complexity to be associated with lower levels of inequality.

To explore the ability of multidimensional complexity to explain variations in income inequality we model an economy's Gini coefficient, a standard measure of inequality. Larger values for the Gini coefficient indicate larger income inequality. We divide the data into four four-year panels: 1996-1999, 2000-2003, 2004-2007, 2008-2011, and 2012-2015 and set up a baseline model given by the Kuznets curve: the idea that as an economy develops market forces first increase and then decrease income inequality[53] (Gini ~ GDP per capita, its square, and time fixed-effects, see SI Supplementary Note 5).

We find that trade and technology ECIs are significant and negative predictors of income inequality with, respectively, adjusted $R^2 = 0.54$ (column 2 of table 2) and $R^2 = 0.48$ (column 3 of table 2). Trade and technology ECIs also outperform measures of concentration (entropy and Herfindahl-Hirschman, see SI Supplementary Note 5). Moreover, they provide an important improvement over the baseline model, which has an adjusted $R^2 = 0.33$ (column 1 of table 2). Research ECI, however, is only a minor predictor of income inequality providing little improvement to the explanatory power (adjusted $R^2 = 0.36$, column 4 of table 2).

Again, the model combining trade and technology provides the best explanatory power (columns 5-11 of table 2). However, the interaction term between trade and technology is not significant, meaning that the two complexities do not behave as substitutes or complements. The multidimensional model is given by column 5 of table 2 (adjusted $R^2 = 0.56$). This model is also robust when including the log of population and log of human capital and outperforms similar models based on production intensity, measures of diversification, and other measures of complexity. The final model—the one that best explains international variations in income inequality—includes the log of population and human capital in addition to the multidimensional ECI and the Kuznets term, (see SI Supplementary Note 5).

## Table 2. Multidimensional Complexity and Income Inequality

| | (1) | (2) | (3) | (4) | (5) | (6) | (7) | (8) | (9) | (10) | (11) | (12) |
|---|---|---|---|---|---|---|---|---|---|---|---|---|
| | *Dependent variable:* Gini coefficient (1996-99, 2000-03, 2004-07, 2008-11, 2012-15) | | | | | | | | | | | |
| ECI (trade) | | -23.543*** | | | -17.902*** | -23.116*** | | -17.778*** | -9.279 | -21.289** | | -18.449 |
| | | (5.285) | | | (5.495) | (5.328) | | (5.565) | (9.547) | (9.295) | | (26.796) |
| ECI (technology) | | | -11.211*** | | -5.269** | | -12.317*** | -6.216** | 1.208 | | -3.964 | 11.923 |
| | | | (2.715) | | (2.685) | | (2.582) | (2.528) | (6.101) | | (5.742) | (25.183) |
| ECI (research) | | | | -7.654 | | -1.336 | 3.400 | 2.783 | | 0.649 | 16.132 | 21.084 |
| | | | | (5.310) | | (4.017) | (3.665) | (3.386) | | (8.981) | (10.310) | (31.233) |
| ECI (trade) x ECI (technology) | | | | | | | | | -11.449 | | | -8.570 |
| | | | | | | | | | (10.851) | | | (41.156) |
| ECI (trade) x ECI (research) | | | | | | | | | | -2.990 | | -0.339 |
| | | | | | | | | | | (13.710) | | (54.046) |
| ECI (research) x ECI (technology) | | | | | | | | | | | -16.026 | -31.788 |
| | | | | | | | | | | | (12.629) | (42.894) |
| ECI (trade) x ECI (research) x ECI (technology) | | | | | | | | | | | | 12.933 |
| | | | | | | | | | | | | (69.991) |
| Controls | ✓ | ✓ | ✓ | ✓ | ✓ | ✓ | ✓ | ✓ | ✓ | ✓ | ✓ | ✓ |
| Log of population F-Statistic | | 30.10*** | 38.50*** | 4.10** | 59.00*** | | | | | | | |
| Log of human capital F-Statistic | | 16.50*** | 19.40*** | 1.90 | 27.10*** | | | | | | | |
| Log of natural resource exports per capita F-Statistic | | 31.30*** | 18.00*** | 2.50 | 37.10*** | | | | | | | |
| Log of production intensity F-Statistic | | 19.70*** | 2.10 | 0.33 | 6.40** | | | | | | | |
| HHI F-Statistic | | 10.40*** | 8.90*** | 1.60 | 5.70* | | | | | | | |
| Entropy F-Statistic | | 9.80*** | 8.90*** | 1.50 | 11.10** | | | | | | | |
| Log of Fitness F-Statistic | | 5.50** | 9.10*** | 0.75 | 10.90*** | | | | | | | |
| i-ECI F-statistic | | 16.00*** | 7.20*** | 0.64 | 23.50*** | | | | | | | |
| Instrumental variables model | | 18.90*** | 11.70*** | 2.50 | 44.10*** | | | | | | | |
| Observations | 332 | 332 | 332 | 332 | 332 | 332 | 332 | 332 | 332 | 332 | 332 | 332 |
| $R^2$ | 0.346 | 0.551 | 0.496 | 0.371 | 0.573 | 0.552 | 0.500 | 0.575 | 0.576 | 0.552 | 0.508 | 0.590 |
| Adjusted $R^2$ | 0.334 | 0.542 | 0.485 | 0.358 | 0.562 | 0.541 | 0.487 | 0.563 | 0.564 | 0.540 | 0.494 | 0.573 |

Notes: Each regression includes period fixed effects. Clustered standard errors in brackets. *p<0.1, **p<0.05, ***p<0.01. The F-statistics for the models in columns 1-3 were estimated using models given in Supplementary Tables 10-12. The F-statistics for the model in column 9 were estimated using models estimated in Supplementary Tables 13-18. The vertical cell borders highlight the multidimensional model.

**Emission intensity:** Trade complexity is known to be associated with lower greenhouse gas emissions per unit of output[30] and better environmental performance[54,55]. The idea is that the emissions required to, for instance, produce a unit of GDP by extracting tin ore are larger than the emissions required to produce a unit of GDP by manufacturing metal cutting machines. Here, we explore whether the technology and research dimensions add to the ability of trade complexity to explain emission intensity by modelling the logarithm of a country's yearly greenhouse gas emissions per unit of GDP (in kilotons of $CO_2$ equivalent per dollar of GDP). Larger values represent larger emission intensity. We divide our analysis into five panels: 1996-1999, 2000-2003, 2004-2007, 2008-2011, 2012-2015, and 2016-2019. The baseline model includes the logs of the GDP per capita (constant PPP dollars), population, human capital, and natural resource exports, as well as time fixed-effects (see SI Supplementary Note 6).

Unlike in the previous two cases, here we find that individual ECI measures do not perform better than metrics of concentration (Entropy, Herfindahl-Hirschman index) and other complexity measures (Fitness) (see SI Supplementary Note 6). Nevertheless, the best multidimensional complexity model is robust and includes the three-way interaction between trade, technology, and research complexity (adjusted $R^2 = 0.40$, column 12 of table 3, Fig. 3 c). This implies that countries that score high in all dimensions (e.g., Sweden, France, Austria) have the lowest emission intensities. The final model also includes all of the variables from the baseline model: measures of population size, human capital, and natural resource exports per capita and production intensity (the model in column 11). This means that the measures of complexity explain variation in emission intensities that go beyond the variation accounted for by the natural resource export intensity of an economy.

Finally, we run an alternative model with embodied emission intensities as the dependent variable (see SI Supplementary Note 7). Embodied emissions add territorial and imported emissions and subtract exported emissions[56,57]. Thus, they are an indicator of the emissions related to local consumption. We expect embodied emissions to behave differently than emission intensities because they are a consumption indicator. Complexity metrics are estimators of productive capacities, and thus, we expect them to correlate with characteristics of an economy's productive sectors. We should not expect, however, ECI to explain consumption patterns, especially those of imported products. As expected, we do not find a robust relationship between complexity and embodied emission intensity (the correlation is positive but not robust).

## Table 3. Multidimensional Complexity and Emission Intensity

| | | | | | | | *Dependent variable:* | | | | | |
|---|---|---|---|---|---|---|---|---|---|---|---|---|
| | | | | | | | GHG emissions per GDP (1996-99, 2000-03, 2004-07, 2008-11, 2012-15, 2016-19) | | | | | |
| | (1) | (2) | (3) | (4) | (5) | (6) | (7) | (8) | (9) | (10) | (11) | (12) |
| ECI (trade) | | -1.223*** | | | -0.954** | -1.123** | | -0.939* | 0.495 | -0.202 | | -5.007*** |
| | | (0.447) | | | (0.475) | (0.455) | | (0.484) | (1.020) | (0.956) | | (1.903) |
| ECI (technology) | | | -0.646*** | | -0.358* | | -0.547** | -0.269 | 0.660 | | 0.049 | -4.059*** |
| | | | (0.217) | | (0.214) | | (0.221) | (0.221) | (0.596) | | (0.441) | (1.370) |
| ECI (research) | | | | -0.647* | | -0.478 | -0.412 | -0.390 | | 0.530 | 0.440 | -5.928*** |
| | | | | (0.340) | | (0.311) | (0.339) | (0.317) | | (0.859) | (0.611) | (1.987) |
| ECI (trade) x ECI (technology) | | | | | | | | | -1.878* | | | 6.666*** |
| | | | | | | | | | (1.097) | | | (2.419) |
| ECI (trade) x ECI (research) | | | | | | | | | | -1.511 | | 10.567*** |
| | | | | | | | | | | (1.258) | | (3.564) |
| ECI (research) x ECI (technology) | | | | | | | | | | | -1.125 | 8.607*** |
| | | | | | | | | | | | (0.799) | (2.662) |
| ECI (trade) x ECI (research) x ECI (technology) | | | | | | | | | | | | -15.268*** |
| | | | | | | | | | | | | (4.293) |
| Controls | ✓ | ✓ | ✓ | ✓ | ✓ | ✓ | ✓ | ✓ | ✓ | ✓ | ✓ | ✓ |
| Log of population F-Statistic | | 8.40 | 2.00 | 1.60 | | | | | | | | 23.30*** |
| Log of human capital F-Statistic | | 11.20 | 2.90* | 2.10 | | | | | | | | 31.90*** |
| Log of natural resource exports per capita F-Statistic | | 2.80* | 0.53 | 0.56 | | | | | | | | 18.30** |
| Log of production intensity F-Statistic | | 3.30* | 2.50 | 2.60 | | | | | | | | 15.10** |
| HHI F-Statistic | | 2.90* | 4.70** | 2.60 | | | | | | | | 24.50*** |
| Entropy F-Statistic | | 2.50 | 4.50** | 2.50 | | | | | | | | 24.10*** |
| Log of Fitness F-Statistic | | 0.50 | 10.90*** | 1.80 | | | | | | | | 21.20*** |
| i-ECI F-Statistic | | 6.90*** | 6.20** | 3.00* | | | | | | | | 33.80*** |
| Country fixed-effects F-Statistic | | 0.99 | 6.0** | 2.20 | | | | | | | | 19.10*** |
| Instrumental variables model F-Statistic | | 6.60** | 6.20** | 4.60** | | | | | | | | 24.40*** |
| Observations | 528 | 528 | 528 | 528 | 528 | 528 | 528 | 528 | 528 | 528 | 528 | 528 |
| $R^2$ | 0.302 | 0.355 | 0.339 | 0.323 | 0.364 | 0.366 | 0.346 | 0.370 | 0.380 | 0.373 | 0.353 | 0.415 |
| Adjusted $R^2$ | 0.290 | 0.342 | 0.326 | 0.309 | 0.350 | 0.352 | 0.332 | 0.356 | 0.366 | 0.359 | 0.338 | 0.397 |

Notes: Each regression includes period fixed effects. Clustered standard errors in brackets. *p<0.1, **p<0.05, ***p<0.01. The F-statistics for the models in columns 1-3 were estimated using models given in Supplementary Tables 19-21. The F-statistics for the model in column 9 were estimated using models estimated in Supplementary Tables 22-27. The vertical cell borders highlight the multidimensional model.

In Fig. 3 we summarize our empirical findings. Adding complexity metrics for technology and research can improve the ability of the regression models to explain variations in economic growth, income inequality, and emission intensity. In fact, our final models explain more than 50% of cross-country variation in economic growth and income inequality, and almost 40% of the variations in emission intensity (Figs. 3a-c), a drastic increase compared to including only trade metrics. Technology complexity adds to the ability of trade complexity to explain economic growth and income inequality, and trade, technology, and research complexity complement each other in their ability to explain greenhouse gas emissions (Figs. 3d-f). We also calculate the overall marginal effect of the different ECI coefficients by creating a multidimensional ECI by weighting each ECI coefficient according to the size of the regression coefficient in the final model, and re-estimating the final model of economic growth, income inequality and emissions intensity. The multidimensional ECI is correlated with increases in economic growth and decreases in income inequality and emissions intensity (Figs. 3g-i).

Nevertheless, we find that the individual effect of different dimensions of complexity is not always linear since complexity estimates interact. In the case of economic growth, the negative interaction suggests a mild substitution between these two variables (high complexity in exports and technology help explain growth, but there is no additional effect of scoring high on both). In the case of inequality, the effects seem to be linear and additive since the interaction term here is not significant. Finally, for emission intensities, we find significance across all interaction terms, meaning that we expect to observe lower emissions in economies that score high in the three complexity metrics. This validates the idea that complexities in different forms of activities combine to explain inclusive green growth. But are these results robust to possible omitted variables?

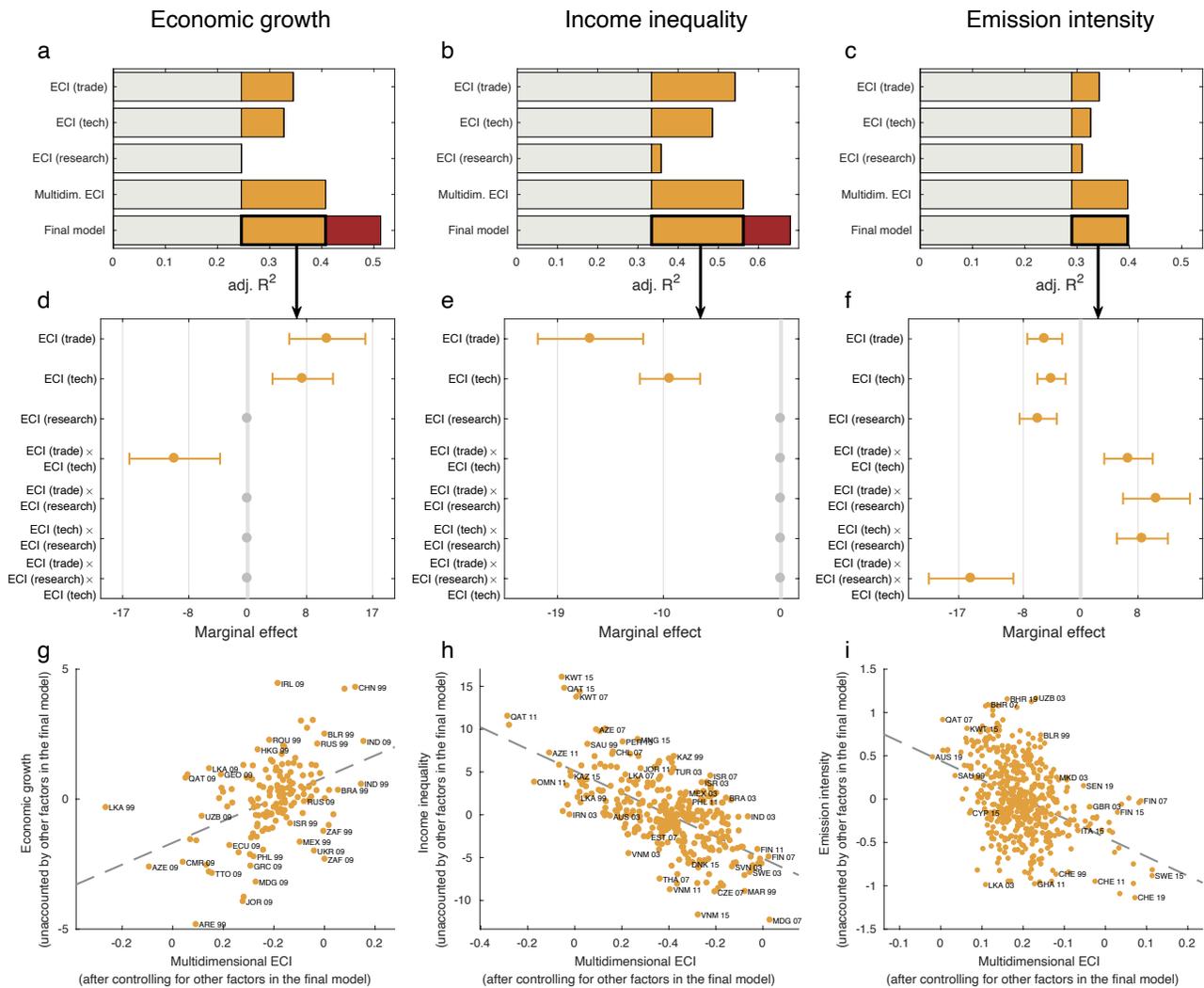

**Fig. 3. Explaining international variations in economic growth, income inequality, and emission intensity with multidimensional economic complexity.** **a-c** Contribution of the baseline, ECIs, and other covariates to the variance explained by various models (adjusted $R^2$) for **a** economic growth, **b** income inequality, and **c** emission intensity. The baseline adjusted $R^2$'s are presented in grey, the contributions of the three individual ECIs and of the multidimensional ECI in orange, and the variance explained by additional factors in the final model is shown in in red. **d-f** Error bars for the marginal effects (with 95% confidence intervals) for the ECI coefficients in the final models for **d**, economic growth (Supplementary Table 4, column 17), **e** income inequality (Supplementary Table 12, column 17) and, **f** emission intensity (Supplementary Table 21, column 14). **g-i** The conditional correlation between the multidimensional ECI (created by weighting each ECI coefficient according to the size of the regression coefficient in the final model) and **g** economic growth, **h** income inequality, and **i** emission intensity. Conditional correlations are obtained by controlling for all other factors included in the final models.

**Instrumental Variable**

To further validate these results, we pursue an instrumental variable approach where we replace a country's complexity values with those of its three most similar non-neighbors (countries with a similar specialization pattern but that do not share a land or maritime border). The idea is that there might be factors that are either local (e.g., culture, geography) or relevant only to certain dependent variables (e.g., country-specific environmental policies for GHG emission intensity) that could drive both complexity and macroeconomic outcomes. To decouple local factors and conditions from our complexity estimates, we replace the complexity values of each country with the average of the three non-neighboring countries with the most similar specialization pattern (based on the conditional probability that two countries are specialized in the same vector of activities[58] (exports, technologies, research areas), see SI Supplementary Note 8). For example, in 2014 Japan's export structure was similar to that of Germany, Great Britain, and Czechia, whereas Australia's technological structure was similar to Great Britain, Spain, and Canada. In SI Supplementary Note 8 we provide a full list of the three most similar economies in 2014 for every country and dimension used in our analysis. We find the results remain virtually unchanged, reducing the risk that the explanatory value of these complexity metrics comes from an omitted local factor (F-statistics for the Wald restriction tests are given in Tables 1-3, see also SI Supplementary Note 8 for first and second stage results).

**Discussion**

Economic complexity methods have become important tools to explain regional and international variations in inclusive green growth[59–66]. Yet, most applied work on economic complexity relies on metrics derived from trade data that are limited in their ability to capture information from non-trade activities. This can lead to distorted estimates for the complexity of certain countries and limited information about how different types of activities combine to explain variations in inclusive green growth.

Here, we combined trade, technology, and research data, to explore the role of complexity metrics in inclusive green growth. We found that technology complexity adds to the ability of trade complexity to explain economic growth and income inequality, and that trade, technology, and research complexity complement each other in their ability to explain greenhouse gas emissions. We also found that complexities expressed in different forms of activities sometimes interact. Trade and technology complexities are partly substitutes in the growth regression, but not in the inequality model. Moreover, in the emission intensities model the highest predictive power was obtained by the model with the triple interaction, meaning that lower emission intensities correlate with countries that score high in all three metrics of complexity.

But what do these results mean?

On the one hand, product exports and patent applications can be easily tied to monetary outcomes such as economic growth or income inequality (e.g., product exports generate revenues, whereas patents generate royalties). Thus, the structure of these activities should contribute directly to monetary outcomes, unlike the geography of research papers which may have a more indirect effect. Emission intensities, on the other hand, seem to correlate negatively with the presence of complexity in trade, technology, and research, suggesting that countries with lower emissions are sophisticated across these three dimensions. For instance, Australia's high emission intensity can be explained by its lack of sophistication in exports[57]. Yet, we should also expect Australia's emission intensity to be relatively low compared to countries with a similar export structure, because of Australia's high complexity in technology and research.

These results are relevant for identifying strategic areas for economic diversification and development, as they provide a more holistic target than the one provided by metrics of trade complexity alone[30,32]. In fact, much of the recent work in smart specialization has focused of single relatedness-complexity diagrams in attempts to identify activities that are both accessible and attractive. Our approach can be used to expand this in two important ways, by evaluating multiple targets and considering multiple diversification options. For instance, beyond complexity, we can evaluate the inequality and emissions implications of a new activity (this was already anticipated in Hartmann et al.[25] for inequality and in Romero and Gramkow[30] for emissions, but it has not been put together). Similarly, we can look at relatedness across a series of activities (e.g. diversification not only in product exports, but in patents and research areas). For instance, some countries may have an easier time climbing the technology ladder than the export ladder. Thus, putting these ideas together suggests a more comprehensive strategic landscape for strategic economic development, balancing multiple targets (growth, inequality, emissions) and opportunities (products, patents, and research). This should be of interest to policy makers using complexity metrics for inclusive green development and reinforce the idea that metrics of economic complexity go beyond measures of trade sophistication[33,34,64,67,68].

Yet, this approach is not without limitations.

First, patent application and research publication data also have limitations. For instance, since patent applications and research documents are usually written in English, these datasets can favor both, English speaking countries (e.g., USA, Australia) and countries with high proficiency in English (e.g., Netherlands, Sweden). Moreover, patent applications may differ from granted patents, and could potentially be used to game patent-based indicators by actors willing to submit spurious patents to increase their reported output in certain technologies. The use of patent applications, however, is common in the geography of innovation literature, and hence, our use of it makes our work comparable to previous research[40,69].

Second, there are plenty of activities that are not captured in either trade, patent, or research publication data—such as services, digital products, and cultural activities. These may capture additional aspects of the complexity of economies that would need to be included in a more comprehensive multidimensional framework[16,70,71]. Unfortunately, the current state of the art does not include internationally comparable fine-grained datasets for these additional activities (e.g. service trade data is too aggregate to approximate the productive structure of an economy, see Ref.[72] and SI Supplementary Note 9).

Third, our research is also limited by differences in the granularity of the three datasets: trade data is the most granular, with about 1,200 unique products, while research publication data involves only about 300 subject categories. This may be one of the reasons why we do not see strong effects from research complexity in economic growth and income inequality, and one of the reasons why combining these datasets into a unified matrix (e.g., by concatenating or multiplying these matrices) is non-trivial.

Fourth, these results cannot be readily generalized to other geographic scales, such as states and provinces. For instance, while future economic growth has been shown to correlate with the complexity of countries[15,17,20,51,66] and regions[70], the relationship between complexity and inequality is known to reverse at the regional scale[21,71–74]. Thus, this approach cannot tell us much about regional effects, which could be different from those observed at the international scale[21,74–77].

Fifth, our analysis is also limited by the potential multicollinearity of the variables (e.g. human capital is correlated with ECI). This multicollinearity, however, should lead to larger standard errors and would play against finding significance. Our results, however, are still robust despite this data limitation.

Finally, spelling out the implications of this multidimensional approach can be challenging. Not only because they lean on multiple targets, but because not all countries may be simultaneously sophisticated. Indeed, the international (and even regional) division of labor push us to question the possibility that all countries and regions could become equally sophisticated. Still, there is the possibility for the world to make progress in that direction. For instance, extractive activities can vary from exploitative and labor-intensive manual operations, to sophisticated and highly automated capital-intensive processes. The same applies for agriculture. Urban transportation systems can also be improved in ways that reduce emissions and travel times (e.g. electric bicycles, rail, etc.). So, while it may be hard for all economies to become sophisticated, there is plenty of room to sophisticate less advanced economies. While these increases in sophistication may not bring them to the top of the complexity ladder, they may still enable more sustainable, inclusive, and prosperous economies in the developing world.

Yet, multidimensional complexity improves upon the state-of-the-art when explaining international differences in economic growth, income inequality, and greenhouse gas emissions. These findings advance our understanding of the role of economic complexity in inclusive green growth and should motivate new research on comprehensive metrics of complexity and sustainable development.

**Methods**

Economic complexity metrics are derived from specialization matrices, summarizing the geography of multiple economic activities (using dimensionality reduction techniques akin to Singular Value Decomposition or Principal Component Analysis)[28,42]. In particular, given an output matrix $X_{cp}$, summarizing the exports, patents, or publications of an economy $c$ in an activity $p$, we can estimate the economic complexity index $ECI_c$ of an economy and the product complexity index $PCI_p$ of an activity, by first normalizing and binarizing this matrix:

$$R_{cp} = \left. X_{cp} X \middle/ X_p X_c \right., $$

$$M_{cp} = \begin{cases} 1 & if\ R_{cp} \geq 1 \\ 0 & otherwise \end{cases}, \qquad (1)$$

where muted indexes have been added over (e.g., $X_p = \sum_c X_{cp}$) and $R_{cp}$ stands for the revealed comparative advantage of economy $c$ in activity $p$. Then, we define the iterative mapping:

$$ECI_c = \frac{1}{M_c} \sum_p M_{cp} PCI_p,$$

$$PCI_p = \frac{1}{M_p} \sum_c M_{cp} ECI_c . \qquad (2)$$

That is, according to (2), the complexity of an economy $c$ is defined as the average complexity of the activities $p$ present in it (and vice-versa). The normalization steps in (1) and (2) are required to make the units of observation comparable (e.g. China and Uruguay are very different in terms of size). The solution of (2) can be obtained by calculating the eigenvector corresponding to the second largest eigenvalue of the matrix:

$$M_{cc\prime} = \sum_{pc\prime} \frac{M_{cp} M_{c\prime p}}{M_c M_p} \qquad (3)$$

Which is a matrix of similarity between economies $c$ and $c'$ normalized by the sum of the rows and columns of the binary specialization matrix $M_{cp}$ (it considers similarity among economies counting more strongly rare coincidences).

To obtain $ECI_c$, the values of the eigenvector are normalized using a z-score transformation (meaning that the average complexity is 0). In regression analyzes we further normalize the values of $ECI_c$ to be non-negative using a max-min technique (i.e., they are between 0 and 1).

We build our results using the standard definition of ECI[13,15] because of multiple reasons. First, because this is a widely used definition, it makes our results more readily comparable with previous research. Second, because it is a definition designed for data on the geography of economic activities, and focused on where activities come from instead of where they are consumed, we can apply it directly to our three datasets (without the need for special adaptations). Nevertheless, throughout the remainder of the paper we also compare our results with two alternative definitions of economic complexity, the fitness index[46,47] and the innovation-adjusted ECI[40] (i-ECI). These controls and their definition are presented in the supplementary information. We find our results to be robust to controlling for these alternative definitions.

**Acknowledgements:**

We thank Eva Coll, Ron Boschma, Koen Frenken, Eduardo Hernández, the attendees of the 2022 Economic Geography and Networks PhD school in Utrecht, and the members of the Center for Collective Learning for valuable feedback. We thank Pierre-Alexandre Balland for facilitating patent data. This project was supported by: ANR-19-PI3A-0004

**Author contribution statement**

V.S. conceptualization, methodology, software, data curation, validation, formal analysis, investigation, writing, P.K. conceptualization, methodology, formal analysis, writing, C.A.H conceptualization, methodology, formal analysis, writing, supervision.

**Competing interests**

CAH is a founder and creator of Datawheel and the OEC (oec.world). VS and PK have no competing interests.

**Data availability statement**

The data that support the findings of this study are available at:
https://doi.org/10.7910/DVN/K4MEFW

**Code availability statement**

The code needed to reproduce the results is available from V.S. upon request.

# Supplementary Information for:

# Multidimensional Economic Complexity and Inclusive Green Growth

Viktor Stojkoski, Philipp Koch, César A. Hidalgo

Center for Collective Learning, ANITI, TSE-R, IAST, IRIT, Université de Toulouse, 31000 Toulouse, France

Table of Contents

# Supplementary Note 1. Data

We analyze the spatial distribution of trade of goods, scientific knowledge, and new technologies, across 150 countries and 24 years (spanning from 1996 up to 2019). For each country and a given year, we approximate the magnitude of trade in a particular product category through the export value, the level of activity in a research area through the number of published articles in that field, and the number of innovations in a technological class via the number of patent applications in the class.

In each year, we restrict our analysis to countries which:

- had population above 1 million;
- had a total product export value of more than 1 billion USD;
- had more than 4 patent applications;
- had more than 30 scientific publications.

Adding a threshold for the minimum intensity in trade, patent applications or scientific publications is a standard in the economic complexity literature. It helps reduce the noise in the data arising from small economies whose specialization structure greatly varies over the years[1–3].

**Product exports data:** We look at country-goods associations by using international trade data with goods disaggregated into categories according to the HS4 classification. With this classification we end up with 1241 product categories. More detailed classifications which disaggregate the goods into more categories, such as the HS6, can also be used. However, then the number of categories is not comparable to the number of scientific fields and patent classes, thus making distortions in the level of the disaggregation of the different dimensions. For each year, we remove from the analysis the products whose total world exports were less than 500 000 USD. The data are taken from the Observatory of Economic Complexity [4].

Supplementary Figure 1 gives the histogram for the distribution of the share of exports to the main export partner for each country (for the period 2010-2019), which was used to argue that Mexico and Czechia are integrated into their neighbors' value chains.

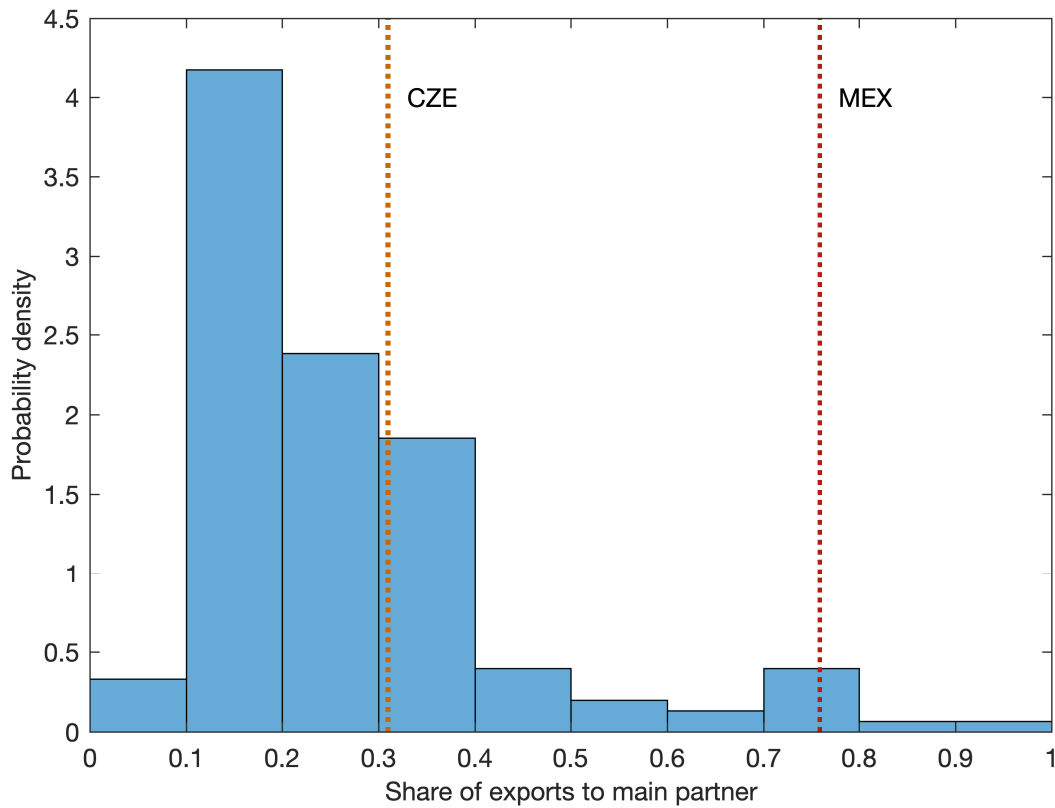

**Supplementary Figure 1. Histogram for the Distribution of the Share of Exports to the Main Export Partner (2010-2019)**

**Patent applications data:** The patent data are gathered from the World Intellectual Property Organization's International Patent System. The patent data in this system are based on the Patent Cooperation Treaty (PCT).

Each patent application under the PCT simultaneously seeks protection for an invention in many countries. This reduces the potential home bias which may arise when using patent data that come from a single Patent Office. We assign each patent to the country of residence of the inventors and to one or more distinct technology classes that reflects the scope of the claims listed in the patent document according to the Cooperative Patent Classification, disaggregated to a 4-digit level. If the inventors are coming from multiple countries, then each inventor residence is assigned as the origin of the patent application. In each year include only patent classes for which there were more than 5 applications. After clearing the data, we end up with 668 distinct technological classes.

**Research publications data:** The research article data come from the SCImago Journal & Country Rank portal[5]. The portal contains country statistics on research documents published in the Scopus database and disaggregates the scientific fields into 313 specific categories according to Scopus Classification. The Scopus database assigns each published scientific document published to each unique country listed in the affiliation of the authors of that document. Also, Scopus assigns each document to one or more categories based on the categories of the Journal in which it was published. Hence, one document can be assigned to multiple countries and multiple categories. For a given year, we set each country-research category pair to be equal to 0 if there were less than 3 documents, or if the citations of all of the published documents in the research category in the subsequent four years was less than 400 (average of 100 citations per year). This helps us reduce the potential bias that may arise due to noise in publishing in research categories with very low activity. Also, in each year we remove from the analysis each research category which had less than 30 publications.

# Supplementary Note 2.     Regression analysis setup

**Individual regressions setup:** In order to assess the ability of each dimension to explain variations in economic growth, income inequality and greenhouse gas emissions, we estimate period fixed effects panel regression models of the form

$$y_{ct} = b_0 + \boldsymbol{a}^T X_{ct} + bECI_{ct}^d + \mu_t + e_{ct},$$

where $y_{ct}$ is the dependent variable (economic growth, income inequality or extent of greenhouse gas emissions) for country $c$ in year $t$, $b_0$ is the intercept term, $X_{ct}$ is a vector of control independent variables that account for observed factors that are not related with the economic complexity. The coefficient $b$ is of particular interest to us as it is an estimate of the marginal effect of the economic complexity of the country in dimension $d$. The $\mu_t$ coefficient are period fixed effects that help to control for any unobserved factors that are period-specific and apply to all countries, and $e_{ct}$ is the error term.

**Interactions regressions setup:** To infer how different forms of output (trade, technology, and research) combine to help explain geographic variations in economic growth, income inequality, and greenhouse gas emissions, we resort to three different specifications of interaction regression analyses. In the first specification, we assume that there is no interaction between the dimensions and that they share an additive relationship in explaining the economic outcome. The regression form of this specification is

$$y_{ct} = b_0 + \boldsymbol{a}^T X_{ct} + \sum_d b_d ECI_{ct}^d + \mu_t + e_{ct},$$

Where $d$ is a superscript used to differentiate between trade ECI, technology ECI, and research ECI. We conduct four different no-interaction regressions, depending on which dimensions are included in the analysis. They are 1) trade and technology, 2) trade and research, 3) technology and research, and 4) trade, technology, and research.

In the second setup, we study the pairwise relationship between two dimensions $d_1$ and $d_2$. This specification includes the interaction term between the two dimensions and is formally written as

$$y_{ct} = b_0 + \boldsymbol{a}^T X_{ct} + b_1 ECI_{ct}^{d_1} + b_2 ECI_{ct}^{d_2} + b_{12} ECI_{ct}^{d_1} \times ECI_{ct}^{d_2} + \mu_t + e_{ct}.$$

The interaction coefficient $b_{12}$ allows us to infer how the two dimensions combine in explaining the variations of growth, inequality, or emission intensity. Specifically, if $b_{12}$ is significant and has the same sign as $b_1$ and $b_2$, then it is said that the two dimensions are complements in explaining the economic outcome (economic growth, income inequality, or greenhouse gas emissions). If $b_{12}$ is significant and negative then the dimensions are substitutes, and if it is insignificant then there is no relationship between the dimensions. Given that we investigate the performance of three dimensions of economic complexity, we end up doing three pairwise regressions: 1) trade and technology, 2) trade and research, and 3) technology and research interactions.

In the third specification, we study the three-way interaction between every dimension. In this case, the regression is specified as

$$y_{ct} = b_0 + \boldsymbol{a}^T X_{ct} + b_1 ECI_{ct}^{d_1} + b_2 ECI_{ct}^{d_2} + b_3 ECI_{ct}^{d_1} + b_{12} ECI_{ct}^{d_1} \times ECI_{ct}^{d_2}$$

$$+ b_{13} ECI_{ct}^{d_1} \times ECI_{ct}^{d_3} + b_{23} ECI_{ct}^{d_2} \times ECI_{ct}^{d_3} + b_{123} ECI_{ct}^{d_1} \times ECI_{ct}^{d_2} \times ECI_{ct}^{d_3}$$

$$+ \mu_t + e_{ct}.$$

In this specification if $b_{kl}$ and $b_{klm}$ are significant and have the same sign as $b_k$ and $b_l$, then it is said that dimensions $k$ and $l$ are complements in explaining the economic outcome (economic growth, income inequality, or greenhouse gas emissions). Otherwise, their relationship is dependent on the third dimension $m$ and may range from complementary to substitute or no relationship.

To compare the performances of different models, we use the adjusted coefficient of determination $R^2$, a standard measure for the explanatory power of a model. It's magnitude ranges between 0 and 1, with higher values implying that one can predict a higher amount of the variation in the dependent variable from the independent variables. We select the model that has the highest adjusted $R^2$ and in which every included complexity metric is significant as the *multidimensional ECI* model.

## Supplementary Note 3.  Robustness check setup

We check the robustness of the multidimensional economic complexity regression model in explaining variations in three different ways.

First, we add 1) the log of the population, 2) the log of the initial human capital to the regressions, and 3) the log of natural resource exports per capita to the regression models. These variables may affect the dependent variable but are not related with the complexity of the economy. Second, we compare the multidimensional economic complexity regression model to alternate models that account for the intensity and concentration of production, and to alternate models for economic complexity. Finally, we also add country-fixed effects in the models but do not expect our results to remain significant in this specification because our models are limited in terms of time-series data availability to only few time periods. Namely, most of our models involve panels with few periods (e.g., 2 periods for growth, 5 for income inequality). In this case, country-fixed effects take too many degrees of freedom away (basically half of them in the case of the growth model) and as a result reduce the variation in the data.

**Additional explanatory variables robustness check setup:** We add three possible additional explanatory variables to the regression specification (separately and together): 1) the log of the population, 2) the log of the initial human capital, and 3) the log of natural resource exports per capita to the regressions.

The first variable, the log of the initial population is our measure for the "size" of the economy. The data for this variable are taken from the World Bank's World Development Indicators database and are available at

https://data.worldbank.org/indicator/SP.POP.TOTL .

The second variable, the log of the initial human capital, is an aggregate measure for the "formal knowledge" in the country. We quantify the human capital using the human capital index provided by the Penn World Tables. The index is based on data for the average years of schooling of the population[6], and is available at

https://www.rug.nl/ggdc/productivity/pwt/?lang=en .

The last variable, the log of natural resource exports per capita, is a measure for the extent to which a country's economy depends on natural resources. Countries that are endowed with more natural resources usually grow faster, but also have larger emissions. This variable is estimated using data from the Observatory of economic complexity and using the methodology given in Ref.[7].

**Comparison with alternate models' setup:** We compare the multidimensional ECI model to five alternate regression models based on: 1) the intensity of a country in exports, patent applications and published documents (Intensity), 2) the diversification of the exports, patent applications and published documents based on the Herfindahl-Hirschman index (HHI), 3) the concentration of the exports, patent applications and published documents based on Shannon's Information entropy (Entropy), 4) the complexity of a country in terms of exports, patent applications and published documents based on the Fitness indicator (Fitness), and 5) the innovation-adjusted Economic Complexity Index (i-ECI). The first three models can go in tandem with economic complexity indicators as they quantify different aspects of economic outputs (production intensity and production concentration), whereas the last two are substitute models that use different mathematical definitions for calculating economic complexity.

The first four models have indicators that are defined on a dimension level. Therefore, we use the same procedure as ECI and estimate a multidimensional Intensity, HHI, Entropy, and Fitness models (the model which has the highest adjusted $R^2$ and all included coefficients are significant) in which the variables are normalized to be between 0 and 1 and compare them with the multidimensional ECI model. The Intensity, HHI, Entropy, and Fitness indicators for each dimension are estimated using the data described in Supplementary Note 1. The last model, the i-ECI, combines patent and trade data into a single indicator for complexity. To be comparable with the data used in the other models, for this model we also add the research ECI as an additional covariate.

In what follows, we describe in short how we define the indicators of the five alternate models.

**Production intensity:** The production intensity describes the "aggregate" output of each dimension. By comparing the multidimensional ECI model to the production intensity allows us to investigate whether the structure of the dimension is more related and/or offers different information than the aggregate output in explaining economic growth, income inequality, and emission intensity.

Formally, we define the production intensity as the log of exports per capita, log of patent applications per capita, and the log of the number of published scientific documents per capita.

**Herfindahl-Hirschman index:** The Herfindahl-Hirschman index $HHI_c$ of country $c$ is defined as

$$HHI_c = \sum_p (S_{cp})^2,$$

where $S_{cp} = X_{cp}/\sum_p X_{cp}$ is the share of output belonging to activity $p$. The index ranges between 0 and 1, with lower HHI values suggesting higher diversification.

**Shannon's Information entropy:** The Entropy $E_c$ of country $c$ is defined as

$$E_c = -\sum_p S_{cp} \log S_{cp}.$$

High entropy values are characteristic of diversified economies, whereas low entropy values are associated with economies whose production is concentrated in a small number of activities.

**Fitness:** The Fitness $F_c$ of a country is an alternate measure for the complexity of an economy and it is estimated as the limiting value of the following coupled equations

$$\widetilde{F}_c^{(n)} = \sum_p M_{cp} Q_p^{(n-1)}; \quad F_c^n = \frac{\widetilde{F}_c^{(n)}}{\text{mean}(\widetilde{F}_c^{(n)})},$$

$$\widetilde{Q}_p^{(n)} = \frac{1}{\sum_p M_{cp} \frac{1}{F_c^{(n-1)}}}; \quad Q_p^n = \frac{\widetilde{Q}_p^{(n)}}{\text{mean}(\widetilde{Q}_p^{(n)})},$$

where $F_c^n$ is the fitness of country $c$ in step $n$ and $Q_p^n$ is the complexity of activity $p$ in step $n$. Because the distribution of Fitness is right-skewed, in regression analysis we use its logarithmic value.

**innovation-adjusted ECI:** The i-ECI is an alternate metric for complexity that adds patent data into the measurement of product-based comparative advantage. The method is premised on the idea that an activity or a location are more technologically sophisticated if the activity itself requires more innovation or if the location is more capable of incorporating innovation into that activity. Therefore, under this approach patent applications are assigned to countries where inventors seek protection from. This is slightly different to our approach, where patent applications are assigned to the countries of the inventors. In fact, with our data we cannot use the same analogy as in the original definition because patent applications under PCT simultaneously seek protection in every country that signed the treaty. Having in mind this limitation of the data, we estimate the i-ECI and use it as an additional robustness check to our results.

i-ECI is calculated using the same formula as the original ECI, just with a slight caveat that "corrects" the value $X_{cp}$ of exports of a country in a product by the attractiveness of that country to patent in the given product. The steps needed for this process are as follows. First, to adjust the attractiveness of a product category for patenting in it we need to concord the patents to be in the same classification as the products. This is done using the correspondence between patent classification (CPC4) and product classification (HS6) developed by Lybbert and Zolas[8]. Second, we use the results to calculate the revealed comparative advantage of patents as

$$R_{cp}^{pat} = \left. X_{cp}^{pat} X^{pat} \middle/ X_p^{pat} X_c^{pat} \right.,$$

where, $X_{cp}^{pat}$ is the number of patents of country $c$ in a product $p$ and muted indexes have been added over. Also, we calculate the share $\rho_p^{pat}$ of total global patent applications received by product $p$ as

$$\rho_p^{pat} = \left. X_p^{pat} \middle/ X^{pat} \right.,$$

We use these two variables to calculate innovation-adjusted trade weights as

$$w_{cp}^{pat} = \frac{R_{cp}^{pat} - \min_c(R_{cp}^{pat})}{\max_c(R_{cp}^{pat}) - \min_c(R_{cp}^{pat})} \times \frac{\rho_p^{pat} - \min_p(\rho_p^{pat})}{\max_p(\rho_p^{pat}) - \min_p(\rho_p^{pat})},$$

where $\min_z$ ($\max_z$) are the minimum (maximum) operation over all items in the set $z$.

The innovation-adjusted exports $\tilde{X}_{cp}$ that are used to calculate the revealed comparative advantage for products and subsequently the i-ECI are

$$\tilde{X}_{cp} = X_{cp} w_{cp}^{pat}.$$

# Supplementary Note 4. Economic growth regression analysis setup and robustness checks

**Economic growth regression setup:** In the economic growth regression analysis, the dependent variable is defined as the 10-year annualized growth rate of the GDP per capita in constant 2017 dollars adjusted for power purchasing parity (PPP), i.e.,

$$y_{ct} = 100 \times \frac{1}{\Delta} \log\left(\frac{GDP_{ct+\Delta}}{GDP_{ct}}\right),$$

where $\Delta = 10$. In the regressions, as a control variable we include only the log of the initial GDP per capita in constant 2017 dollars (PPP), $GDP_{ct}$. In the economic development literature, this variable is known as the Solow term[9]. It is used to account for the "convergence" phenomenon which suggests that, on average, poorer countries should grow faster than rich countries and thus catch up. The data for the GDP per capita were taken from the World Bank's World Development Indicators database and are available at

https://data.worldbank.org/indicator/NY.GDP.PCAP.PP.KD .

The regression analysis is focused on two time periods 1999-2009 and 2009-2019. We focus on these years as they are they encompass the longest periods for which all the data are available. Moreover, this way we are the closest to constructing regressions that are consistent with the "original" growth regression analysis performed in Ref.[2] with trade data (there the periods were 1978-1988,1988-1998,1998-2008).

**Economic growth individual regressions:** In Supplementary Tables 1-3, we present the results for the individual economic growth regressions for ECI (trade), ECI (technology), and ECI (research), respectively. These results were used to estimate the F-statistics for testing the robustness of the individual regressions discussed in the main text.

**Supplementary Table 1. ECI (trade) economic growth regressions**

|  | Dependent variable: | | | | | | | | | | |
|---|---|---|---|---|---|---|---|---|---|---|---|
|  | Annualized GDP pc growth (1999-09, 2009-19) | | | | | | | | | | |
|  | (1) | (2) | (3) | (4) | (5) | (6) | (7) | (8) | (9) | (10) | (11) |
| ECI (trade) | 5.658*** | 6.218*** | 3.569*** | 7.323*** | 4.865*** | 5.352*** | 4.365*** | 4.339*** | 4.843*** | 6.459*** | 6.142*** |
|  | (1.172) | (1.064) | (1.277) | (1.284) | (1.309) | (1.114) | (1.458) | (1.479) | (1.859) | (2.161) | (1.391) |
| Log of initial population |  | -0.158 |  |  | 0.096 |  |  |  |  |  |  |
|  |  | (0.123) |  |  | (0.123) |  |  |  |  |  |  |
| Log of initial human capital |  |  | 3.300*** |  | 3.001*** |  |  |  |  |  |  |
|  |  |  | (0.731) |  | (0.806) |  |  |  |  |  |  |
| Log of natural resource exports per capita |  |  |  | 0.882*** | 0.767*** |  |  |  |  |  |  |
|  |  |  |  | (0.246) | (0.253) |  |  |  |  |  |  |
| Intensity (trade) |  |  |  |  |  | 3.701** |  |  |  |  |  |
|  |  |  |  |  |  | (1.678) |  |  |  |  |  |
| HHI (trade) |  |  |  |  |  |  | -1.532* |  |  |  |  |
|  |  |  |  |  |  |  | (0.857) |  |  |  |  |
| Entropy (trade) |  |  |  |  |  |  |  | 1.411* |  |  |  |
|  |  |  |  |  |  |  |  | (0.831) |  |  |  |
| Log of Fitness (trade) |  |  |  |  |  |  |  |  | 0.962 |  |  |
|  |  |  |  |  |  |  |  |  | (1.565) |  |  |
| i-ECI |  |  |  |  |  |  |  |  |  | -0.779 |  |
|  |  |  |  |  |  |  |  |  |  | (1.915) |  |
| Log of initial GDP per capita | -1.825*** | -2.000*** | -2.201*** | -3.262*** | -3.311*** | -2.574*** | -1.700*** | -1.697*** | -1.764*** | -1.819*** | -1.883*** |
|  | (0.259) | (0.261) | (0.256) | (0.503) | (0.450) | (0.449) | (0.266) | (0.266) | (0.271) | (0.257) | (0.282) |
| Observations | 152 | 152 | 152 | 152 | 152 | 152 | 152 | 152 | 152 | 152 | 152 |
| $R^2$ | 0.358 | 0.370 | 0.447 | 0.430 | 0.494 | 0.378 | 0.366 | 0.366 | 0.360 | 0.359 | 0.358 |
| Adjusted $R^2$ | 0.345 | 0.353 | 0.432 | 0.415 | 0.473 | 0.361 | 0.349 | 0.349 | 0.343 | 0.342 | 0.345 |

Notes: Each regression includes period fixed effects. Clustered standard errors in brackets. *p<0.1, **p<0.05, ***p<0.01.
The last column includes the instrumental variables estimate (see Supplementary Note 8).

## Supplementary Table 2. ECI (technology) economic growth regressions

| | Dependent variable: | | | | | | | | | | |
|---|---|---|---|---|---|---|---|---|---|---|---|
| | Annualized GDP pc growth (1999-09, 2009-19) | | | | | | | | | | |
| | (1) | (2) | (3) | (4) | (5) | (6) | (7) | (8) | (9) | (10) | (11) |
| ECI (technology) | 2.577*** | 3.487*** | 1.786** | 2.945*** | 2.113*** | 2.043* | 1.888** | 1.863** | 0.876 | 1.860** | 2.577*** |
| | (0.745) | (0.739) | (0.789) | (0.758) | (0.749) | (1.094) | (0.841) | (0.851) | (1.505) | (0.901) | (0.745) |
| Log of initial population | | -0.292** | | | 0.007 | | | | | | |
| | | (0.114) | | | (0.115) | | | | | | |
| Log of initial human capital | | | 3.523*** | | 3.316*** | | | | | | |
| | | | (0.762) | | (0.786) | | | | | | |
| Log of natural resource exports per capita | | | | 0.659*** | 0.543** | | | | | | |
| | | | | (0.237) | (0.227) | | | | | | |
| Intensity (technology) | | | | | | 1.231 | | | | | |
| | | | | | | (1.437) | | | | | |
| HHI (technology) | | | | | | | -1.291 | | | | |
| | | | | | | | (0.868) | | | | |
| Entropy (technology) | | | | | | | | 1.292 | | | |
| | | | | | | | | (0.880) | | | |
| Log of Fitness (technology) | | | | | | | | | 1.815* | | |
| | | | | | | | | | (1.098) | | |
| i-ECI | | | | | | | | | | 2.116* | |
| | | | | | | | | | | (1.262) | |
| Log of initial GDP per capita | -1.568*** | -1.914*** | -2.108*** | -2.553*** | -2.878*** | -1.785*** | -1.612*** | -1.612*** | -1.571*** | -1.728*** | -1.568*** |
| | (0.248) | (0.267) | (0.267) | (0.462) | (0.410) | (0.363) | (0.251) | (0.250) | (0.252) | (0.258) | (0.248) |
| Observations | 152 | 152 | 152 | 152 | 152 | 152 | 152 | 152 | 152 | 152 | 152 |
| R² | 0.341 | 0.374 | 0.450 | 0.384 | 0.478 | 0.346 | 0.348 | 0.348 | 0.357 | 0.355 | 0.341 |
| Adjusted R² | 0.327 | 0.357 | 0.435 | 0.367 | 0.457 | 0.328 | 0.330 | 0.330 | 0.340 | 0.337 | 0.327 |

Notes: Each regression includes period fixed effects. Clustered standard errors in brackets. *p<0.1, **p<0.05, ***p<0.01.
The last column includes the instrumental variables estimate (see Supplementary Note 8).

# Supplementary Table 3. ECI (research) economic growth regressions

|  | *Dependent variable:* | | | | | | | | | | |
|---|---|---|---|---|---|---|---|---|---|---|---|
|  | Annualized GDP pc growth (1999-09, 2009-19) | | | | | | | | | | |
|  | (1) | (2) | (3) | (4) | (5) | (6) | (7) | (8) | (9) | (10) | (11) |
| ECI (research) | 1.184 | 1.715 | 1.054 | 2.052 | 0.960 | -1.133 | 0.873 | 0.930 | 3.726 | -0.159 | 1.119 |
|  | (1.724) | (1.827) | (1.489) | (1.697) | (1.578) | (1.617) | (1.744) | (1.734) | (2.451) | (1.608) | (1.622) |
| Log of initial population |  | -0.089 |  |  | 0.163 |  |  |  |  |  |  |
|  |  | (0.144) |  |  | (0.142) |  |  |  |  |  |  |
| Log of initial human capital |  |  | 4.055*** |  | 4.182*** |  |  |  |  |  |  |
|  |  |  | (0.711) |  | (0.785) |  |  |  |  |  |  |
| Log of natural resource exports per capita |  |  |  | 0.559** | 0.569** |  |  |  |  |  |  |
|  |  |  |  | (0.224) | (0.232) |  |  |  |  |  |  |
| Intensity (research) |  |  |  |  |  | 4.231** |  |  |  |  |  |
|  |  |  |  |  |  | (2.123) |  |  |  |  |  |
| HHI (research) |  |  |  |  |  |  | -1.437 |  |  |  |  |
|  |  |  |  |  |  |  | (2.197) |  |  |  |  |
| Entropy (research) |  |  |  |  |  |  |  | 0.997 |  |  |  |
|  |  |  |  |  |  |  |  | (2.194) |  |  |  |
| Log of Fitness (research) |  |  |  |  |  |  |  |  | -3.398 |  |  |
|  |  |  |  |  |  |  |  |  | (2.799) |  |  |
| i-ECI |  |  |  |  |  |  |  |  |  | 3.921*** |  |
|  |  |  |  |  |  |  |  |  |  | (1.270) |  |
| Log of initial GDP per capita | -1.311*** | -1.445*** | -2.063*** | -2.216*** | -2.762*** | -1.705*** | -1.294*** | -1.298*** | -1.379*** | -1.639*** | -1.302*** |
|  | (0.359) | (0.389) | (0.336) | (0.513) | (0.466) | (0.450) | (0.361) | (0.360) | (0.354) | (0.338) | (0.361) |
| Observations | 152 | 152 | 152 | 152 | 152 | 152 | 152 | 152 | 152 | 152 | 152 |
| $R^2$ | 0.260 | 0.264 | 0.416 | 0.290 | 0.445 | 0.292 | 0.262 | 0.262 | 0.269 | 0.325 | 0.260 |
| Adjusted $R^2$ | 0.245 | 0.243 | 0.400 | 0.271 | 0.422 | 0.273 | 0.242 | 0.241 | 0.249 | 0.306 | 0.245 |

Notes: Each regression includes period fixed effects. Clustered standard errors in brackets. *p<0.1, **p<0.05, ***p<0.01.
The last column includes the instrumental variables estimate (see Supplementary Note 8)

**Economic growth multidimensional regression model additional explanatory variables robustness check:** In Supplementary Table 4 we reproduce the main results for the economic growth regression analysis (Columns (1-12)) and test the robustness of the multidimensional economic growth regression by adding the log of the population, the log of the initial human capital, and the log of natural resource exports per capita as additional explanatory variables, separately (Columns (13-15)) and together (Column (16)). In each case the product ECI, technology ECI, and their interaction term remain significant predictors of economic growth, thus confirming the robustness of the regression results. More importantly, the human capital and natural resources appear also significant. Therefore, in column (17) we re-estimate the economic growth model by including only these two as additional explanatory variables. This is our final economic growth model, i.e., the model that has the best explanatory power out of all economic growth regression analyses (adjusted $R^2 = 0.513$). In column (18) we add country-fixed effects to the model but observe that the adjusted $R^2$ of this model is very low, suggesting that these effects constrain the variation in the data. Supplementary Figure 2 gives the correlation between the variables used in these regressions.

**Economic growth multidimensional regression model Production intensity robustness check:** In Supplementary Table 5, Columns (2-12), we estimate the production intensity economic growth models. We find that the multidimensional model includes only the technological intensity (adjusted $R^2 = 0.308$). By comparing this model directly to the multidimensional ECI model (Column (13)), in a regression specification which also all of the potential explanatory variables (Column (14)), and in the final model regression specification (Column (15)), we find that the multidimensional ECI model clearly outperforms the multidimensional production intensity model as the coefficients of the former remain highly significant, whereas the coefficients of the later lose significance. The country-fixed effects (Column (16)) model does not adequately represent the variation in our data. Supplementary Figure 3 gives the correlations between the variables used in these regressions.

**Economic growth multidimensional regression model HHI robustness check:** In Supplementary Table 6, Columns (2-12), we estimate the HHI economic growth models. We find that the multidimensional model includes the trade and technological HHI, but not their interaction term (adjusted $R^2 = 0.345$). By comparing this model directly to the multidimensional ECI model (Column (13)), in a regression specification which also all of the potential explanatory variables (Column (14)), and in the final model regression specification (Column (15)), we find that the multidimensional ECI model clearly outperforms the multidimensional HHI model as the coefficients of the former remain highly significant, whereas the coefficients of the later lose significance. The country-fixed effects (Column (16)) model does not adequately represent the variation in our data. Supplementary Figure 4 gives the correlations between the variables used in these regressions.

**Economic growth multidimensional regression model Entropy robustness check:** In Supplementary Table 7, Columns (2-12), we estimate the Entropy economic growth models. Identically to the HHI case, we find that the multidimensional model includes the trade and technological Entropy, but not their interaction term (adjusted $R^2 = 0.345$). By comparing this model directly to the multidimensional ECI model (Column (13)), in a regression specification which also all of the potential explanatory variables (Column (14)), and in the final model regression specification (Column (15)), we find that the multidimensional ECI model clearly outperforms the multidimensional Entropy model as the coefficients of the former remain highly significant, whereas the coefficients of the later lose significance. The country-fixed effects (Column (16)) model does not adequately represent the variation in our data. Supplementary Figure 5 gives the correlations between the variables used in these regressions.

**Economic growth multidimensional regression model Fitness robustness check:** In Supplementary Table 8, Columns (2-12), we estimate the Fitness economic growth models. We find that the multidimensional Fitness model includes the trade, technological, and research Fitness, but not their interaction terms (adjusted $R^2 = 0.380$). By comparing this model directly to the multidimensional ECI model (Column (13)), in a regression specification which also all of the potential explanatory variables (Column (14)), and in the final model regression specification (Column (15)), we find that the multidimensional ECI model outperforms the multidimensional Fitness model as the coefficients of the former remain highly significant, whereas the coefficients of the later lose significance. The country-fixed effects (Column (16)) model does not adequately represent the variation in our data. Supplementary Figure 6 gives the correlations between the variables used in these regressions.

**Economic growth multidimensional regression model i-ECI robustness check:** In Supplementary Table 9, Columns (2-4), we estimate the i-ECI economic growth models. We find that the multidimensional i-ECI model includes the i-ECI, the research ECI, and their interaction (adjusted $R^2 = 0.334$). By comparing this model directly to the multidimensional ECI model (Column (5)), in a regression specification which also all of the potential explanatory variables (Column (6)), and in the final model regression specification (Column (7)), we find that the multidimensional ECI model outperforms the i-ECI model as the coefficients of the former remain highly significant, whereas the coefficients of the later lose significance. The country-fixed effects (Column (8)) model does not adequately represent the variation in our data. Supplementary Figure 7 gives the correlations between the variables used in these regressions.

# Supplementary Table 4. Economic Growth Regressions: Additional Explanatory Variables Robustness Check

| | | | | | | | | | *Dependent variable:* | | | | | | | | | |
|---|---|---|---|---|---|---|---|---|---|---|---|---|---|---|---|---|---|---|
| | | | | | | | | | Annualized GDP pc growth (1999-09, 2009-19) | | | | | | | | | |
| | (1) | (2) | (3) | (4) | (5) | (6) | (7) | (8) | (9) | (10) | (11) | (12) | (13) | (14) | (15) | (16) | (17) | (18) |
| ECI (trade) | | 5.658*** | | | 4.006*** | 5.981*** | | 4.022*** | 12.255*** | 12.134*** | | 17.331 | 11.471*** | 8.597*** | 14.030*** | 10.891*** | 10.832*** | 3.114 |
| | | (1.172) | | | (1.405) | (1.274) | | (1.469) | (2.955) | (3.863) | | (10.986) | (2.854) | (2.776) | (2.988) | (3.079) | (3.008) | (5.748) |
| ECI (technology) | | | 2.577*** | | 1.351 | | 3.323*** | 2.098** | 9.099*** | | 5.483** | 12.756 | 9.243*** | 7.033*** | 9.141*** | 7.409*** | 7.490*** | -3.903 |
| | | | (0.745) | | (0.893) | | (0.765) | (0.928) | (2.497) | | (2.647) | (10.129) | (2.521) | (2.296) | (2.555) | (2.365) | (2.450) | (4.871) |
| ECI (research) | | | | 1.184 | | -0.890 | -2.541 | -2.563 | | 6.318 | 0.380 | -5.469 | | | | | | |
| | | | | (1.724) | | (1.568) | (1.617) | (1.607) | | (4.847) | (4.282) | (12.688) | | | | | | |
| ECI (trade) x ECI (technology) | | | | | | | | | -12.260*** | | | -22.692 | -11.254*** | -9.208*** | -12.368*** | -9.987*** | -9.922*** | 10.462 |
| | | | | | | | | | (3.656) | | | (15.524) | (3.692) | (3.279) | (3.623) | (3.478) | (3.410) | (7.513) |
| ECI (trade) x ECI (research) | | | | | | | | | | -10.111* | | -3.392 | | | | | | |
| | | | | | | | | | | (5.831) | | (20.029) | | | | | | |
| ECI (research) x ECI (technology) | | | | | | | | | | | -3.856 | 0.435 | | | | | | |
| | | | | | | | | | | | (4.556) | (16.737) | | | | | | |
| ECI (trade) x ECI (research) x ECI (technology) | | | | | | | | | | | | 9.443 | | | | | | |
| | | | | | | | | | | | | (25.142) | | | | | | |
| Log of initial population | | | | | | | | | | | | | -0.239** | | | 0.036 | | |
| | | | | | | | | | | | | | (0.102) | | | (0.109) | | |
| Log of initial human capital | | | | | | | | | | | | | | 2.799*** | | 2.309*** | 2.227*** | -4.184 |
| | | | | | | | | | | | | | | (0.719) | | (0.808) | (0.773) | (3.088) |
| Log of natural resource exports per capita | | | | | | | | | | | | | | | 0.885*** | 0.766*** | 0.742*** | -0.556 |
| | | | | | | | | | | | | | | | (0.244) | (0.251) | (0.234) | (0.839) |
| Log of initial GDP per capita | -1.147*** | -1.825*** | -1.568*** | -1.311*** | -1.848*** | -1.740*** | -1.337*** | -1.616*** | -1.877*** | -1.809*** | -1.376*** | -1.612*** | -2.151*** | -2.186*** | -3.317*** | -3.337*** | -3.331*** | -6.325*** |
| | (0.208) | (0.259) | (0.248) | (0.359) | (0.252) | (0.321) | (0.321) | (0.313) | (0.239) | (0.320) | (0.322) | (0.337) | (0.248) | (0.251) | (0.476) | (0.431) | (0.429) | (0.853) |
| Fixed effects | No | No | No | No | No | No | No | No | No | No | No | No | No | No | No | No | No | Yes |
| Observations | 152 | 152 | 152 | 152 | 152 | 152 | 152 | 152 | 152 | 152 | 152 | 152 | 152 | 152 | 152 | 152 | 152 | 152 |
| $R^2$ | 0.256 | 0.358 | 0.341 | 0.260 | 0.373 | 0.361 | 0.355 | 0.388 | 0.427 | 0.377 | 0.361 | 0.452 | 0.449 | 0.487 | 0.499 | 0.536 | 0.536 | 0.640 |
| Adjusted $R^2$ | 0.246 | 0.345 | 0.327 | 0.245 | 0.356 | 0.343 | 0.338 | 0.367 | 0.407 | 0.356 | 0.339 | 0.417 | 0.426 | 0.466 | 0.479 | 0.510 | 0.513 | 0.028 |

Notes: Each regression includes period fixed effects. Clustered standard errors in brackets. *p<0.1, **p<0.05, ***p<0.01.

# Supplementary Table 5. Economic Growth Regressions: Production Intensity Robustness Check

|  | \multicolumn{16}{c}{Dependent variable:} |
| --- |
|  | Annualized GDP pc growth (1999-09, 2009-19) |||||||||||||||||
|  | (1) | (2) | (3) | (4) | (5) | (6) | (7) | (8) | (9) | (10) | (11) | (12) | (13) | (14) | (15) | (16) |
| --- | --- | --- | --- | --- | --- | --- | --- | --- | --- | --- | --- | --- | --- | --- | --- | --- |
| ECI (trade) | 12.255*** |  |  |  |  |  |  |  |  |  |  |  | 13.898*** | 10.928*** | 10.826*** | 5.871 |
|  | (2.955) |  |  |  |  |  |  |  |  |  |  |  | (2.770) | (3.154) | (3.005) | (6.176) |
| ECI (patents) | 9.099*** |  |  |  |  |  |  |  |  |  |  |  | 10.682*** | 7.380*** | 7.339*** | -0.303 |
|  | (2.497) |  |  |  |  |  |  |  |  |  |  |  | (2.399) | (2.393) | (2.346) | (5.723) |
| ECI (trade) x ECI (patents) | -12.260*** |  |  |  |  |  |  |  |  |  |  |  | -15.510*** | -9.716** | -9.456*** | 2.777 |
|  | (3.656) |  |  |  |  |  |  |  |  |  |  |  | (3.573) | (3.825) | (3.406) | (9.179) |
| Intensity (trade) |  | 4.678** |  |  | 2.939* | 3.955** |  | 2.983 | 4.821*** | 4.863 |  | -3.041 | 4.984*** | 1.205 | 1.122 | -0.470 |
|  |  | (1.882) |  |  | (1.740) | (1.684) |  | (1.817) | (1.865) | (4.441) |  | (10.646) | (1.678) | (2.236) | (2.176) | (2.543) |
| Intensity (technology) |  |  | 3.207*** |  | 2.735*** |  | 3.089* | 2.474 | 5.131** |  | 10.236*** | 22.587** | 0.776 | -0.716 | -0.856 | 6.305* |
|  |  |  | (0.939) |  | (0.915) |  | (1.581) | (1.718) | (2.151) |  | (3.629) | (11.076) | (1.417) | (1.444) | (1.327) | (3.814) |
| Intensity (research) |  |  |  | 3.480* |  | 2.979 | 0.232 | 0.500 |  | 3.660 | 2.692 | 2.684 |  |  |  |  |
|  |  |  |  | (2.019) |  | (1.954) | (2.868) | (3.003) |  | (4.466) | (3.123) | (9.807) |  |  |  |  |
| Intensity (trade) x Intensity (technology) |  |  |  |  |  |  |  |  | -3.368 |  |  | -10.103 |  |  |  |  |
|  |  |  |  |  |  |  |  |  | (2.680) |  |  | (16.352) |  |  |  |  |
| Intensity (trade) x Intensity (research) |  |  |  |  |  |  |  |  |  | -1.096 |  | 7.600 |  |  |  |  |
|  |  |  |  |  |  |  |  |  |  | (5.358) |  | (16.923) |  |  |  |  |
| Intensity (research) x Intensity (technology) |  |  |  |  |  |  |  |  |  |  | -7.731** | -25.996* |  |  |  |  |
|  |  |  |  |  |  |  |  |  |  |  | (3.638) | (14.717) |  |  |  |  |
| Intensity trade x Intensity (research) x Intensity (technology) |  |  |  |  |  |  |  |  |  |  |  | 14.050 |  |  |  |  |
|  |  |  |  |  |  |  |  |  |  |  |  | (20.571) |  |  |  |  |
| Log of initial population |  |  |  |  |  |  |  |  |  |  |  |  |  | 0.025 |  |  |
|  |  |  |  |  |  |  |  |  |  |  |  |  |  | (0.123) |  |  |
| Log of initial human capital |  |  |  |  |  |  |  |  |  |  |  |  |  | 2.388*** | 2.360*** | -8.898* |
|  |  |  |  |  |  |  |  |  |  |  |  |  |  | (0.830) | (0.824) | (4.697) |
| Log of natural resource exports per capita |  |  |  |  |  |  |  |  |  |  |  |  |  | 0.687** | 0.677** | -0.978 |
|  |  |  |  |  |  |  |  |  |  |  |  |  |  | (0.292) | (0.285) | (0.793) |
| Log of initial GDP per capita | -1.877*** | -2.139*** | -1.940*** | -1.736*** | -2.447*** | -2.490*** | -1.950*** | -2.476*** | -2.574*** | -2.522*** | -2.059*** | -2.421*** | -2.991*** | -3.376*** | -3.351*** | -6.416*** |
|  | (0.239) | (0.510) | (0.372) | (0.435) | (0.522) | (0.604) | (0.405) | (0.584) | (0.501) | (0.629) | (0.407) | (0.605) | (0.427) | (0.511) | (0.473) | (1.168) |
| Fixed effects | No | No | No | No | No | No | No | No | No | No | No | No | No | No | No | Yes |
| Observations | 152 | 152 | 152 | 152 | 152 | 152 | 152 | 152 | 152 | 152 | 152 | 152 | 152 | 152 | 152 | 152 |
| $R^2$ | 0.427 | 0.287 | 0.315 | 0.289 | 0.326 | 0.311 | 0.315 | 0.326 | 0.333 | 0.312 | 0.337 | 0.369 | 0.464 | 0.538 | 0.538 | 0.658 |
| Adjusted $R^2$ | 0.407 | 0.273 | 0.301 | 0.275 | 0.308 | 0.293 | 0.296 | 0.303 | 0.310 | 0.288 | 0.314 | 0.329 | 0.438 | 0.505 | 0.509 | 0.043 |

Notes: Each regression includes period fixed effects. Clustered standard errors in brackets. *p<0.1, **p<0.05, ***p<0.01.

# Supplementary Table 6. Economic Growth Regressions: Herfindahl-Hirschman Index Robustness Check

|  | \multicolumn{16}{c}{*Dependent variable:*} |
|---|---|

| | (1) | (2) | (3) | (4) | (5) | (6) | (7) | (8) | (9) | (10) | (11) | (12) | (13) | (14) | (15) | (16) |
|---|---|---|---|---|---|---|---|---|---|---|---|---|---|---|---|---|
| | \multicolumn{16}{c}{Annualized GDP pc growth (1999-09, 2009-19)} |

| | (1) | (2) | (3) | (4) | (5) | (6) | (7) | (8) | (9) | (10) | (11) | (12) | (13) | (14) | (15) | (16) |
|---|---|---|---|---|---|---|---|---|---|---|---|---|---|---|---|---|
| ECI (trade) | 12.255*** | | | | | | | | | | | | 11.777*** | 8.612** | 8.690** | -0.129 |
| | (2.955) | | | | | | | | | | | | (3.544) | (3.555) | (3.640) | (5.008) |
| ECI (patents) | 9.099*** | | | | | | | | | | | | 8.765*** | 6.026** | 6.199** | -5.219 |
| | (2.497) | | | | | | | | | | | | (2.883) | (2.746) | (2.893) | (3.833) |
| ECI (trade) x ECI (patents) | -12.260*** | | | | | | | | | | | | -11.869*** | -8.304** | -8.262** | 12.345* |
| | (3.656) | | | | | | | | | | | | (4.058) | (3.865) | (3.889) | (6.374) |
| HHI (trade) | | -3.601*** | | | -2.776*** | -3.558*** | | -2.819*** | -3.513** | -3.113*** | | -4.077* | -0.241 | -1.424 | -1.218 | -4.191* |
| | | (0.760) | | | (0.756) | (0.720) | | (0.738) | (1.465) | (0.925) | | (2.140) | (1.000) | (0.988) | (0.982) | (2.484) |
| HHI (technology) | | | -2.877*** | | -2.119** | | -2.940*** | -2.237*** | -2.343** | | -2.403** | -1.853 | -0.163 | -0.315 | -0.418 | -1.004 |
| | | | (0.837) | | (0.846) | | (0.849) | (0.868) | (1.000) | | (0.999) | (1.661) | (0.939) | (0.928) | (0.886) | (0.746) |
| HHI (research) | | | | -2.004 | | -0.530 | 0.536 | 1.096 | | 0.686 | 4.361 | 5.955 | | | | |
| | | | | (2.299) | | (2.130) | (2.062) | (2.131) | | (2.243) | (3.203) | (5.837) | | | | |
| HHI (trade) x HHI (technology) | | | | | | | | | 2.545 | | | 4.003 | | | | |
| | | | | | | | | | (4.825) | | | (9.162) | | | | |
| HHI (trade) x HHI (research) | | | | | | | | | | | -8.796 | -1.308 | | | | |
| | | | | | | | | | | | (9.931) | (33.180) | | | | |
| HHI (research) x HHI (technology) | | | | | | | | | | | -11.363 | -15.844 | | | | |
| | | | | | | | | | | | (7.063) | (29.198) | | | | |
| HHI trade x HHI (research) x HHI (technology) | | | | | | | | | | | | 7.854 | | | | |
| | | | | | | | | | | | | (143.950) | | | | |
| Log of initial population | | | | | | | | | | | | | | 0.073 | | |
| | | | | | | | | | | | | | (0.120) | | | |
| Log of initial human capital | | | | | | | | | | | | | | 2.456*** | 2.299*** | -4.402 |
| | | | | | | | | | | | | | (0.787) | (0.748) | (3.227) |
| Log of natural resource exports per capita | | | | | | | | | | | | | | 0.826*** | 0.770*** | -0.631 |
| | | | | | | | | | | | | | (0.246) | (0.223) | (0.842) |
| Log of initial GDP per capita | -1.877*** | -1.216*** | -1.496*** | -1.183*** | -1.457*** | -1.225*** | -1.494*** | -1.452*** | -1.456*** | -1.219*** | -1.458*** | -1.404*** | -1.860*** | -3.297*** | -3.292*** | -5.989*** |
| | (0.239) | (0.205) | (0.232) | (0.216) | (0.222) | (0.209) | (0.233) | (0.222) | (0.220) | (0.210) | (0.243) | (0.238) | (0.254) | (0.428) | (0.431) | (0.834) |
| Fixed effects | No | No | No | No | No | No | No | No | No | No | No | No | No | No | No | Yes |
| Observations | 152 | 152 | 152 | 152 | 152 | 152 | 152 | 152 | 152 | 152 | 152 | 152 | 152 | 152 | 152 | 152 |
| $R^2$ | 0.427 | 0.330 | 0.323 | 0.260 | 0.362 | 0.330 | 0.323 | 0.364 | 0.363 | 0.332 | 0.330 | 0.374 | 0.427 | 0.542 | 0.541 | 0.662 |
| Adjusted $R^2$ | 0.407 | 0.316 | 0.309 | 0.245 | 0.345 | 0.312 | 0.305 | 0.342 | 0.342 | 0.309 | 0.307 | 0.334 | 0.399 | 0.510 | 0.512 | 0.055 |

Notes: Each regression includes period fixed effects. Clustered standard errors in brackets. *p<0.1, **p<0.05, ***p<0.01.

# Supplementary Table 7. Economic Growth Regressions: Entropy Robustness Check

|  | \multicolumn{16}{c}{*Dependent variable:*} |
|---|---|

|  | (1) | (2) | (3) | (4) | (5) | (6) | (7) | (8) | (9) | (10) | (11) | (12) | (13) | (14) | (15) | (16) |
|---|---|---|---|---|---|---|---|---|---|---|---|---|---|---|---|---|
| \multicolumn{17}{l}{Annualized GDP pc growth (1999-09, 2009-19)} |
| ECI (trade) | 12.255*** |  |  |  |  |  |  |  |  |  |  |  | 11.897*** | 8.564** | 8.649** | 0.019 |
|  | (2.955) |  |  |  |  |  |  |  |  |  |  |  | (3.554) | (3.555) | (3.646) | (5.090) |
| ECI (patents) | 9.099*** |  |  |  |  |  |  |  |  |  |  |  | 8.838*** | 6.012** | 6.187** | -5.290 |
|  | (2.497) |  |  |  |  |  |  |  |  |  |  |  | (2.894) | (2.752) | (2.902) | (3.916) |
| ECI (trade) x ECI (patents) | -12.260*** |  |  |  |  |  |  |  |  |  |  |  | -11.964*** | -8.304** | -8.256** | 12.396* |
|  | (3.656) |  |  |  |  |  |  |  |  |  |  |  | (4.053) | (3.858) | (3.884) | (6.479) |
| Entropy (trade) |  | 3.346*** |  |  | 2.527*** | 3.350*** |  | 2.606*** | 1.604 | 12.105 |  | -1.660 | 0.163 | 1.390 | 1.181 | 3.410 |
|  |  | (0.714) |  |  | (0.726) | (0.668) |  | (0.702) | (3.341) | (7.874) |  | (78.832) | (0.965) | (0.947) | (0.937) | (2.267) |
| Entropy (technology) |  |  | 2.863*** |  | 2.058** |  | 3.021*** | 2.277** | 0.878 |  | 12.665* | 5.275 | 0.130 | 0.265 | 0.381 | 0.995 |
|  |  |  | (0.834) |  | (0.855) |  | (0.854) | (0.886) | (3.963) |  | (6.808) | (79.319) | (0.943) | (0.933) | (0.887) | (0.769) |
| Entropy (research) |  |  |  | 1.607 |  | -0.044 | -1.122 | -1.734 |  | 7.716 | 5.553 | 4.392 |  |  |  |  |
|  |  |  |  | (2.286) |  | (2.152) | (2.126) | (2.212) |  | (7.331) | (4.987) | (62.720) |  |  |  |  |
| Entropy (trade) x Entropy (technology) |  |  |  |  |  |  |  |  | 1.322 |  |  | 9.596 |  |  |  |  |
|  |  |  |  |  |  |  |  |  | (4.472) |  |  | (103.283) |  |  |  |  |
| Entropy (trade) x Entropy (research) |  |  |  |  |  |  |  |  |  | -9.302 |  | 2.264 |  |  |  |  |
|  |  |  |  |  |  |  |  |  |  | (8.370) |  | (84.324) |  |  |  |  |
| Entropy (research) x Entropy (technology) |  |  |  |  |  |  |  |  |  |  | -10.231 | -6.167 |  |  |  |  |
|  |  |  |  |  |  |  |  |  |  |  | (7.371) | (84.876) |  |  |  |  |
| Entropy trade x Entropy (research) x Entropy (technology) |  |  |  |  |  |  |  |  |  |  |  | -6.860 |  |  |  |  |
|  |  |  |  |  |  |  |  |  |  |  |  | (110.046) |  |  |  |  |
| Log of initial population |  |  |  |  |  |  |  |  |  |  |  |  |  | 0.076 |  |  |
|  |  |  |  |  |  |  |  |  |  |  |  |  |  | (0.121) |  |  |
| Log of initial human capital |  |  |  |  |  |  |  |  |  |  |  |  |  | 2.464*** | 2.300*** | -4.536 |
|  |  |  |  |  |  |  |  |  |  |  |  |  |  | (0.786) | (0.747) | (3.245) |
| Log of natural resource exports per capita |  |  |  |  |  |  |  |  |  |  |  |  |  | 0.836*** | 0.776*** | -0.621 |
|  |  |  |  |  |  |  |  |  |  |  |  |  |  | (0.246) | (0.222) | (0.846) |
| Log of initial GDP per capita | -1.877*** | -1.218*** | -1.503*** | -1.182*** | -1.457*** | -1.217*** | -1.498*** | -1.447*** | -1.457*** | -1.208*** | -1.460*** | -1.400*** | -1.865*** | -3.299*** | -3.294*** | -6.023*** |
|  | (0.239) | (0.205) | (0.233) | (0.218) | (0.223) | (0.210) | (0.233) | (0.223) | (0.221) | (0.212) | (0.245) | (0.242) | (0.255) | (0.427) | (0.431) | (0.850) |
| Fixed effects | No | No | No | No | No | No | No | No | No | No | No | No | No | No | No | Yes |
| Observations | 152 | 152 | 152 | 152 | 152 | 152 | 152 | 152 | 152 | 152 | 152 | 152 | 152 | 152 | 152 | 152 |
| R² | 0.427 | 0.331 | 0.324 | 0.259 | 0.362 | 0.331 | 0.326 | 0.365 | 0.362 | 0.334 | 0.332 | 0.375 | 0.427 | 0.542 | 0.541 | 0.659 |
| Adjusted R² | 0.407 | 0.317 | 0.311 | 0.244 | 0.345 | 0.313 | 0.308 | 0.344 | 0.340 | 0.311 | 0.309 | 0.335 | 0.399 | 0.510 | 0.512 | 0.047 |

Notes: Each regression includes period fixed effects. Clustered standard errors in brackets. *p<0.1, **p<0.05, ***p<0.01.

## Supplementary Table 8. Economic Growth Regressions: Fitness Robustness Check

| | (1) | (2) | (3) | (4) | (5) | (6) | (7) | (8) | (9) | (10) | (11) | (12) | (13) | (14) | (15) | (16) |
|---|---|---|---|---|---|---|---|---|---|---|---|---|---|---|---|---|
| | | | | | | | *Dependent variable:* Annualized GDP pc growth (1999-09, 2009-19) | | | | | | | | | |
| ECI (trade) | 12.255*** | | | | | | | | | | | | 11.371*** | 9.039** | 8.892** | 4.185 |
| | (2.955) | | | | | | | | | | | | (3.962) | (3.851) | (3.807) | (6.467) |
| ECI (patents) | 9.099*** | | | | | | | | | | | | 9.000*** | 7.296** | 7.284** | -0.794 |
| | (2.497) | | | | | | | | | | | | (3.282) | (3.076) | (3.105) | (6.024) |
| ECI (trade) x ECI (patents) | -12.260*** | | | | | | | | | | | | -11.004*** | -9.176** | -8.961** | 8.080 |
| | (3.656) | | | | | | | | | | | | (4.058) | (3.924) | (3.870) | (8.555) |
| Log of fitness (trade) | | 4.136*** | | | 1.946 | 5.575*** | | 3.391*** | 4.836* | 16.705*** | | 10.070 | 0.062 | 2.286 | 2.202 | -1.478 |
| | | (0.981) | | | (1.241) | (1.053) | | (1.149) | (2.635) | (4.998) | | (15.355) | (1.955) | (2.011) | (1.951) | (3.977) |
| Log of fitness (technology) | | | 2.461*** | | 1.885*** | | 2.977*** | 2.219*** | 5.918 | | 6.110** | -2.335 | 0.282 | -0.387 | -0.301 | -1.937* |
| | | | (0.547) | | (0.647) | | (0.647) | (0.664) | (3.911) | | (3.042) | (19.387) | (0.910) | (0.890) | (0.866) | (1.090) |
| Log of fitness (research) | | | | 0.218 | | -3.481* | -3.002 | -4.432** | | 8.843 | 0.743 | 5.217 | -3.876** | -2.245 | -1.829 | 4.813 |
| | | | | (2.033) | | (1.958) | | (1.825) | | (5.657) | (3.416) | (14.303) | (1.823) | (1.976) | (1.747) | (4.057) |
| Log of fitness (trade) x Log of fitness (technology) | | | | | | | | | | | | | | 2.282*** | 2.129*** | -1.232 |
| | | | | | | | | | | | | | | (0.837) | (0.808) | (3.437) |
| Log of fitness (trade) x Log of fitness (research) | | | | | | | | | | | | | | 0.834*** | 0.783*** | -0.744 |
| | | | | | | | | | | | | | | (0.251) | (0.229) | (0.689) |
| Log of fitness (research) x Log of fitness (technology) | | | | | | | | | | | -4.893 | 4.075 | | | | |
| | | | | | | | | | | | (4.522) | (25.506) | | | | |
| Log of fitness trade x Log of fitness (research) x Log of fitness (technology) | | | | | | | | | | | | -7.360 | | | | |
| | | | | | | | | | | | | (33.812) | | | | |
| Log of initial population | | | | | | | | | -5.226 | | | 7.134 | | | | |
| | | | | | | | | | (5.015) | | | (25.086) | | | | |
| Log of initial human capital | | | | | | | | | | | -16.073** | -10.865 | | | | |
| | | | | | | | | | | | (7.051) | (23.843) | | | | |
| Log of natural resource export per capita | | | | | | | | | | | | | | | 0.087 | |
| | | | | | | | | | | | | | | | (0.130) | |
| Log of initial GDP per capita | -1.877*** | -1.302*** | -1.528*** | -1.165*** | -1.512*** | -1.063*** | -1.355*** | -1.245*** | -1.466*** | -0.996*** | -1.316*** | -1.171*** | -1.703*** | -3.137*** | -3.160*** | -6.534*** |
| | (0.239) | (0.211) | (0.216) | (0.306) | (0.210) | (0.273) | (0.210) | (0.260) | (0.241) | (0.286) | (0.294) | (0.277) | (0.333) | (0.454) | (0.452) | (0.669) |
| Fixed effects | No | No | No | No | No | No | No | No | No | No | No | No | No | No | No | Yes |
| Observations | 152 | 152 | 152 | 152 | 152 | 152 | 152 | 152 | 152 | 152 | 152 | 152 | 152 | 152 | 152 | 152 |
| $R^2$ | 0.427 | 0.327 | 0.354 | 0.256 | 0.364 | 0.349 | 0.373 | 0.400 | 0.372 | 0.368 | 0.380 | 0.417 | 0.454 | 0.546 | 0.544 | 0.680 |
| Adjusted $R^2$ | 0.407 | 0.313 | 0.341 | 0.241 | 0.347 | 0.332 | 0.356 | 0.380 | 0.350 | 0.346 | 0.358 | 0.380 | 0.423 | 0.510 | 0.512 | 0.090 |

Notes: Each regression includes period fixed effects. Clustered standard errors in brackets. *p<0.1, **p<0.05, ***p<0.01.

## Supplementary Table 9. Economic Growth Regressions: i-ECI Robustness Check

|  | *Dependent variable:* | | | | | | | |
|---|---|---|---|---|---|---|---|---|
|  | Annualized GDP pc growth (1999-09, 2009-19) | | | | | | | |
|  | (1) | (2) | (3) | (4) | (5) | (6) | (7) | (8) |
| ECI (trade) | 12.255*** | | | | 13.795*** | 14.026*** | 13.782*** | 4.169 |
|  | (2.955) | | | | (3.246) | (3.264) | (3.131) | (6.663) |
| ECI (patents) | 9.099*** | | | | 11.947*** | 8.944*** | 9.131*** | -2.876 |
|  | (2.497) | | | | (3.340) | (3.157) | (3.158) | (5.435) |
| ECI (trade) x ECI (patents) | -12.260*** | | | | -15.411*** | -11.776*** | -11.780*** | 9.192 |
|  | (3.656) | | | | (4.580) | (4.280) | (4.276) | (8.171) |
| i-ECI | | 3.881*** | 3.921*** | 12.270*** | -3.486 | -5.627 | -5.389 | -2.918 |
|  | | (1.197) | (1.270) | (3.896) | (5.254) | (5.041) | (4.995) | (6.756) |
| ECI (research) | | | -0.159 | 6.096* | -5.564 | -3.866 | -3.695 | -2.930 |
|  | | | (1.608) | (3.466) | (4.666) | (4.579) | (4.558) | (6.420) |
| i-ECI x ECI (research) | | | | -12.801** | 6.254 | 5.710 | 5.691 | 4.365 |
|  | | | | (5.592) | (7.315) | (7.487) | (7.440) | (8.691) |
| Log of initial population | | | | | | 0.071 | | |
|  | | | | | | (0.122) | | |
| Log of initial human capital | | | | | | 2.363*** | 2.199*** | -3.639 |
|  | | | | | | (0.864) | (0.796) | (3.395) |
| Log of natural resource export per capita | | | | | | 0.809*** | 0.764*** | -0.528 |
|  | | | | | | (0.238) | (0.220) | (0.805) |
| Log of initial GDP per capita | -1.877*** | -1.656*** | -1.639*** | -1.771*** | -1.620*** | -3.229*** | -3.235*** | -6.335*** |
|  | (0.239) | (0.268) | (0.338) | (0.332) | (0.336) | (0.481) | (0.480) | (0.922) |
| Fixed effects | No | No | No | No | No | No | No | Yes |
| Observations | 152 | 152 | 152 | 152 | 152 | 152 | 152 | 152 |
| $R^2$ | 0.427 | 0.325 | 0.325 | 0.356 | 0.444 | 0.544 | 0.543 | 0.642 |
| Adjusted $R^2$ | 0.407 | 0.311 | 0.306 | 0.334 | 0.412 | 0.508 | 0.510 | -0.020 |

Notes: Each regression includes period fixed effects. Clustered standard errors in brackets. *p<0.1, **p<0.05, ***p<0.01.

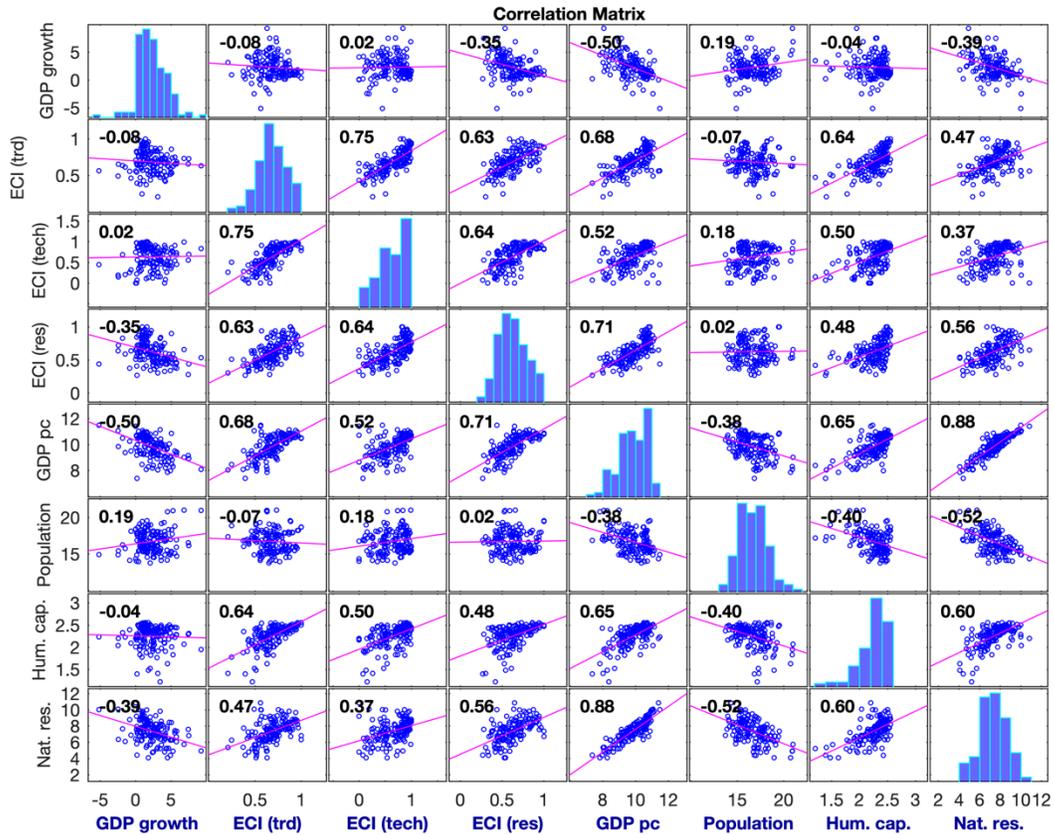

**Supplementary Figure 2. Correlations Between the Variables used in the Additional Explanatory Variables Economic Growth Robustness Check Regression Analysis**

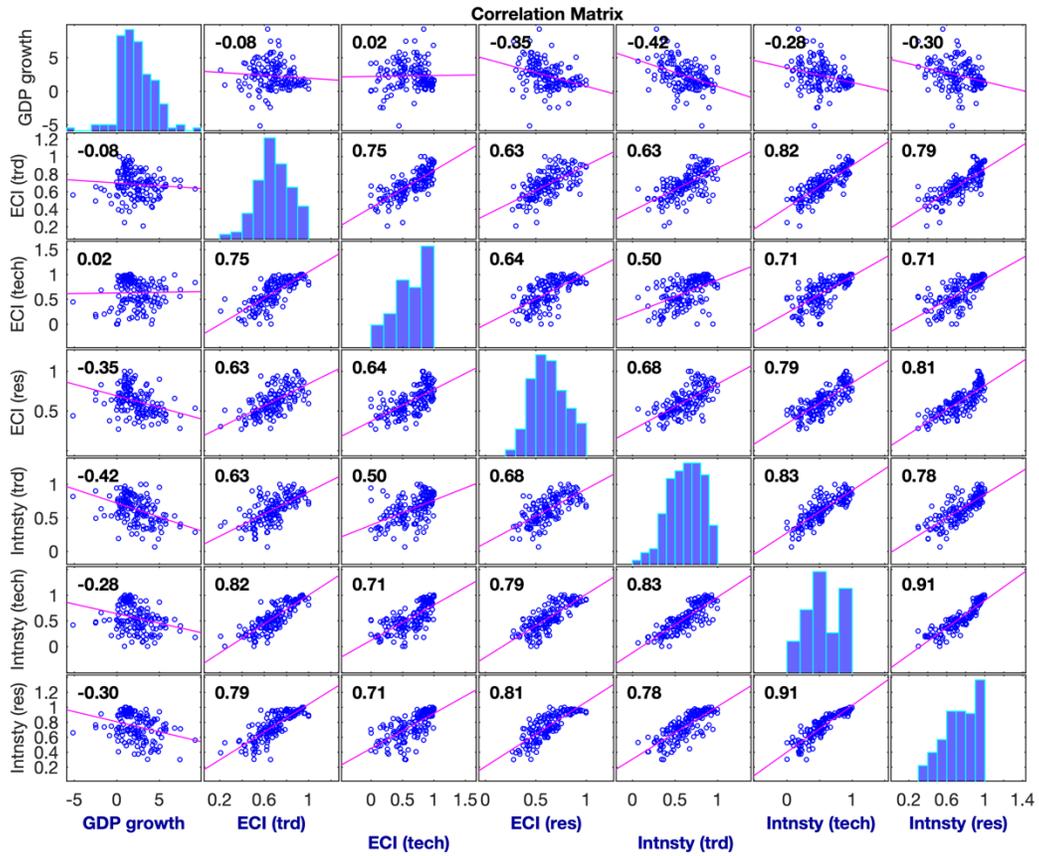

**Supplementary Figure 3. Correlations Between the Variables used in the Production Intensity Economic Growth Robustness Check Regression Analysis**

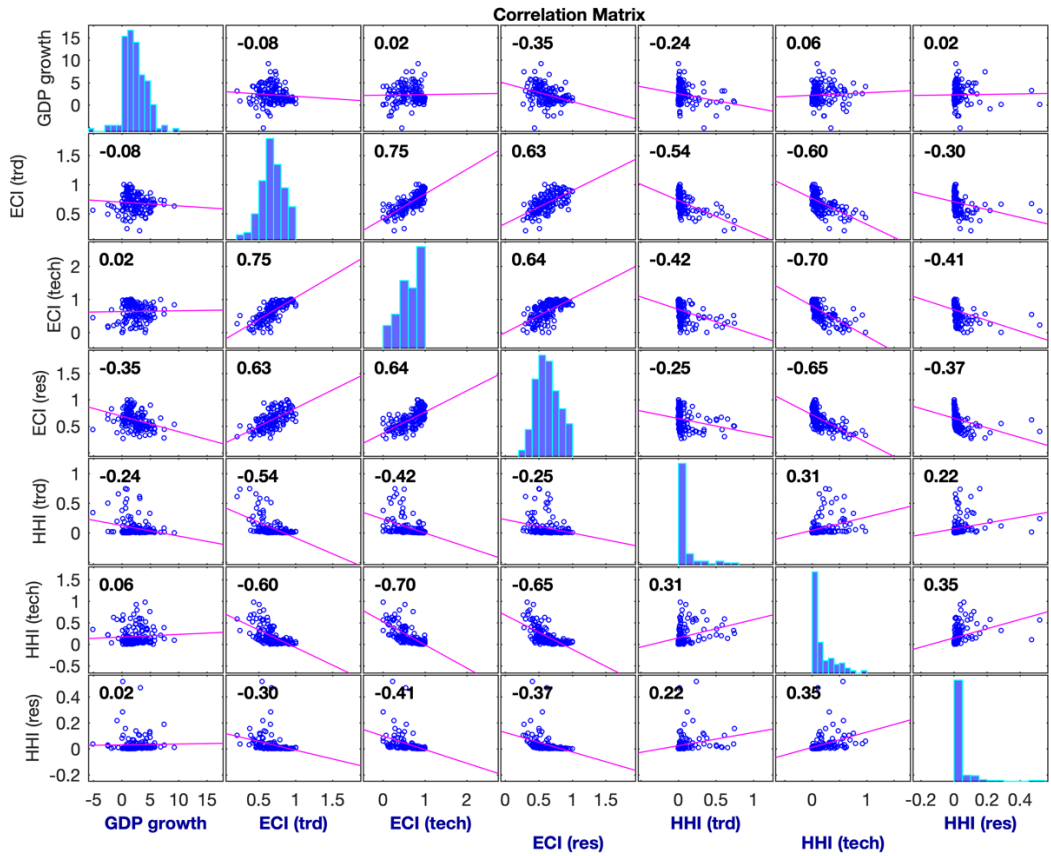

**Supplementary Figure 4. Correlations Between the Variables used in the HHI Economic Growth Robustness Check Regression Analysis**

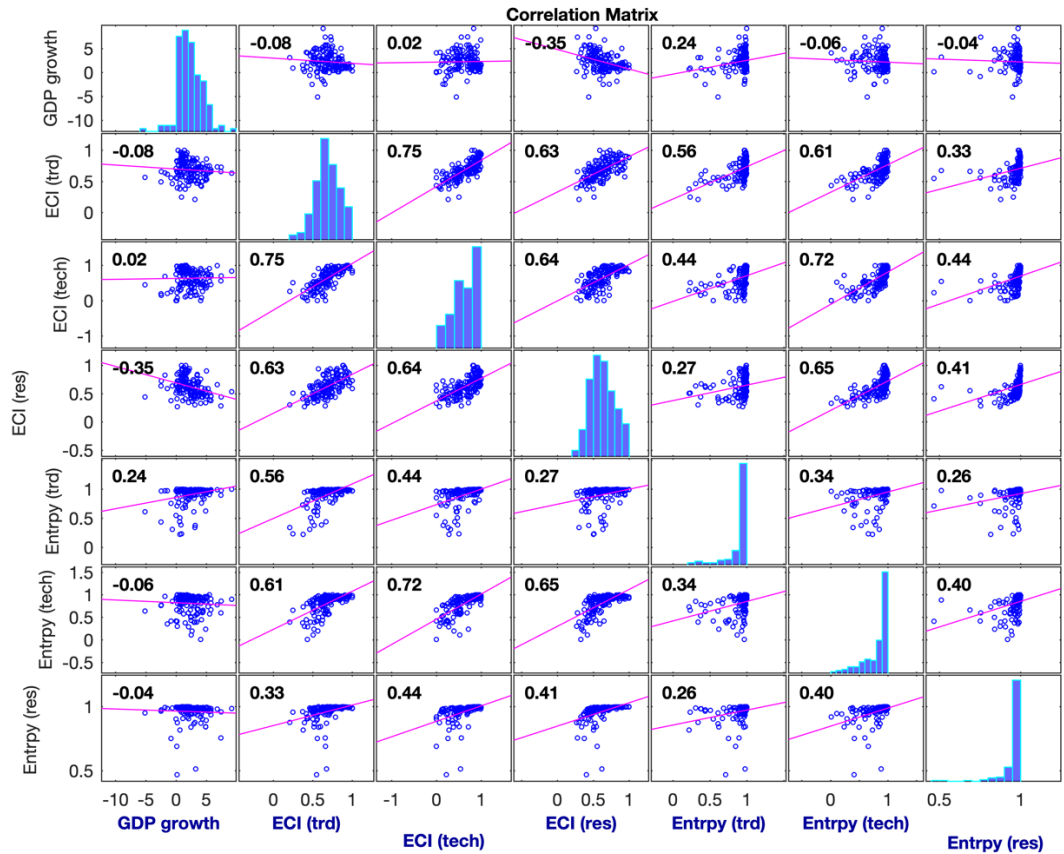

**Supplementary Figure 5. Correlations Between the Variables used in the Entropy Economic Growth Robustness Check Regression Analysis**

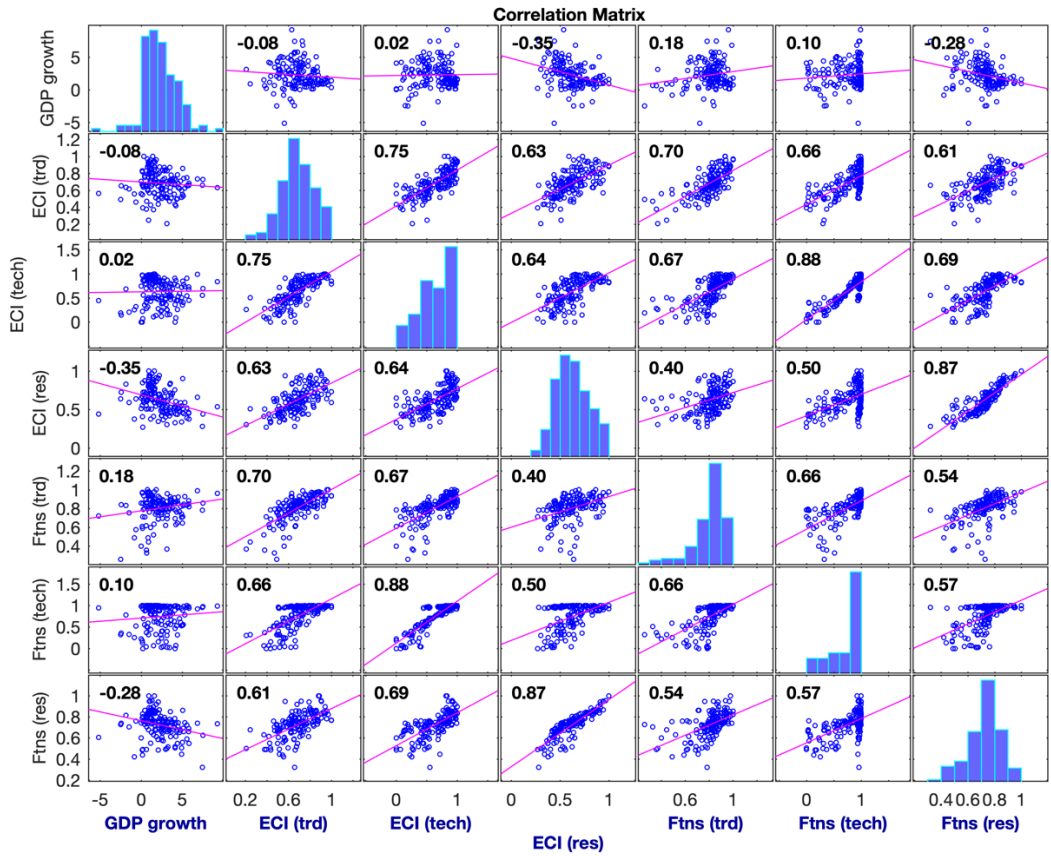

**Supplementary Figure 6. Correlations Between the Variables used in the Fitness Economic Growth Robustness Check Regression Analysis**

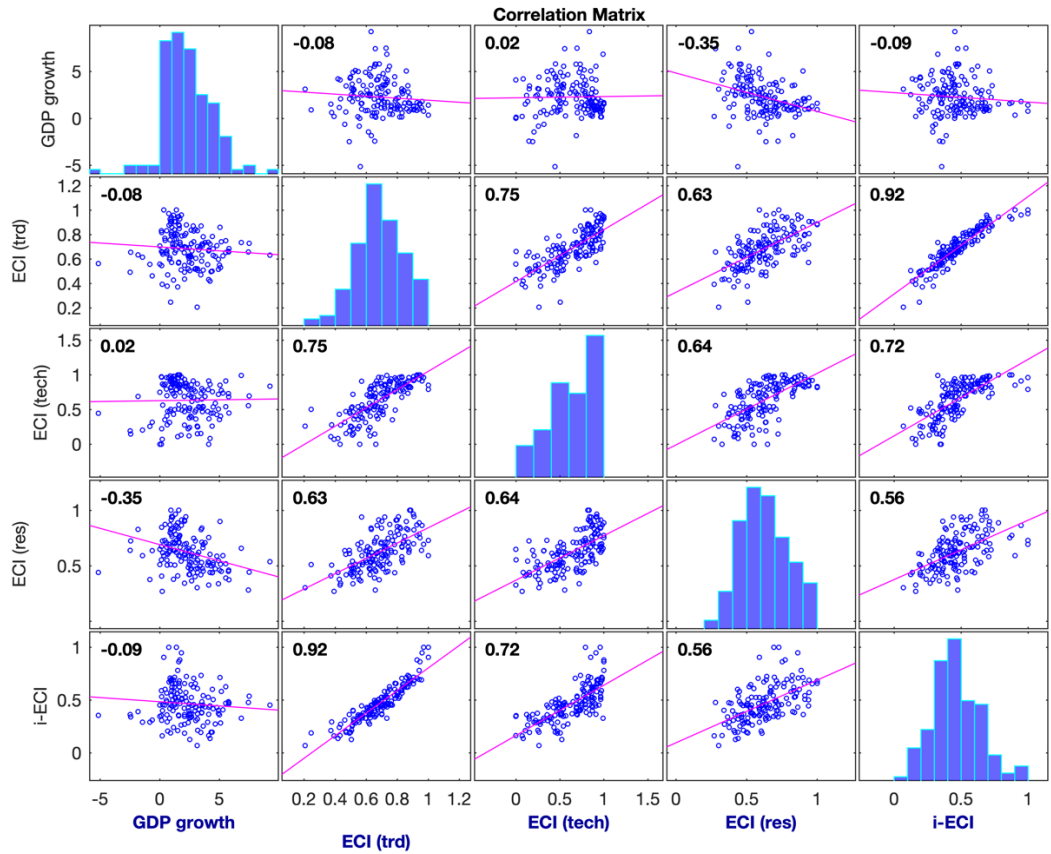

**Supplementary Figure 7. Correlations Between the Variables used in the i-ECI Economic Growth Robustness Check Regression Analysis**

# Supplementary Note 5. Income inequality regression analysis

**Income inequality regression setup:** In the income inequality regression analysis, the dependent variable is quantified through the Gini coefficient $GINI_{ct}$, i.e.,

$$y_{ct} = GINI_{ct}.$$

The Gini coefficient quantifies the extent to which the observed income distribution differs from the line of perfect equality, i.e., the income distribution in a hypothetical society where every individual has the same income[10]. The coefficient is a normalized quantity whose values are between 0 and 1, with larger values indicating higher income inequality. The data for the Gini coefficient were taken from the Estimated Household Income Inequality dataset and are available at

https://utip.gov.utexas.edu/datasets.html .

In the regressions, as control variables we include the log of the initial GDP per capita in constant 2017 dollars (PPP) and its squared value. In the income inequality literature, these two terms are known as the Kuznets hypothesis[11]. According to this hypothesis, as an economy develops, market forces first increase and then decrease economic inequality.

The regression analysis is focused on the periods 1996-1999, 2000-2003, 2004-2007, 2008-2011 and 2012-2015. Because of the sparseness of the Gini dataset and slow temporal changes in the coefficients within a country, we follow[12] and use average values for each panel.

**Income inequality individual regressions:** In Supplementary Tables 10-12, we present the results for the individual income inequality regressions for ECI (trade), ECI (technology), and ECI (research), respectively. These results were used to estimate the F-statistics for testing the robustness of the individual regressions discussed in the main text.

# Supplementary Table 10. ECI (trade) Income Inequality regressions

|  | \multicolumn{11}{c}{*Dependent variable:*} |
| --- | --- | --- | --- | --- | --- | --- | --- | --- | --- | --- | --- |
|  | \multicolumn{11}{c}{Gini coefficient (1996-99, 2000-03, 2004-07, 2008-11, 2012-15)} |
|  | (1) | (2) | (3) | (4) | (5) | (6) | (7) | (8) | (9) | (10) | (11) |
| ECI (trade) | -23.543*** | -27.481*** | -19.151*** | -26.876*** | -24.144*** | -22.173*** | -17.531*** | -17.015*** | -13.296** | -31.648*** | -25.528*** |
|  | (5.285) | (5.006) | (4.717) | (4.803) | (4.018) | (4.999) | (5.427) | (5.422) | (5.651) | (7.923) | (5.873) |
| Log of population |  | 0.973*** |  |  | 0.600** |  |  |  |  |  |  |
|  |  | (0.246) |  |  | (0.263) |  |  |  |  |  |  |
| Log of human capital |  |  | -9.073*** |  | -6.730*** |  |  |  |  |  |  |
|  |  |  | (2.437) |  | (2.551) |  |  |  |  |  |  |
| Log of natural resource exports per capita |  |  |  | -1.742** | -0.748 |  |  |  |  |  |  |
|  |  |  |  | (0.738) | (0.739) |  |  |  |  |  |  |
| Intensity (trade) |  |  |  |  |  | -17.753*** |  |  |  |  |  |
|  |  |  |  |  |  | (4.466) |  |  |  |  |  |
| HHI (trade) |  |  |  |  |  |  | 8.897** |  |  |  |  |
|  |  |  |  |  |  |  | (4.432) |  |  |  |  |
| Entropy (trade) |  |  |  |  |  |  |  | -8.636** |  |  |  |
|  |  |  |  |  |  |  |  | (4.062) |  |  |  |
| Fitness (trade) |  |  |  |  |  |  |  |  | -13.424** |  |  |
|  |  |  |  |  |  |  |  |  | (5.380) |  |  |
| i-ECI |  |  |  |  |  |  |  |  |  | 8.061 |  |
|  |  |  |  |  |  |  |  |  |  | (7.854) |  |
| Log of GDP per capita | -8.070 | -2.789 | 4.605 | -6.778 | 5.142 | -7.788 | -7.267 | -6.841 | -7.667 | -7.951 | -7.906 |
|  | (10.761) | (9.665) | (9.669) | (10.328) | (8.866) | (9.583) | (10.271) | (10.207) | (10.049) | (10.784) | (10.639) |
| Log of GDP per capita, squared | 0.361 | 0.152 | -0.235 | 0.452 | -0.171 | 0.547 | 0.278 | 0.253 | 0.288 | 0.354 | 0.366 |
|  | (0.569) | (0.517) | (0.504) | (0.539) | (0.465) | (0.510) | (0.540) | (0.536) | (0.518) | (0.570) | (0.561) |
| Observations | 332 | 332 | 332 | 332 | 332 | 332 | 332 | 332 | 332 | 332 | 332 |
| $R^2$ | 0.551 | 0.600 | 0.609 | 0.581 | 0.637 | 0.605 | 0.578 | 0.580 | 0.588 | 0.557 | 0.551 |
| Adjusted $R^2$ | 0.542 | 0.590 | 0.599 | 0.570 | 0.626 | 0.595 | 0.567 | 0.570 | 0.578 | 0.546 | 0.541 |

Notes: Each regression includes period fixed effects. Clustered standard errors in brackets. *p<0.1, **p<0.05, ***p<0.01. The last column includes the instrumental variables estimate (see Supplementary Note 8).

## Supplementary Table 11. ECI (technology) Income Inequality regressions

|  | Dependent variable: | | | | | | | | | | |
|---|---|---|---|---|---|---|---|---|---|---|---|
|  | Gini coefficient (1996-99, 2000-03, 2004-07, 2008-11, 2012-15) | | | | | | | | | | |
|  | (1) | (2) | (3) | (4) | (5) | (6) | (7) | (8) | (9) | (10) | (11) |
| ECI (technology) | -11.211*** | -16.507*** | -9.385*** | -11.673*** | -13.896*** | -3.673 | -11.284*** | -11.341*** | -12.420*** | -7.543*** | -10.786*** |
|  | (2.715) | (2.660) | (2.130) | (2.749) | (2.212) | (2.526) | (3.791) | (3.797) | (4.114) | (2.808) | (3.158) |
| Log of population |  | 1.482*** |  |  | 1.239*** |  |  |  |  |  |  |
|  |  | (0.271) |  |  | (0.270) |  |  |  |  |  |  |
| Log of human capital |  |  | -11.094*** |  | -8.344*** |  |  |  |  |  |  |
|  |  |  | (2.687) |  | (2.209) |  |  |  |  |  |  |
| Log of natural resource documents per capita |  |  |  | -0.769 | 0.615 |  |  |  |  |  |  |
|  |  |  |  | (0.658) | (0.611) |  |  |  |  |  |  |
| Intensity (technology) |  |  |  |  |  | -16.469*** |  |  |  |  |  |
|  |  |  |  |  |  | (4.424) |  |  |  |  |  |
| HHI (technology) |  |  |  |  |  |  | -0.120 |  |  |  |  |
|  |  |  |  |  |  |  | (2.683) |  |  |  |  |
| Entropy (technology) |  |  |  |  |  |  |  | 0.208 |  |  |  |
|  |  |  |  |  |  |  |  | (2.686) |  |  |  |
| Fitness (technology) |  |  |  |  |  |  |  |  | 1.502 |  |  |
|  |  |  |  |  |  |  |  |  | (2.691) |  |  |
| i-ECI |  |  |  |  |  |  |  |  |  | -10.637* |  |
|  |  |  |  |  |  |  |  |  |  | (5.461) |  |
| Log of GDP per capita | -3.795 | 6.686 | 11.132 | -3.091 | 15.631* | -11.642 | -3.863 | -3.921 | -4.842 | -5.104 | -4.032 |
|  | (11.358) | (9.323) | (9.274) | (11.350) | (8.148) | (9.636) | (11.762) | (11.775) | (11.651) | (10.955) | (11.285) |
| Log of GDP per capita, squared | 0.071 | -0.372 | -0.606 | 0.098 | -0.830** | 0.638 | 0.074 | 0.077 | 0.124 | 0.184 | 0.079 |
|  | (0.595) | (0.493) | (0.480) | (0.584) | (0.416) | (0.517) | (0.617) | (0.618) | (0.612) | (0.574) | (0.593) |
| Observations | 332 | 332 | 332 | 332 | 332 | 332 | 332 | 332 | 332 | 332 | 332 |
| $R^2$ | 0.496 | 0.589 | 0.589 | 0.503 | 0.639 | 0.586 | 0.496 | 0.496 | 0.497 | 0.533 | 0.496 |
| Adjusted $R^2$ | 0.485 | 0.579 | 0.579 | 0.490 | 0.628 | 0.576 | 0.484 | 0.484 | 0.485 | 0.521 | 0.485 |

Notes: Each regression includes period fixed effects. Clustered standard errors in brackets. *p<0.1, **p<0.05, ***p<0.01. The last column includes the instrumental variables estimate (see Supplementary Note 8).

## Supplementary Table 12. ECI (research) Income Inequality regressions

| | Dependent variable: | | | | | | | | | | |
|---|---|---|---|---|---|---|---|---|---|---|---|
| | Gini coefficient (1996-99, 2000-03, 2004-07, 2008-11, 2012-15) | | | | | | | | | | |
| | (1) | (2) | (3) | (4) | (5) | (6) | (7) | (8) | (9) | (10) | (11) |
| ECI (research) | -7.654 | -11.330** | -6.067 | -8.095 | -7.660* | 2.943 | -6.997 | -6.898 | -6.715 | -3.352 | -8.428 |
| | (5.310) | (5.589) | (4.352) | (5.126) | (4.580) | (3.848) | (5.579) | (5.686) | (7.771) | (4.204) | (5.320) |
| Log of population | | 0.704** | | | 0.381 | | | | | | |
| | | (0.304) | | | (0.324) | | | | | | |
| Log of human capital | | | -13.035*** | | -12.512*** | | | | | | |
| | | | (2.999) | | (3.098) | | | | | | |
| Log of natural resource documents per capita | | | | -0.379 | 0.397 | | | | | | |
| | | | | (0.711) | (0.756) | | | | | | |
| Intensity (publications) | | | | | | -24.325*** | | | | | |
| | | | | | | (5.232) | | | | | |
| HHI (research) | | | | | | | 5.666 | | | | |
| | | | | | | | (13.291) | | | | |
| Entropy (research) | | | | | | | | -5.002 | | | |
| | | | | | | | | (11.508) | | | |
| Fitness (research) | | | | | | | | | -1.484 | | |
| | | | | | | | | | (8.511) | | |
| i-ECI | | | | | | | | | | -16.518*** | |
| | | | | | | | | | | (5.670) | |
| Log of GDP per capita | -14.997 | -13.806 | 4.766 | -15.063 | 4.689 | -5.600 | -14.633 | -14.576 | -14.909 | -11.069 | -15.500 |
| | (15.911) | (15.394) | (11.987) | (15.919) | (11.757) | (12.434) | (15.771) | (15.788) | (15.837) | (12.653) | (15.839) |
| Log of GDP per capita, squared | 0.616 | 0.609 | -0.289 | 0.652 | -0.295 | 0.293 | 0.596 | 0.592 | 0.610 | 0.497 | 0.648 |
| | (0.866) | (0.842) | (0.651) | (0.852) | (0.635) | (0.681) | (0.859) | (0.860) | (0.862) | (0.685) | (0.862) |
| Observations | 332 | 332 | 332 | 332 | 332 | 332 | 332 | 332 | 332 | 332 | 332 |
| $R^2$ | 0.371 | 0.394 | 0.504 | 0.373 | 0.509 | 0.494 | 0.372 | 0.372 | 0.372 | 0.490 | 0.371 |
| Adjusted $R^2$ | 0.358 | 0.379 | 0.492 | 0.357 | 0.494 | 0.482 | 0.357 | 0.357 | 0.356 | 0.478 | 0.358 |

Notes: Each regression includes period fixed effects. Clustered standard errors in brackets. *p<0.1, **p<0.05, ***p<0.01. The last column includes the instrumental variables estimate (see Supplementary Note 8).

**Income inequality multidimensional regression model additional explanatory variables robustness check:** In Supplementary Table 13 we reproduce the main results for the income inequality regression analysis (Columns (1-12)) and test the robustness of the multidimensional income inequality regression by adding the log of the population, the log of the initial human capital, and the log of natural resource exports as additional explanatory variables, separately (Columns (13-15)) and together (Column (16)). In each case the product ECI and the technology ECI remain significant predictors of income inequality, thus confirming the robustness of the regression results. More importantly, the human capital and populations appear also significant. Therefore, in column (17) we re-estimate the income inequality model by including only these two as additional explanatory variables. This is our final model, i.e., the model that has the best explanatory power out of all income inequality regression analyses (adjusted $R^2 = 0.680$). The country-fixed effects model (Column (18)) does not adequately represent the variation in our data. Supplementary Figure 8 gives the correlations between the variables used in these regressions.

**Income inequality multidimensional regression model production intensity robustness check:** In Supplementary Table 14, Columns (2-12), we estimate the production intensity income inequality models. We find that the multidimensional model includes the trade, technological, and research intensities but not their interaction terms (adjusted $R^2 = 0.608$). By comparing this model directly to the multidimensional ECI model (Column (13)), in a regression specification which also all of the potential explanatory variables (Column (14)), and in the final model regression specification (Column (15)), we find that the multidimensional ECI model coefficients remain significant. Only, in the model in Column (13), the ECI (technology) coefficient loses significance, but it regains it when we include the variables of the final model specification (Column (15)). The country-fixed effects model (Column (16)) does not adequately represent the variation in our data. Hence the multidimensional ECI income inequality model is robust against the model of production intensity. Supplementary Figure 9 gives the correlations between the variables used in these regressions.

**Income inequality multidimensional regression model HHI robustness check:** In Supplementary Table 15, Columns (2-12), we estimate the HHI income inequality models. We find that the multidimensional model includes the trade and technological HHI, but not their interaction term (adjusted $R^2 = 0.516$). By comparing this model directly to the multidimensional ECI model (Column (13)), in a regression specification which also all of the potential explanatory variables (Column (14)), and in the final model regression specification (Column (15)), we find that the multidimensional ECI model clearly outperforms the multidimensional HHI model as the coefficients of the former remain highly significant, whereas the coefficients of the later lose significance. Only, in the model in Column (13), the ECI (technology) coefficient loses significance, but it regains it when we include the variables of the final model specification (Column (15)). The country-fixed effects model (Column (16)) does not adequately represent the variation in our data. Supplementary Figure 10 gives the correlations between the variables used in these regressions.

**Income inequality multidimensional regression model Entropy robustness check:** In Supplementary Table 16, Columns (2-12), we estimate the Entropy income inequality models. Identically to the HHI case, we find that the multidimensional model includes the trade and technological Entropy, but not their interaction term (adjusted $R^2 = 0.522$). By comparing this model directly to the multidimensional ECI model (Column (13)), in a regression specification which also all of the potential explanatory variables (Column (14)), and in the final model regression specification (Column (15)), we find that the multidimensional ECI model clearly outperforms the production intensity model as the coefficients of the former remain highly significant, whereas the coefficients of the later lose significance. Only, in the model in Column (13), the ECI (technology) coefficient loses significance, but it regains it when we include the variables of the final model specification (Column (15)). The country-fixed effects model (Column (16)) does not adequately represent the variation in our data. Supplementary Figure 11 gives the correlations between the variables used in these regressions.

**Income inequality multidimensional regression model Fitness robustness check:** In Supplementary Table 17, Columns (2-12), we estimate the Fitness income inequality models. We find that the multidimensional Fitness model includes the trade, technological, and research Fitness, but not their interaction terms (adjusted $R^2 = 0.576$). By comparing this model directly to the multidimensional ECI model (Column (13)), in a regression specification which also all of the potential explanatory variables (Column (14)), and in the final model regression specification (Column (15)), we find that the multidimensional ECI model coefficients remain significant. Hence, the multidimensional ECI income inequality model is robust against the multidimensional Fitness income inequality model. The country-fixed effects model (Column (16)) does not adequately represent the variation in our data. Supplementary Figure 12 gives the correlations between the variables used in these regressions.

**Income inequality multidimensional regression model i-ECI robustness check:** In Supplementary Table 17, Columns (2-4), we estimate the i-ECI income inequality models. We find that the multidimensional i-ECI model includes just the i-ECI (adjusted $R^2 = 0.474$). By comparing this model directly to the multidimensional ECI model (Column (5)), in a regression specification which also all of the potential explanatory variables (Column (6)), and in the final model regression specification (Column (7)), we find that the multidimensional ECI model coefficients remain significant. Hence, the multidimensional ECI income inequality model is robust against the multidimensional i-ECI income inequality model. The country-fixed effects model (Column (8)) does not adequately represent the variation in our data. Supplementary Figure 13 gives the correlations between the variables used in these regressions.

**Supplementary Table 13. Income Inequality Regressions: Additional Explanatory Variables Robustness Check**

| | *Dependent variable:* | | | | | | | | | | | | | | | | | |
|---|---|---|---|---|---|---|---|---|---|---|---|---|---|---|---|---|---|---|
| | Gini coefficient (1996-99, 2000-03, 2004-07, 2008-11, 2012-15) | | | | | | | | | | | | | | | | | |
| | (1) | (2) | (3) | (4) | (5) | (6) | (7) | (8) | (9) | (10) | (11) | (12) | (13) | (14) | (15) | (16) | (17) | (18) |
| ECI (trade) | | -23.543*** | | | -17.902*** | -23.116*** | | -17.778*** | -9.279 | -21.289** | | -18.449 | -18.418*** | -13.587*** | -21.293*** | -16.195*** | -15.680*** | -3.264 |
| | | (5.285) | | | (5.495) | (5.328) | | (5.565) | (9.547) | (9.295) | | (26.796) | (4.091) | (5.050) | (4.464) | (3.466) | (3.810) | (3.561) |
| ECI (technology) | | | -11.211*** | | -5.269** | | -12.317*** | -6.216** | 1.208 | | -3.964 | 11.923 | -10.520*** | -5.214** | -5.176** | -9.442*** | -9.611*** | -1.274 |
| | | | (2.715) | | (2.685) | | (2.582) | (2.528) | (6.101) | | (5.742) | (25.183) | (2.388) | (2.297) | (2.527) | (2.264) | (2.328) | (2.216) |
| ECI (research) | | | | -7.654 | | -1.336 | 3.400 | 2.783 | | 0.649 | 16.132 | 21.084 | | | | | | |
| | | | | (5.310) | | (4.017) | (3.665) | (3.386) | | (8.981) | (10.310) | (31.233) | | | | | | |
| ECI (trade) x ECI (technology) | | | | | | | | | -11.449 | | | -8.570 | | | | | | |
| | | | | | | | | | (10.851) | | | (41.156) | | | | | | |
| ECI (trade) x ECI (research) | | | | | | | | | | -2.990 | | -0.339 | | | | | | |
| | | | | | | | | | | (13.710) | | (54.046) | | | | | | |
| ECI (research) x ECI (technology) | | | | | | | | | | | -16.026 | -31.788 | | | | | | |
| | | | | | | | | | | | (12.629) | (42.894) | | | | | | |
| ECI (trade) x ECI (research) x ECI (technology) | | | | | | | | | | | | 12.933 | | | | | | |
| | | | | | | | | | | | | (69.991) | | | | | | |
| Log of population | | | | | | | | | | | | | 1.517*** | | | 1.219*** | 1.264*** | 1.942 |
| | | | | | | | | | | | | | (0.259) | | | (0.278) | (0.259) | (2.080) |
| Log of human capital | | | | | | | | | | | | | | -9.038*** | | -5.449*** | -5.554*** | -11.103 |
| | | | | | | | | | | | | | (2.422) | | (1.995) | (1.938) | (9.509) |
| Log of natural resource exports per capita | | | | | | | | | | | | | | | -1.720** | -0.244 | | |
| | | | | | | | | | | | | | | (0.715) | (0.614) | | |
| Log of GDP per capita | -10.019 | -8.070 | -3.795 | -14.997 | -5.612 | -8.974 | -0.970 | -3.287 | -9.554 | -9.697 | -3.620 | -6.090 | 5.067 | 6.988 | -4.380 | 10.741 | 11.031 | -9.927 |
| | (14.782) | (10.761) | (11.358) | (15.911) | (10.702) | (11.690) | (12.768) | (11.733) | (11.611) | (12.686) | (12.734) | (12.355) | (8.584) | (9.484) | (10.162) | (8.025) | (8.161) | (11.821) |
| Log of GDP per capita, squared | 0.300 | 0.361 | 0.071 | 0.616 | 0.239 | 0.415 | -0.092 | 0.104 | 0.442 | 0.452 | 0.042 | 0.252 | -0.210 | -0.353 | 0.331 | -0.466 | -0.499 | 0.291 |
| | (0.780) | (0.569) | (0.595) | (0.866) | (0.557) | (0.626) | (0.683) | (0.621) | (0.608) | (0.679) | (0.683) | (0.654) | (0.447) | (0.488) | (0.522) | (0.411) | (0.420) | (0.620) |
| Fixed effects | No | No | No | No | No | No | No | No | No | No | No | No | No | No | No | No | No | Yes |
| Observations | 332 | 332 | 332 | 332 | 332 | 332 | 332 | 332 | 332 | 332 | 332 | 332 | 332 | 332 | 332 | 332 | 332 | 332 |
| $R^2$ | 0.346 | 0.551 | 0.496 | 0.371 | 0.573 | 0.552 | 0.500 | 0.575 | 0.576 | 0.552 | 0.508 | 0.590 | 0.670 | 0.630 | 0.601 | 0.690 | 0.689 | 0.204 |
| Adjusted $R^2$ | 0.334 | 0.542 | 0.485 | 0.358 | 0.562 | 0.541 | 0.487 | 0.563 | 0.564 | 0.540 | 0.494 | 0.573 | 0.661 | 0.620 | 0.590 | 0.679 | 0.680 | -0.089 |

Notes: Each regression includes period fixed effects. Clustered standard errors in brackets. *p<0.1, **p<0.05, ***p<0.01.

## Supplementary Table 14. Income Inequality Regressions: Production Intensity Robustness Check

| | *Dependent variable:* | | | | | | | | | | | | | | | |
|---|---|---|---|---|---|---|---|---|---|---|---|---|---|---|---|---|
| | Gini coefficient (1996-99, 2000-03, 2004-07, 2008-11, 2012-15) | | | | | | | | | | | | | | | |
| | (1) | (2) | (3) | (4) | (5) | (6) | (7) | (8) | (9) | (10) | (11) | (12) | (13) | (14) | (15) | (16) |
| ECI (trade) | -17.902*** | | | | | | | | | | | | -10.748** | -11.182*** | -13.670*** | -2.186 |
| | (5.495) | | | | | | | | | | | | (4.885) | (3.689) | (3.957) | (3.506) |
| ECI (technology) | -5.269** | | | | | | | | | | | | -3.017 | -9.119*** | -8.575*** | -1.034 |
| | (2.685) | | | | | | | | | | | | (2.589) | (2.525) | (2.437) | (2.684) |
| Intensity (trade) | | -21.455*** | | | -14.519*** | -19.247*** | | -15.104*** | -4.776 | 22.323 | | 25.613 | -16.230*** | -15.248*** | -10.307*** | -2.619 |
| | | (4.640) | | | (4.663) | (4.658) | | (4.748) | (7.141) | (15.136) | | (26.907) | (4.968) | (4.934) | (4.467) | (3.719) |
| Intensity (technology) | | | -19.928*** | | -18.307*** | | -17.212*** | -14.854*** | -6.191 | | 1.830 | 16.722 | -7.109 | 1.998 | 1.837 | -1.840 |
| | | | (3.696) | | (3.868) | | (4.331) | (4.453) | (5.757) | | (9.315) | (17.106) | (5.279) | (5.120) | (5.195) | (5.615) |
| Intensity (research) | | | | -22.407*** | | -21.222*** | -5.574 | -6.950* | | 10.873 | 1.578 | 23.169 | -4.873 | -5.277 | -5.506 | -5.161 |
| | | | | (5.459) | | (5.243) | (4.032) | (3.821) | | (8.934) | (5.291) | (16.131) | (3.656) | (3.750) | (3.571) | (5.836) |
| Intensity (trade) x Intensity (technology) | | | | | | | | | -17.888* | | | -49.815 | | | | |
| | | | | | | | | | (9.430) | | | (32.977) | | | | |
| Intensity (trade) x Intensity (research) | | | | | | | | | | -52.374*** | | -52.405 | | | | |
| | | | | | | | | | | (17.818) | | (36.646) | | | | |
| Intensity (research) x Intensity (technology) | | | | | | | | | | | -21.497* | -36.498 | | | | |
| | | | | | | | | | | | (11.099) | (25.021) | | | | |
| Intensity (trade) x Intensity (research) x Intensity (technology) | | | | | | | | | | | | 58.659 | | | | |
| | | | | | | | | | | | | (38.116) | | | | |
| Log of population | | | | | | | | | | | | | | 1.055*** | 0.968*** | 2.147 |
| | | | | | | | | | | | | | | (0.301) | (0.296) | (2.391) |
| Log of human capital | | | | | | | | | | | | | | -5.988*** | -5.693*** | -8.481 |
| | | | | | | | | | | | | | | (1.828) | (1.857) | (10.886) |
| Log of natural resource exports per capita | | | | | | | | | | | | | | 1.034 | | |
| | | | | | | | | | | | | | | (0.742) | | |
| Log of GDP per capita | -5.612 | -9.542 | -14.450 | -7.712 | -13.766 | -7.405 | -13.272 | -12.270 | -28.783** | -39.790** | -25.113** | -35.952** | -8.172 | 10.671 | 10.188 | -6.433 |
| | (10.702) | (13.907) | (9.384) | (11.140) | (8.809) | (9.888) | (9.253) | (8.627) | (13.720) | (15.912) | (11.259) | (14.068) | (8.988) | (8.873) | (8.956) | (10.773) |
| Log of GDP per capita, squared | 0.239 | 0.529 | 0.800 | 0.405 | 0.915* | 0.605 | 0.758 | 0.867* | 1.666** | 2.210*** | 1.367** | 2.055*** | 0.641 | -0.393 | -0.335 | 0.171 |
| | (0.557) | (0.745) | (0.508) | (0.604) | (0.477) | (0.542) | (0.501) | (0.465) | (0.717) | (0.825) | (0.605) | (0.718) | (0.470) | (0.469) | (0.476) | (0.561) |
| Fixed effects | No | No | No | No | No | No | No | No | No | No | No | No | No | No | No | Yes |
| Observations | 332 | 332 | 332 | 332 | 332 | 332 | 332 | 332 | 332 | 332 | 332 | 332 | 332 | 332 | 332 | 332 |
| $R^2$ | 0.573 | 0.425 | 0.577 | 0.491 | 0.611 | 0.554 | 0.581 | 0.619 | 0.623 | 0.594 | 0.595 | 0.639 | 0.654 | 0.712 | 0.707 | 0.220 |
| Adjusted $R^2$ | 0.562 | 0.412 | 0.568 | 0.480 | 0.602 | 0.543 | 0.571 | 0.608 | 0.613 | 0.582 | 0.584 | 0.624 | 0.642 | 0.699 | 0.695 | -0.081 |

Notes: Each regression includes period fixed effects. Clustered standard errors in brackets. *p<0.1, **p<0.05, ***p<0.01.

# Supplementary Table 15. Income Inequality Regressions: Herfindahl-Hirschman Index Robustness Check

| | Dependent variable: | | | | | | | | | | | | | | | |
|---|---|---|---|---|---|---|---|---|---|---|---|---|---|---|---|---|
| | Gini coefficient (1996-99, 2000-03, 2004-07, 2008-11, 2012-15) | | | | | | | | | | | | | | | |
| | (1) | (2) | (3) | (4) | (5) | (6) | (7) | (8) | (9) | (10) | (11) | (12) | (13) | (14) | (15) | (16) |
| ECI (trade) | -17.902*** | | | | | | | | | | | | -14.351** | -14.745*** | -14.117*** | -5.603 |
| | (5.495) | | | | | | | | | | | | (5.699) | (3.765) | (4.127) | (3.440) |
| ECI (technology) | -5.269** | | | | | | | | | | | | -3.594 | -7.256*** | -8.035*** | -0.459 |
| | (2.685) | | | | | | | | | | | | (3.237) | (2.800) | (2.831) | (2.021) |
| HHI (trade) | | 17.627*** | | | 14.905*** | 17.744*** | | 15.320*** | 13.758*** | 17.575** | | 18.041*** | 7.315+ | 4.365 | 3.190 | -8.193** |
| | | (5.125) | | | (4.684) | (5.287) | | (4.849) | (5.253) | (6.847) | | (7.557) | (4.006) | (4.478) | (3.889) | (3.337) |
| HHI (technology) | | | 10.444*** | | 6.103*** | | 10.386*** | 6.586*** | 5.292* | | 10.765*** | 8.966** | 0.646 | 1.397 | 1.281 | 1.669 |
| | | | (2.487) | | (2.013) | | (2.270) | (1.945) | (2.866) | | (3.434) | (3.666) | (2.516) | (2.476) | (2.519) | (2.060) |
| HHI (research) | | | | 18.060 | | -1.858 | 0.961 | -9.979 | | -2.425 | 3.529 | 3.464 | | | | |
| | | | | (16.517) | | (11.535) | (11.818) | (10.487) | | (10.957) | (15.470) | (11.555) | | | | |
| HHI (trade) x HHI (technology) | | | | | | | | | 5.187 | | | -13.822 | | | | |
| | | | | | | | | | (10.303) | | | (14.198) | | | | |
| HHI (trade) x HHI (research) | | | | | | | | | | 4.656 | | -116.491 | | | | |
| | | | | | | | | | | (80.243) | | (81.853) | | | | |
| HHI (research) x HHI (technology) | | | | | | | | | | | -11.774 | -98.861 | | | | |
| | | | | | | | | | | | (57.293) | (65.543) | | | | |
| HHI (trade) x HHI (research) x HHI (technology) | | | | | | | | | | | | 524.349** | | | | |
| | | | | | | | | | | | | (238.467) | | | | |
| Log of population | | | | | | | | | | | | | | 1.061*** | 1.198*** | 2.617 |
| | | | | | | | | | | | | | | (0.282) | (0.251) | (2.180) |
| Log of human capital | | | | | | | | | | | | | | -5.402*** | -5.610*** | -12.043 |
| | | | | | | | | | | | | | | (1.888) | (1.875) | (9.201) |
| Log of natural resource exports per capita | | | | | | | | | | | | | | | -0.581 | |
| | | | | | | | | | | | | | | | (0.694) | |
| Log of GDP per capita | -5.612 | -7.442 | -0.620 | -10.222 | -2.348 | -7.404 | -0.684 | -1.741 | -2.621 | -7.398 | -1.007 | -1.635 | -5.185 | 10.630 | 11.387 | -5.588 |
| | (10.702) | (11.387) | (12.195) | (14.253) | (10.793) | (11.527) | (12.489) | (11.081) | (10.749) | (11.502) | (12.463) | (10.222) | (10.494) | (8.173) | (8.436) | (13.312) |
| Log of GDP per capita, squared | 0.239 | 0.166 | -0.135 | 0.321 | -0.067 | 0.163 | -0.131 | -0.102 | -0.057 | 0.162 | -0.113 | -0.111 | 0.181 | -0.462 | -0.537 | 0.125 |
| | (0.557) | (0.597) | (0.645) | (0.756) | (0.564) | (0.608) | (0.662) | (0.582) | (0.561) | (0.607) | (0.662) | (0.538) | (0.545) | (0.422) | (0.436) | (0.688) |
| Fixed effects | No | No | No | No | No | No | No | No | No | No | No | No | No | No | No | Yes |
| Observations | 332 | 332 | 332 | 332 | 332 | 332 | 332 | 332 | 332 | 332 | 332 | 332 | 332 | 332 | 332 | 332 |
| $R^2$ | 0.573 | 0.502 | 0.430 | 0.355 | 0.527 | 0.502 | 0.430 | 0.530 | 0.528 | 0.502 | 0.430 | 0.538 | 0.589 | 0.695 | 0.693 | 0.235 |
| Adjusted $R^2$ | 0.562 | 0.492 | 0.418 | 0.341 | 0.516 | 0.490 | 0.416 | 0.517 | 0.515 | 0.489 | 0.414 | 0.520 | 0.577 | 0.682 | 0.681 | -0.055 |

Notes: Each regression includes period fixed effects. Clustered standard errors in brackets. *p<0.1, **p<0.05, ***p<0.01.

# Supplementary Table 16. Income Inequality Regressions: Entropy Robustness Check

| | (1) | (2) | (3) | (4) | (5) | (6) | (7) | (8) | (9) | (10) | (11) | (12) | (13) | (14) | (15) | (16) |
|---|---|---|---|---|---|---|---|---|---|---|---|---|---|---|---|---|
| | *Dependent variable:* | | | | | | | | | | | | | | | |
| | Gini coefficient (1996-99, 2000-03, 2004-07, 2008-11, 2012-15) | | | | | | | | | | | | | | | |
| ECI (trade) | -17.902*** | | | | | | | | | | | | -13.992** | -14.387*** | -13.738*** | -5.655 |
| | (5.495) | | | | | | | | | | | | (5.704) | (3.778) | (4.148) | (3.449) |
| ECI (technology) | -5.269** | | | | | | | | | | | | -3.556 | -6.891** | -7.768*** | -0.362 |
| | (2.685) | | | | | | | | | | | | (3.236) | (2.793) | (2.833) | (2.034) |
| Entropy (trade) | | -16.592*** | | | -14.066*** | -16.815*** | | -14.572*** | -18.637** | -25.472 | | 277.069** | -7.159* | -4.760 | -3.502 | 7.692** |
| | | (4.572) | | | (4.213) | (4.785) | | (4.415) | (8.486) | (56.120) | | (130.475) | (3.695) | (4.203) | (3.620) | (3.222) |
| Entropy (technology) | | | -10.379*** | | -5.759*** | | -10.389*** | -6.411*** | -10.662 | | 7.725 | -266.731** | -0.515 | -1.524 | -1.413 | -1.828 |
| | | | (2.516) | | (1.991) | | (2.302) | (1.925) | (7.718) | | (44.193) | (119.991) | (2.527) | (2.499) | (2.557) | (2.035) |
| Entropy (research) | | | | -15.885 | | 2.887 | 0.122 | 10.262 | | -4.804 | 14.320 | -192.646* | | | | |
| | | | | (13.330) | | (9.798) | (10.262) | (9.234) | | (56.192) | (38.660) | (104.124) | | | | |
| Entropy (trade) x Entropy (technology) | | | | | | | | | 5.966 | | | 338.822** | | | | |
| | | | | | | | | | (9.489) | | | (155.384) | | | | |
| Entropy (trade) x Entropy (research) | | | | | | | | | | 9.062 | | 271.120* | | | | |
| | | | | | | | | | | (60.271) | | (139.188) | | | | |
| Entropy (research) x Entropy (technology) | | | | | | | | | | | -18.871 | 268.347** | | | | |
| | | | | | | | | | | | (47.062) | (128.961) | | | | |
| Entropy (trade) x Entropy (research) x Entropy (technology) | | | | | | | | | | | | -349.203** | | | | |
| | | | | | | | | | | | | (165.740) | | | | |
| Log of population | | | | | | | | | | | | | | 1.034*** | 1.187*** | 2.610 |
| | | | | | | | | | | | | | | (0.283) | (0.251) | (2.170) |
| Log of human capital | | | | | | | | | | | | | | -5.421*** | -5.644*** | -11.964 |
| | | | | | | | | | | | | | | (1.864) | (1.856) | (9.188) |
| Log of natural resource exports per capita | | | | | | | | | | | | | | | -0.650 | |
| | | | | | | | | | | | | | | | (0.700) | |
| Log of GDP per capita | -5.612 | -6.619 | -0.143 | -10.244 | -1.657 | -6.533 | -0.132 | -0.788 | -1.976 | -6.514 | -0.836 | -0.970 | -4.929 | 10.857 | 11.648 | -5.604 |
| | (10.702) | (11.110) | (12.127) | (14.218) | (10.638) | (11.289) | (12.479) | (10.953) | (10.597) | (11.228) | (12.525) | (10.175) | (10.459) | (8.220) | (8.499) | (13.249) |
| Log of GDP per capita, squared | 0.239 | 0.124 | -0.157 | 0.324 | -0.102 | 0.117 | -0.158 | -0.152 | -0.090 | 0.116 | -0.119 | -0.146 | 0.165 | -0.473 | -0.554 | 0.127 |
| | (0.557) | (0.582) | (0.641) | (0.755) | (0.555) | (0.596) | (0.661) | (0.575) | (0.552) | (0.592) | (0.666) | (0.535) | (0.543) | (0.424) | (0.439) | (0.685) |
| Fixed effects | No | No | No | No | No | No | No | No | No | No | No | No | No | No | No | Yes |
| Observations | 332 | 332 | 332 | 332 | 332 | 332 | 332 | 332 | 332 | 332 | 332 | 332 | 332 | 332 | 332 | 332 |
| $R^2$ | 0.573 | 0.511 | 0.432 | 0.357 | 0.534 | 0.511 | 0.432 | 0.537 | 0.535 | 0.511 | 0.433 | 0.546 | 0.591 | 0.697 | 0.694 | 0.235 |
| Adjusted $R^2$ | 0.562 | 0.500 | 0.420 | 0.343 | 0.522 | 0.499 | 0.418 | 0.524 | 0.522 | 0.498 | 0.417 | 0.527 | 0.578 | 0.684 | 0.682 | -0.055 |

Notes: Each regression includes period fixed effects. Clustered standard errors in brackets. *p<0.1, **p<0.05, ***p<0.01.

## Supplementary Table 17. Income Inequality Regressions: Fitness Robustness Check

| | Dependent variable: | | | | | | | | | | | | | | | |
|---|---|---|---|---|---|---|---|---|---|---|---|---|---|---|---|---|
| | Gini coefficient (1996-99, 2000-03, 2004-07, 2008-11, 2012-15) | | | | | | | | | | | | | | | |
| | (1) | (2) | (3) | (4) | (5) | (6) | (7) | (8) | (9) | (10) | (11) | (12) | (13) | (14) | (15) | (16) |
| ECI (trade) | -17.902*** | | | | | | | | | | | | -10.105* | -8.029** | -7.106* | -4.897 |
| | (5.495) | | | | | | | | | | | | (5.930) | (3.750) | (4.143) | (3.506) |
| ECI (technology) | -5.269** | | | | | | | | | | | | -6.342* | -7.000*** | -8.351*** | -0.106 |
| | (2.685) | | | | | | | | | | | | (3.319) | (2.328) | (2.365) | (2.212) |
| Log of fitness (trade) | | -22.434*** | | | -20.455*** | -27.655*** | | -25.241*** | -43.974*** | -47.991** | | -18.194 | -16.112*** | -18.753*** | -16.213*** | 4.409 |
| | | (4.278) | | | (4.810) | (4.844) | | (5.183) | (9.154) | (19.581) | | (49.347) | (5.343) | (4.859) | (4.360) | (5.584) |
| Log of fitness (technology) | | | -8.974*** | | -2.183 | | -8.908*** | -3.420** | -24.667*** | | 14.045 | 33.249 | 1.853 | 1.450 | 1.802 | -1.319 |
| | | | (2.309) | | (1.720) | | (1.963) | (1.605) | (8.336) | | (11.977) | (40.431) | (2.272) | (1.668) | (1.645) | (1.583) |
| Log of fitness (research) | | | | -9.247 | | 11.978** | -0.317 | 13.554*** | | -8.855 | 27.488* | 64.110 | 13.588*** | 2.684 | 1.922 | -4.522 |
| | | | | (6.479) | | (4.778) | (5.873) | (4.887) | | (20.264) | (15.242) | (54.128) | (4.797) | (4.344) | (4.379) | (6.636) |
| Log of fitness (trade) x Log of fitness (technology) | | | | | | | | | | | | | | -5.317*** | -5.805*** | -10.937 |
| | | | | | | | | | | | | | | (1.508) | (1.529) | (8.853) |
| Log of fitness (trade) x Log of fitness (research) | | | | | | | | | | | | | | -1.064 | | |
| | | | | | | | | | | | | | | (0.657) | | |
| Log of fitness (research) x Log of fitness (technology) | | | | | | | | | | | -34.298* | -86.583 | | | | |
| | | | | | | | | | | | (18.946) | (63.542) | | | | |
| Log of fitness (trade) x Log of fitness (research) x Log of fitness (technology) | | | | | | | | | | | | 85.540 | | | | |
| | | | | | | | | | | | | (91.520) | | | | |
| Log of population | | | | | | | | | 30.937*** | | | -28.039 | | | | |
| | | | | | | | | | (11.313) | | | (58.024) | | | | |
| Log of human capital | | | | | | | | | 27.834 | | | -45.708 | | | | |
| | | | | | | | | | (26.143) | | | (80.922) | | | | |
| Log of natural resource exports per capita | | | | | | | | | | | | | | 1.171*** | 1.360*** | 3.130 |
| | | | | | | | | | | | | | | (0.278) | (0.278) | (2.326) |
| Log of GDP per capita | -5.612 | -7.928 | 0.239 | -13.276 | -5.617 | -3.223 | 0.051 | 1.016 | -2.426 | -1.828 | -4.941 | 1.461 | -1.493 | 9.791 | 10.596 | -8.034 |
| | (10.702) | (10.007) | (13.340) | (15.279) | (11.011) | (9.926) | (15.191) | (11.279) | (10.397) | (9.931) | (15.574) | (10.698) | (10.259) | (7.564) | (7.816) | (11.537) |
| Log of GDP per capita, squared | 0.239 | 0.221 | -0.166 | 0.508 | 0.115 | -0.067 | -0.156 | -0.271 | -0.074 | -0.147 | 0.113 | -0.316 | -0.069 | -0.430 | -0.540 | 0.231 |
| | (0.557) | (0.519) | (0.696) | (0.824) | (0.565) | (0.522) | (0.807) | (0.586) | (0.534) | (0.523) | (0.832) | (0.560) | (0.532) | (0.396) | (0.404) | (0.604) |
| Fixed effects | No | No | No | No | No | No | No | No | No | No | No | No | No | No | No | Yes |
| Observations | 332 | 332 | 332 | 332 | 332 | 332 | 332 | 332 | 332 | 332 | 332 | 332 | 332 | 332 | 332 | 332 |
| $R^2$ | 0.573 | 0.556 | 0.438 | 0.366 | 0.560 | 0.579 | 0.438 | 0.588 | 0.578 | 0.584 | 0.454 | 0.617 | 0.621 | 0.739 | 0.731 | 0.220 |
| Adjusted $R^2$ | 0.562 | 0.547 | 0.426 | 0.353 | 0.549 | 0.568 | 0.424 | 0.576 | 0.566 | 0.572 | 0.439 | 0.601 | 0.608 | 0.727 | 0.720 | -0.080 |

Notes: Each regression includes period fixed effects. Clustered standard errors in brackets. *p<0.1, **p<0.05, ***p<0.01.

## Supplementary Table 18. Income Inequality Regressions: i-ECI Robustness Check

|  | *Dependent variable:* | | | | | | | |
|---|---|---|---|---|---|---|---|---|
|  | Gini coefficient (1996-99, 2000-03, 2004-07, 2008-11, 2012-15) | | | | | | | |
|  | (1) | (2) | (3) | (4) | (5) | (6) | (7) | (8) |
| ECI (trade) | -17.902*** |  |  |  | -26.654*** | -27.749*** | -25.952*** | -9.702* |
|  | (5.495) |  |  |  | (6.914) | (5.989) | (6.039) | (5.067) |
| ECI (technology) | -5.269** |  |  |  | -5.423** | -9.190*** | -9.526*** | -1.396 |
|  | (2.685) |  |  |  | (2.648) | (2.103) | (2.152) | (2.188) |
| i-ECI |  | -17.332*** | -16.518*** | -16.252* | 8.868 | 11.768** | 10.919** | 8.139** |
|  |  | (5.848) | (5.670) | (9.203) | (6.814) | (5.098) | (5.192) | (3.777) |
| ECI (research) |  |  | -3.352 | -3.147 |  |  |  |  |
|  |  |  | (4.204) | (6.194) |  |  |  |  |
| i-ECI x ECI (research) |  |  |  | -0.431 |  |  |  |  |
|  |  |  |  | (13.042) |  |  |  |  |
| Log of population |  |  |  |  |  | 1.093*** | 1.187*** | 1.133 |
|  |  |  |  |  |  | (0.256) | (0.228) | (1.972) |
| Log of human capital |  |  |  |  |  | -6.432*** | -6.556*** | -11.847 |
|  |  |  |  |  |  | (1.935) | (1.895) | (9.300) |
| Log of natural resource export per capita |  |  |  |  |  | -0.473 |  |  |
|  |  |  |  |  |  | (0.590) |  |  |
| Log of GDP per capita | 0.239 | 0.362 | 0.497 | 0.503 | 0.227 | -0.496 | -0.556 | 0.019 |
|  | (0.557) | (0.617) | (0.685) | (0.749) | (0.559) | (0.417) | (0.427) | (0.605) |
| Log of GDP per capita, squared | -5.612 | -8.833 | -11.069 | -11.181 | -5.409 | 11.655 | 12.133 | -4.783 |
|  | (10.702) | (11.541) | (12.653) | (13.869) | (10.747) | (8.182) | (8.346) | (11.580) |
| Fixed effects | No | No | No | No | No | No | Yes |  |
| Observations | 332 | 332 | 332 | 332 | 332 | 332 | 332 | 332 |
| $R^2$ | 0.573 | 0.486 | 0.490 | 0.490 | 0.580 | 0.701 | 0.699 | 0.227 |
| Adjusted $R^2$ | 0.562 | 0.474 | 0.478 | 0.476 | 0.568 | 0.690 | 0.689 | -0.062 |

Notes: Each regression includes period fixed effects. Clustered standard errors in brackets. *p<0.1, **p<0.05, ***p<0.01.

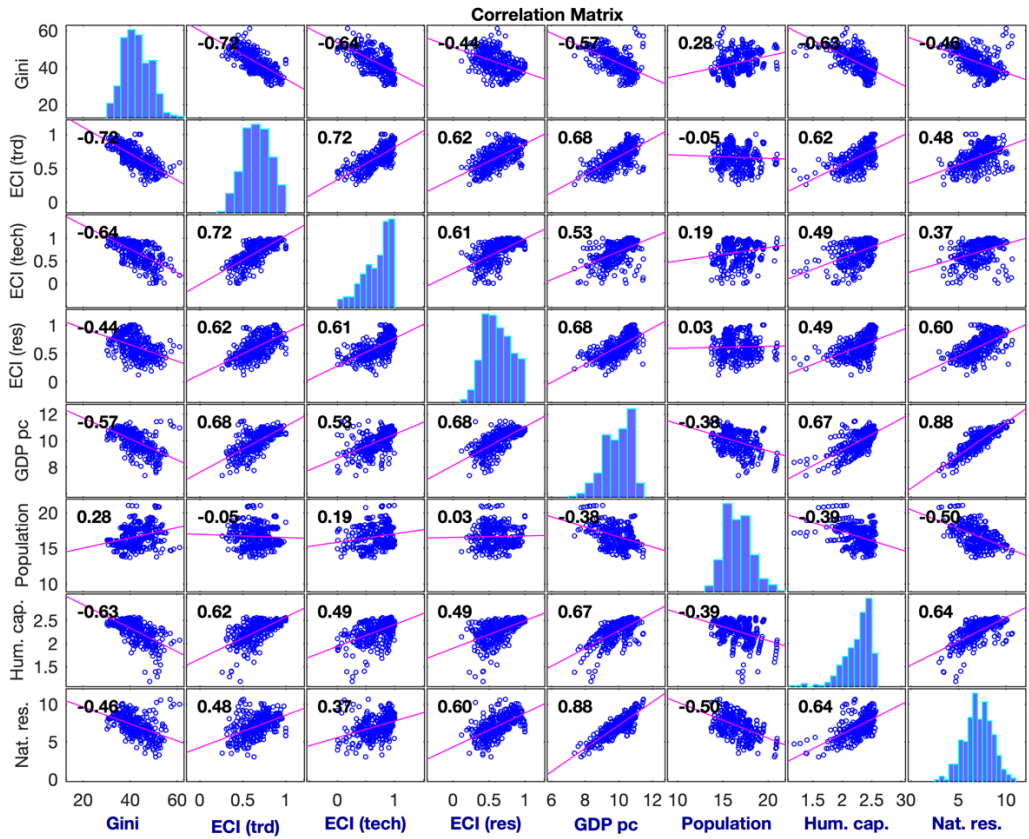

**Supplementary Figure 8. Correlations Between the Variables used in the Additional Explanatory Variables Income Inequality Robustness Check Regression Analysis**

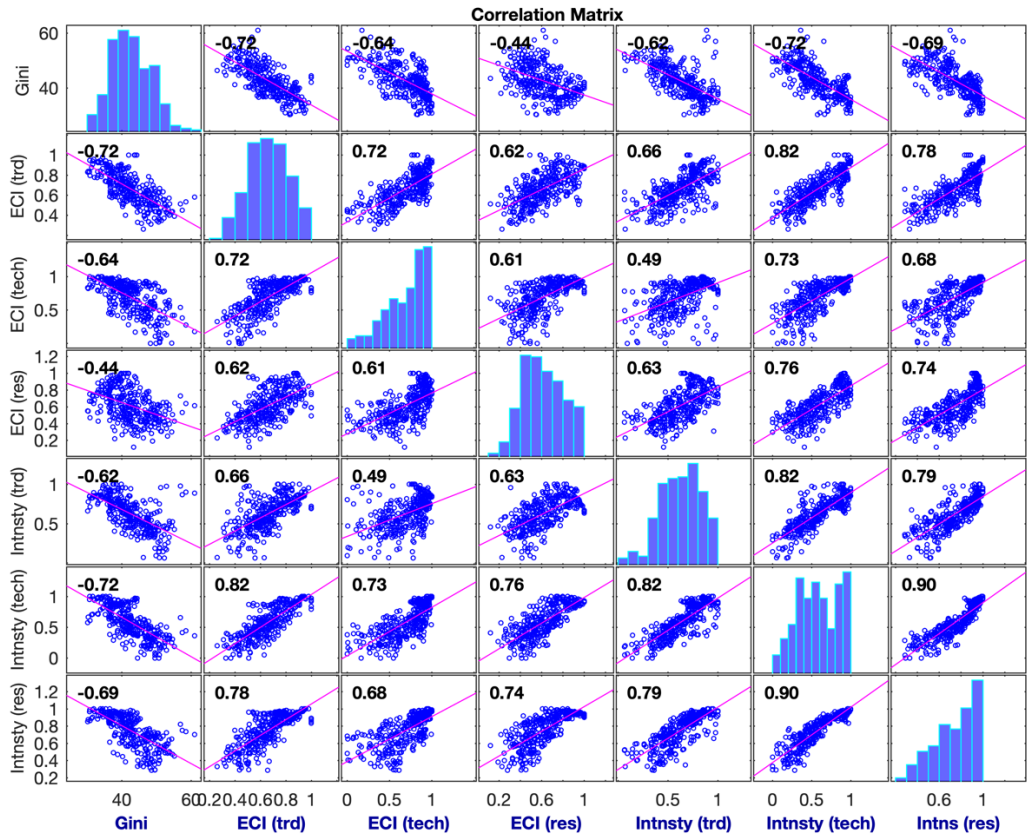

**Supplementary Figure 9. Correlations Between the Variables used in the Production Intensity Income Inequality Robustness Check Regression Analysis**

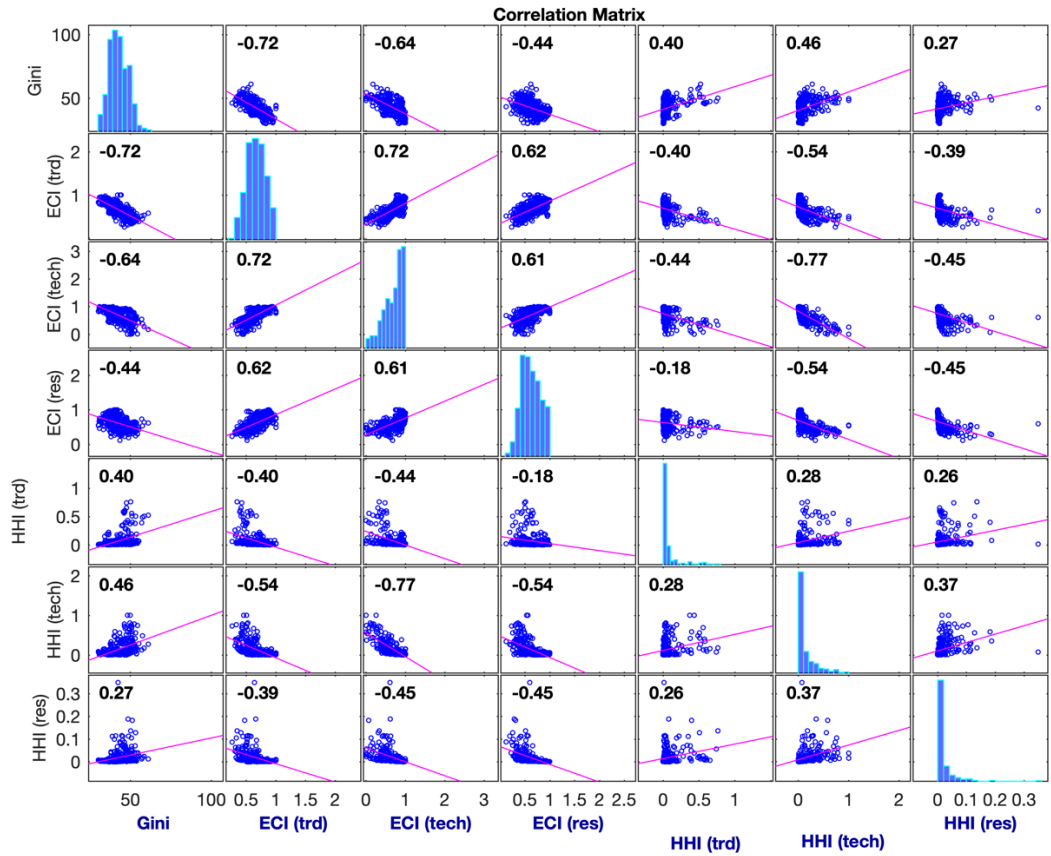

**Supplementary Figure 10. Correlations Between the Variables used in the HHI Income Inequality Robustness Check Regression Analysis**

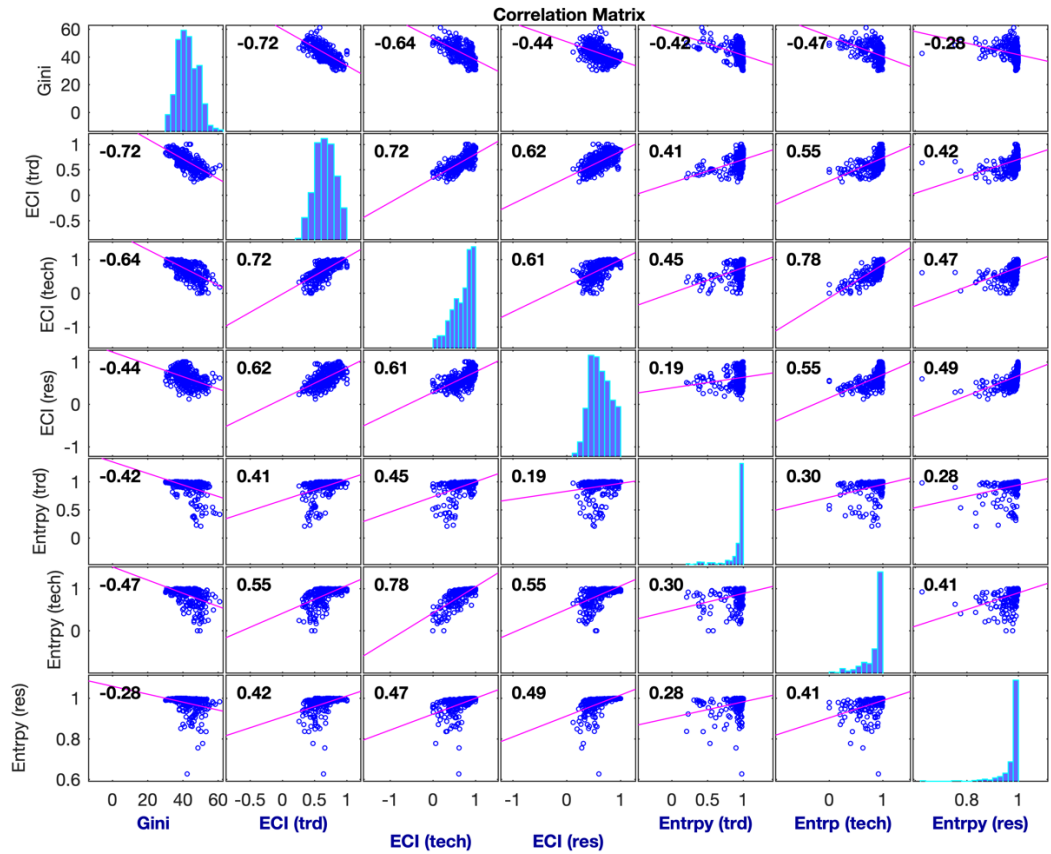

**Supplementary Figure 11. Correlations Between the Variables used in the Entropy Income Inequality Robustness Check Regression Analysis**

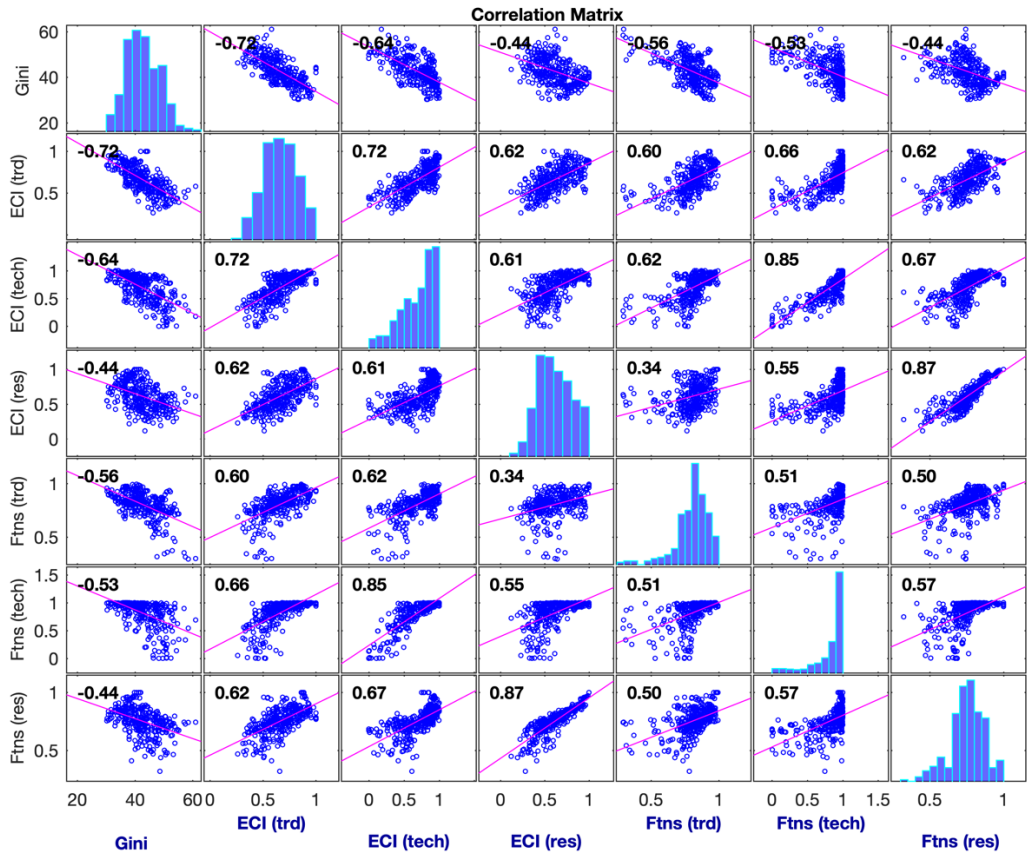

**Supplementary Figure 12. Correlations Between the Variables used in the Fitness Income Inequality Robustness Check Regression Analysis**

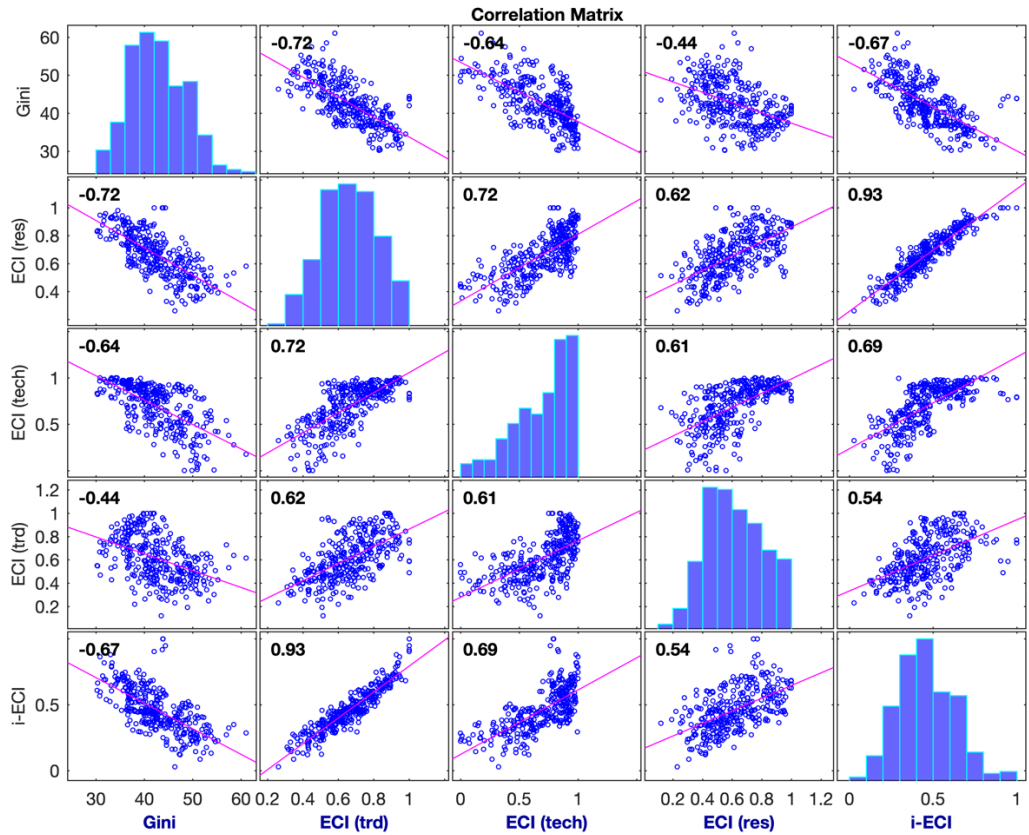

**Supplementary Figure 13. Correlations Between the Variables used in the i-ECI Income Inequality Robustness Check Regression Analysis**

# Supplementary Note 6. Emission intensity regression analysis

**Emission intensity regression setup:** In the emission intensity regression analysis, the dependent variable is defined as the log of the amount of greenhouse gas emissions (in kilotons of CO2 equivalent, CO2e), $GHG_{ct}$, as a share of GDP, $GDP_{ct}$, i.e.,

$$y_{ct} = \log\left(\frac{GHG_{ct}}{GDP_{ct}^{per\ captia} \times POP_{ct}}\right),$$

where $POP_{ct}$ is the population of country $c$ in year $t$. A larger value implies higher emission intensity. The data for greenhouse gas emissions were taken from the World Bank's World Development Indicators, and are available at

https://data.worldbank.org/indicator/EN.ATM.GHGT.KT.CE .

In the regressions, as control variables we include the log of the GDP per capita, log of population, log of human capital, and log of natural resources.

The regression analysis is focused on the periods 1996-1999, 2000-2003, 2004-2007, 2008-2011, 2012-2015 and 2016-19. Because of the sparseness of the greenhouse gas emissions dataset and slow temporal changes in the coefficients within a country, we follow [13] and use average values for each panel.

**Emission intensity individual regressions:** In Supplementary Tables 19-21, we present the results for the individual emission intensity regressions for ECI (trade), ECI (technology), and ECI (research), respectively. These results were used to estimate the F-statistics for testing the robustness of the individual regressions discussed in the main text.

## Supplementary Table 19. ECI (trade) Emission Intensity Regressions

| | \multicolumn{11}{c}{*Dependent variable:*} |
|---|---|---|---|---|---|---|---|---|---|---|---|
| | \multicolumn{11}{c}{GHG emissions per GDP (1996-99, 2000-03, 2004-07, 2008-11, 2012-15, 2016-19)} |
| | (1) | (2) | (3) | (4) | (5) | (6) | (7) | (8) | (9) | (10) | (11) |
| ECI (trade) | -1.015*** | -1.340*** | -0.619* | -1.223*** | -0.911* | -0.872* | -0.829 | -0.393 | -1.829*** | 0.207 | -1.389*** |
| | (0.350) | (0.400) | (0.372) | (0.447) | (0.502) | (0.515) | (0.523) | (0.556) | (0.696) | (0.208) | (0.539) |
| Intensity (trade) | | | | | -1.092* | | | | | | |
| | | | | | (0.625) | | | | | | |
| HHI (trade) | | | | | | 0.510 | | | | | |
| | | | | | | (0.365) | | | | | |
| Entropy (trade) | | | | | | | -0.507 | | | | |
| | | | | | | | (0.335) | | | | |
| log of Fitness (trade) | | | | | | | | -1.127*** | | | |
| | | | | | | | | (0.429) | | | |
| i-ECI | | | | | | | | | 0.662 | | |
| | | | | | | | | | (0.557) | | |
| Log of population | 0.012 | | | 0.067** | 0.063** | 0.056* | 0.056* | 0.083** | 0.059* | 0.200** | 0.070** |
| | (0.031) | | | (0.032) | (0.032) | (0.032) | (0.032) | (0.033) | (0.034) | (0.094) | (0.033) |
| Log of human capital | | 0.678*** | | 0.729*** | 0.688*** | 0.732*** | 0.729*** | 0.756*** | 0.688*** | 1.000*** | 0.769*** |
| | | (0.244) | | (0.264) | (0.267) | (0.258) | (0.256) | (0.250) | (0.265) | (0.212) | (0.273) |
| Log of natural resource exports per capita | | | 0.169** | 0.186*** | 0.280*** | 0.156** | 0.153** | 0.147** | 0.174** | 0.038 | 0.178** |
| | | | (0.069) | (0.072) | (0.101) | (0.070) | (0.069) | (0.067) | (0.070) | (0.039) | (0.072) |
| Log of GDP per capita | -0.117* | -0.217*** | -0.426*** | -0.480*** | -0.420*** | -0.485*** | -0.485*** | -0.482*** | -0.462*** | -0.455*** | -0.451*** |
| | (0.070) | (0.080) | (0.131) | (0.125) | (0.121) | (0.123) | (0.123) | (0.120) | (0.123) | (0.076) | (0.129) |
| Fixed effects | No | No | No | No | No | No | No | No | No | Yes | No |
| Observations | 528 | 528 | 528 | 528 | 528 | 528 | 528 | 528 | 528 | 528 | 528 |
| $R^2$ | 0.252 | 0.303 | 0.290 | 0.355 | 0.366 | 0.365 | 0.366 | 0.383 | 0.360 | 0.426 | 0.354 |
| Adjusted $R^2$ | 0.240 | 0.292 | 0.279 | 0.342 | 0.353 | 0.351 | 0.353 | 0.370 | 0.346 | 0.260 | 0.342 |

Notes: Each regression includes period fixed effects. Clustered standard errors in brackets. *p<0.1, **p<0.05, ***p<0.01.
The last column includes the instrumental variables estimate (see Supplementary Note 8).

## Supplementary Table 20. ECI (technology) Emission Intensity Regressions

| | Dependent variable: | | | | | | | | | | |
|---|---|---|---|---|---|---|---|---|---|---|---|
| | GHG emissions per GDP (1996-99, 2000-03, 2004-07, 2008-11, 2012-15, 2016-19) | | | | | | | | | | |
| | (1) | (2) | (3) | (4) | (5) | (6) | (7) | (8) | (9) | (10) | (11) |
| ECI (technology) | -0.335 | -0.403* | -0.153 | -0.646*** | -0.375 | -0.573** | -0.569** | -0.899*** | -0.518** | 0.192** | -0.642** |
| | (0.239) | (0.236) | (0.209) | (0.078) | (0.236) | (0.263) | (0.267) | (0.272) | (0.208) | (0.078) | (0.257) |
| Intensity (technology) | | | | | -0.528 | | | | | | |
| | | | | | (0.329) | | | | | | |
| HHI (technology) | | | | | | 0.139 | | | | | |
| | | | | | | (0.257) | | | | | | |
| Entropy (technology) | | | | | | | -0.142 | | | | |
| | | | | | | | (0.260) | | | | |
| Log of Fitness (technology) | | | | | | | | 0.351* | | | |
| | | | | | | | | (0.191) | | | |
| i-ECI | | | | | | | | | -0.480 | | |
| | | | | | | | | | (0.374) | | |
| Log of population | 0.012 | | | 0.098 | 0.082** | 0.099*** | 0.099*** | 0.095*** | 0.102*** | 0.170* | 0.098*** |
| | (0.032) | | | (0.094) | (0.035) | (0.033) | (0.033) | (0.033) | (0.034) | (0.094) | (0.035) |
| Log of human capital | | 0.495** | | 0.672*** | 0.736*** | 0.682*** | 0.682*** | 0.635*** | 0.761*** | 0.915*** | 0.671*** |
| | | (0.231) | | (0.212) | (0.241) | (0.234) | (0.235) | (0.232) | (0.265) | (0.212) | (0.238) |
| Log of natural resource exports per capita | | | 0.201*** | 0.255*** | 0.246*** | 0.252*** | 0.252*** | 0.267*** | 0.241*** | 0.035 | 0.255*** |
| | | | (0.064) | (0.037) | (0.062) | (0.062) | (0.062) | (0.062) | (0.065) | (0.037) | (0.061) |
| Log of GDP per capita | -0.200*** | -0.296*** | -0.533*** | -0.613*** | -0.535*** | -0.607*** | -0.606*** | -0.635*** | -0.565*** | -0.443*** | -0.613*** |
| | (0.074) | (0.090) | (0.113) | (0.076) | (0.116) | (0.106) | (0.106) | (0.108) | (0.113) | (0.076) | (0.109) |
| Fixed effects | No | No | No | No | No | No | No | No | No | Yes | No |
| Observations | 528 | 528 | 528 | 528 | 528 | 528 | 528 | 528 | 528 | 528 | 528 |
| $R^2$ | 0.216 | 0.244 | 0.276 | 0.339 | 0.349 | 0.340 | 0.340 | 0.345 | 0.347 | 0.439 | 0.339 |
| Adjusted $R^2$ | 0.204 | 0.233 | 0.265 | 0.326 | 0.335 | 0.326 | 0.326 | 0.331 | 0.333 | 0.277 | 0.326 |

Notes: Each regression includes period fixed effects. Clustered standard errors in brackets. *p<0.1, **p<0.05, ***p<0.01.
The last column includes the instrumental variables estimate (see Supplementary Note 8).

## Supplementary Table 21. ECI (research) Emission Intensity Regressions

| | *Dependent variable:* | | | | | | | | | | |
|---|---|---|---|---|---|---|---|---|---|---|---|
| | GHG emissions per GDP (1996-99, 2000-03, 2004-07, 2008-11, 2012-15, 2016-19) | | | | | | | | | | |
| | (1) | (2) | (3) | (4) | (5) | (6) | (7) | (8) | (9) | (10) | (11) |
| ECI (research) | -0.493 | -0.506 | -0.252 | -0.647* | -0.660 | -0.574 | -0.562 | -0.621 | -0.561* | -0.177 | -0.799** |
| | (0.388) | (0.352) | (0.335) | (0.340) | (0.412) | (0.353) | (0.357) | (0.462) | (0.322) | (0.118) | (0.371) |
| Intensity (research) | | | | | 0.022 | | | | | | |
| | | | | | (0.444) | | | | | | |
| HHI (research) | | | | | | 0.339 | | | | | |
| | | | | | | (0.429) | | | | | |
| Entropy (research) | | | | | | | -0.329 | | | | |
| | | | | | | | (0.421) | | | | |
| Log of Fitness (research) | | | | | | | | -0.042 | | | |
| | | | | | | | | (0.556) | | | |
| i-ECI | | | | | | | | | -0.692* | | |
| | | | | | | | | | (0.381) | | |
| Log of population | 0.009 | | | 0.073** | 0.073** | 0.076** | 0.077** | 0.074** | 0.090*** | 0.214** | 0.080** |
| | (0.033) | | | (0.030) | (0.031) | (0.031) | (0.031) | (0.034) | (0.033) | (0.087) | (0.031) |
| Log of human capital | | 0.423* | | 0.516** | 0.513** | 0.521** | 0.522** | 0.517** | 0.701** | 1.048*** | 0.535** |
| | | (0.237) | | (0.234) | (0.250) | (0.234) | (0.234) | (0.235) | (0.275) | (0.214) | (0.230) |
| Log of natural resource exports per capita | | | 0.201*** | 0.248*** | 0.248*** | 0.251*** | 0.252*** | 0.248*** | 0.230*** | 0.042 | 0.248*** |
| | | | (0.065) | (0.064) | (0.065) | (0.064) | (0.063) | (0.064) | (0.067) | (0.038) | (0.064) |
| Log of GDP per capita | -0.189** | -0.278*** | -0.522*** | -0.601*** | -0.603*** | -0.608*** | -0.608*** | -0.601*** | -0.522*** | -0.435*** | -0.580*** |
| | (0.079) | (0.097) | (0.115) | (0.121) | (0.129) | (0.120) | (0.120) | (0.121) | (0.126) | (0.075) | (0.124) |
| Fixed effects | No | No | No | No | No | No | No | No | No | Yes | No |
| Observations | 528 | 528 | 528 | 528 | 528 | 528 | 528 | 528 | 528 | 528 | 528 |
| $R^2$ | 0.216 | 0.239 | 0.277 | 0.323 | 0.323 | 0.324 | 0.325 | 0.323 | 0.342 | 0.427 | 0.322 |
| Adjusted $R^2$ | 0.204 | 0.227 | 0.266 | 0.309 | 0.308 | 0.310 | 0.310 | 0.308 | 0.328 | 0.262 | 0.308 |

Notes: Each regression includes period fixed effects. Clustered standard errors in brackets. *p<0.1, **p<0.05, ***p<0.01.
The last column includes the instrumental variables estimate (see Supplementary Note 8).

**Emission intensity multidimensional regression model additional explanatory variables robustness check:** In Supplementary Table 20 we reproduce the main results for the emission intensity regression analysis (Columns (1-12)). Here, because we already use the additional explanatory variables as part of our baseline model, we test the robustness of the multidimensional emission intensity regression by only using the log of the population, the log of the human capital, and the log of natural resource exports as separate explanatory variables to a model that only includes the log of the GDP per capita and the final multidimensional ECI coefficients (Columns (13-15)). In each case the ECI (trade), ECI (technology), ECI (research), their interactions terms, and the three-way term remain significant predictors of emission intensity, thus suggesting the robustness of the multidimensional regression model. In this case, the country-fixed effects do not take out all the variation in our data, and the model including them (Column (16)) has a representative adjusted $R^2$ (adjusted $R^2 = 0.314$). In it, the three-way interaction term remains significant thus further indicating the robustness of our results. The model in Column (12) has the highest adjusted $R^2$ (adjusted $R^2 = 0.397$), and in it all of the variables are statistically significant. Therefore, this is our final emission intensity model. Supplementary Figure 14 gives the correlations between the variables used in these regressions.

**Emission intensity multidimensional regression model production intensity robustness check:** In Supplementary Table 23, Columns (2-12), we estimate the production intensity emission intensity models. We find that the best model includes all the intensities (trade, technology, and research), their pairwise interaction terms, and the three-way interaction term (adjusted $R^2 = 0.448$). By comparing this model directly to the multidimensional ECI model (Column (13)) and in a model with country-fixed effects (Column (14)) we find that the coefficients of the multidimensional intensity model remain statistically significant in specifications that include both models. Supplementary Figure 15 gives the correlation between the variables used in these regressions.

**Emission intensity multidimensional regression model HHI robustness check:** In Supplementary Table 24, Columns (2-12), we estimate the HHI emission intensity models. We find that the multidimensional model includes only the trade HHI (adjusted $R^2 = 0.334$). By comparing this model directly to the multidimensional ECI model (Column (13)) and in a regression specification which also includes country-fixed effects (Column (14)), we find that the multidimensional ECI model coefficients remain significant in specifications that include both models. Hence the multidimensional ECI emission intensity model is robust against the multidimensional HHI model. Supplementary Figure 16 gives the correlation between the variables used in these regressions.

**Emission intensity multidimensional regression model Entropy robustness check:** In Supplementary Table 25, Columns (2-12), we estimate the Entropy emission intensity models. Identically to the HHI case, we find that the multidimensional model includes only the trade (adjusted $R^2 = 0.337$). By comparing this model directly to the multidimensional ECI model (Column (13)) and in a regression specification which also includes country fixed effects (Column (14)), we find that the multidimensional ECI model coefficients remain significant in specifications that include both models. Hence the multidimensional ECI emission intensity model is robust against the multidimensional Entropy model. Supplementary Figure 17 gives the correlation between the variables used in these regressions.

**Emission intensity multidimensional regression model Fitness robustness check:** In Supplementary Table 26, Columns (2-12), we estimate the Fitness emission intensity models. We find that the multidimensional Fitness model includes only the trade Fitness (adjusted $R^2 = 0.368$). By comparing this model directly to the multidimensional ECI model (Column (13)) and in a regression specification which also all of the additional explanatory variables (Column (14)), we find that the multidimensional ECI model coefficients remain significant in specifications that include both models. Hence the multidimensional ECI emission intensity model is robust against the multidimensional Fitness model. Supplementary Figure 18 gives the correlation between the variables used in these regressions.

**Emission intensity multidimensional regression model i-ECI robustness check:** In Supplementary Table 27, Columns (2-12), we estimate the i-ECI emission intensity models. We find that the multidimensional i-ECI model includes both the i-ECI and the ECI (research) (adjusted $R^2 = 0.328$). By comparing this model directly to the multidimensional ECI model (Column (13)) and in a regression specification which also all of the additional explanatory variables (Column (14)), we find that the multidimensional ECI model coefficients remain significant in specifications that include both models. Hence the multidimensional ECI emission intensity model is robust against the multidimensional i-ECI model. Supplementary Figure 19 gives the correlation between the variables used in these regressions.

**Supplementary Table 22. Emission Intensity Regressions: Additional Explanatory Variables Robustness Check**

| | \multicolumn{16}{c}{*Dependent variable:*} |
|---|---|---|---|---|---|---|---|---|---|---|---|---|---|---|---|---|
| | \multicolumn{16}{c}{GHG emissions per GDP (1996-99, 2000-03, 2004-07, 2008-11, 2012-15, 2016-19)} |
| | (1) | (2) | (3) | (4) | (5) | (6) | (7) | (8) | (9) | (10) | (11) | (12) | (13) | (14) | (15) | (16) |
| ECI (trade) | | -1.223*** | | | -0.954** | -1.123** | | -0.939* | 0.495 | -0.202 | | -5.007*** | -3.695** | -5.667*** | -3.604* | -0.719 |
| | | (0.447) | | | (0.475) | (0.455) | | (0.484) | (1.020) | (0.956) | | (1.903) | (1.806) | (1.939) | (1.893) | (0.631) |
| ECI (technology) | | | -0.646*** | | -0.358* | | -0.547** | -0.269 | 0.660 | | 0.049 | -4.059*** | -2.912* | -3.940*** | -2.555 | -1.184** |
| | | | (0.217) | | (0.214) | | (0.221) | (0.221) | (0.596) | | (0.441) | (1.370) | (1.648) | (1.394) | (1.616) | (0.466) |
| ECI (research) | | | | -0.647* | | -0.478 | -0.412 | -0.390 | | 0.530 | 0.440 | -5.928*** | -4.053** | -4.960** | -4.142** | -1.170* |
| | | | | (0.340) | | (0.311) | (0.339) | (0.317) | | (0.859) | (0.611) | (1.987) | (1.913) | (1.974) | (2.098) | (0.608) |
| ECI (trade) x ECI (technology) | | | | | | | | | -1.878* | | | 6.666*** | 5.166* | 7.312*** | 5.091* | 2.116*** |
| | | | | | | | | | (1.097) | | | (2.419) | (2.772) | (2.517) | (2.749) | (0.821) |
| ECI (trade) x ECI (research) | | | | | | | | | -1.511 | | | 10.567*** | 7.132** | 9.257** | 8.298** | 1.403 |
| | | | | | | | | | (1.258) | | | (3.564) | (3.601) | (3.761) | (4.024) | (1.077) |
| ECI (research) x ECI (technology) | | | | | | | | | | | -1.125 | 8.607*** | 6.694** | 7.880*** | 6.557** | 2.349*** |
| | | | | | | | | | | | (0.799) | (2.662) | (2.919) | (2.652) | (2.974) | (0.849) |
| ECI (trade) x ECI (research) x ECI (technology) | | | | | | | | | | | | -15.268*** | -11.597** | -14.195*** | -12.360** | -3.324** |
| | | | | | | | | | | | | (4.293) | (4.632) | (4.415) | (4.838) | (1.364) |
| Log of population | 0.041 | 0.067** | 0.098*** | 0.073** | 0.093*** | 0.088*** | 0.109*** | 0.104*** | 0.102*** | 0.086*** | 0.107*** | 0.113*** | 0.021 | | | 0.139 |
| | (0.031) | (0.032) | (0.033) | (0.030) | (0.033) | (0.032) | (0.034) | (0.034) | (0.034) | (0.032) | (0.034) | (0.033) | (0.033) | | | (0.091) |
| Log of human capital | 0.433* | 0.729*** | 0.672*** | 0.516** | 0.797*** | 0.766*** | 0.688*** | 0.810*** | 0.719*** | 0.716** | 0.655*** | 0.810*** | | 0.694*** | | 0.921*** |
| | (0.245) | (0.264) | (0.229) | (0.234) | (0.254) | (0.264) | (0.230) | (0.256) | (0.269) | (0.283) | (0.235) | (0.271) | | (0.256) | | (0.211) |
| Log of natural resource exports per capita | 0.247*** | 0.186*** | 0.255*** | 0.248*** | 0.204*** | 0.192*** | 0.255*** | 0.205*** | 0.232*** | 0.198*** | 0.264*** | 0.233*** | | | 0.191*** | 0.036 |
| | (0.064) | (0.072) | (0.061) | (0.064) | (0.071) | (0.071) | (0.061) | (0.071) | (0.072) | (0.074) | (0.063) | (0.070) | | | (0.065) | (0.034) |
| Log of GDP per capita | -0.690*** | -0.480*** | -0.613*** | -0.601*** | -0.483*** | -0.431*** | -0.568*** | -0.443*** | -0.522*** | -0.446*** | -0.583*** | -0.498*** | -0.098 | -0.209** | -0.472*** | -0.480*** |
| | (0.108) | (0.125) | (0.108) | (0.121) | (0.120) | (0.131) | (0.120) | (0.127) | (0.117) | (0.135) | (0.121) | (0.119) | (0.081) | (0.085) | (0.121) | (0.076) |
| Fixed effects | No | No | No | No | No | No | No | No | No | No | No | No | No | No | No | Yes |
| Observations | 528 | 528 | 528 | 528 | 528 | 528 | 528 | 528 | 528 | 528 | 528 | 528 | 528 | 528 | 528 | 528 |
| $R^2$ | 0.302 | 0.355 | 0.339 | 0.323 | 0.364 | 0.366 | 0.346 | 0.370 | 0.380 | 0.373 | 0.353 | 0.415 | 0.290 | 0.339 | 0.334 | 0.476 |
| Adjusted $R^2$ | 0.290 | 0.342 | 0.326 | 0.309 | 0.350 | 0.352 | 0.332 | 0.356 | 0.366 | 0.359 | 0.338 | 0.397 | 0.271 | 0.321 | 0.316 | 0.314 |

Notes: Each regression includes period fixed effects. Clustered standard errors in brackets. *p<0.1, **p<0.05, ***p<0.01.

## Supplementary Table 23. Emission Intensity Regressions: Production Intensity Robustness Check

| | Dependent variable: GHG emissions per GDP (1996-99, 2000-03, 2004-07, 2008-11, 2012-15, 2016-19) | | | | | | | | | | | | | | |
|---|---|---|---|---|---|---|---|---|---|---|---|---|---|---|---|
| | (1) | (2) | (3) | (4) | (5) | (6) | (7) | (8) | (9) | (10) | (11) | (12) | (13) | (14) | (15) |
| ECI (trade) | -5.007*** | | | | | | | | | | | | -4.012** | -4.012** | -0.489 |
| | (1.903) | | | | | | | | | | | | (1.895) | (1.895) | (0.551) |
| ECI (technology) | -4.059*** | | | | | | | | | | | | -2.787* | -2.787* | -0.723* |
| | (1.370) | | | | | | | | | | | | (1.613) | (1.613) | (0.416) |
| ECI (research) | -5.928*** | | | | | | | | | | | | -5.031*** | -5.031*** | -1.171** |
| | (1.987) | | | | | | | | | | | | (1.757) | (1.757) | (0.552) |
| ECI (trade) x ECI (technology) | 6.666*** | | | | | | | | | | | | 4.110 | 4.110 | 1.122 |
| | (2.419) | | | | | | | | | | | | (2.796) | (2.796) | (0.791) |
| ECI (trade) x ECI (research) | 10.567*** | | | | | | | | | | | | 8.155** | 8.155** | 1.484 |
| | (3.564) | | | | | | | | | | | | (3.174) | (3.174) | (0.965) |
| ECI (research) x ECI (technology) | 8.607*** | | | | | | | | | | | | 5.843** | 5.843** | 1.778** |
| | (2.662) | | | | | | | | | | | | (2.774) | (2.774) | (0.760) |
| ECI (trade) x ECI (research) x ECI (technology) | -15.268*** | | | | | | | | | | | | -9.760** | -9.760** | -2.596** |
| | (4.293) | | | | | | | | | | | | (4.302) | (4.302) | (1.254) |
| Intensity (trade) | | -1.829*** | | | -1.352** | -1.742*** | | -1.354** | -0.181 | 0.831 | | -5.987*** | -5.037** | -5.037** | 1.027* |
| | | (0.586) | | | (0.582) | (0.586) | | (0.578) | (0.787) | (1.331) | | (1.953) | (2.036) | (2.036) | (0.613) |
| Intensity (technology) | | | -0.805*** | | -0.577* | | -0.925*** | -0.698** | 0.851 | | 1.558 | -6.080*** | -4.699** | -4.699** | -0.775 |
| | | | (0.299) | | (0.303) | | (0.332) | (0.332) | (0.752) | | (1.001) | (1.860) | (2.045) | (2.045) | (0.676) |
| Intensity (research) | | | | -0.414 | | -0.204 | 0.235 | 0.239 | | 1.569 | 1.059** | -2.786* | -1.555 | -1.555 | 1.366*** |
| | | | | (0.371) | | (0.349) | (0.390) | (0.373) | | (0.958) | (0.527) | (1.593) | (1.659) | (1.659) | (0.485) |
| Intensity (trade) x Intensity (technology) | | | | | | | | | -1.925** | | | 14.442*** | 12.998*** | 12.998*** | 3.060*** |
| | | | | | | | | | (0.904) | | | (3.456) | (3.538) | (3.538) | (1.095) |
| Intensity (trade) x Intensity (research) | | | | | | | | | -2.973** | | | 6.719** | 5.225 | 5.225 | -1.948* |
| | | | | | | | | | (1.420) | | | (3.022) | (3.250) | (3.250) | (1.006) |
| Intensity (research) x Intensity (technology) | | | | | | | | | | | -2.623** | 8.282*** | 6.367** | 6.367** | 1.157 |
| | | | | | | | | | | | (1.039) | (2.416) | (2.683) | (2.683) | (1.047) |
| Intensity (trade) x Intensity (research) x Intensity (technology) | | | | | | | | | | | | -18.561*** | -15.780*** | -15.780*** | -2.585 |
| | | | | | | | | | | | | (3.794) | (4.043) | (4.043) | (1.610) |
| Log of GDP per population | 0.113*** | 0.045 | 0.053* | 0.048 | 0.053* | 0.049 | 0.051 | 0.050 | 0.059** | 0.050* | 0.066** | 0.063** | 0.112*** | 0.112*** | 0.136* |
| | (0.033) | (0.031) | (0.031) | (0.030) | (0.031) | (0.030) | (0.031) | (0.031) | (0.030) | (0.028) | (0.029) | (0.028) | (0.036) | (0.036) | (0.076) |
| Log of human capital | 0.810*** | 0.491** | 0.683*** | 0.511** | 0.656*** | 0.527** | 0.676*** | 0.648*** | 0.525* | 0.466 | 0.531* | 0.531** | 0.694** | 0.694** | 0.633*** |
| | (0.271) | (0.233) | (0.244) | (0.250) | (0.240) | (0.242) | (0.249) | (0.245) | (0.286) | (0.292) | (0.297) | (0.265) | (0.281) | (0.281) | (0.166) |
| Log of natural resource exports per capita | 0.233*** | 0.378*** | 0.238*** | 0.243*** | 0.338*** | 0.370*** | 0.239*** | 0.339*** | 0.335*** | 0.348*** | 0.269*** | 0.388*** | 0.356*** | 0.356*** | 0.0002 |
| | (0.070) | (0.085) | (0.064) | (0.064) | (0.086) | (0.088) | (0.065) | (0.085) | (0.090) | (0.088) | (0.068) | (0.081) | (0.095) | (0.095) | (0.042) |
| Log of GDP per capita | -0.498*** | -0.501*** | -0.522*** | -0.621*** | -0.430*** | -0.476*** | -0.536*** | -0.444*** | -0.519*** | -0.561*** | -0.612*** | -0.605*** | -0.584*** | -0.584*** | -0.583*** |
| | (0.119) | (0.112) | (0.121) | (0.126) | (0.121) | (0.125) | (0.128) | (0.128) | (0.122) | (0.124) | (0.127) | (0.119) | (0.116) | (0.116) | (0.060) |
| Fixed effects | No | No | No | No | No | No | No | No | No | No | No | No | No | No | Yes |
| Observations | 528 | 528 | 528 | 528 | 528 | 528 | 528 | 528 | 528 | 528 | 528 | 528 | 528 | 528 | 528 |
| R² | 0.415 | 0.344 | 0.342 | 0.310 | 0.361 | 0.345 | 0.343 | 0.363 | 0.390 | 0.384 | 0.382 | 0.465 | 0.496 | 0.496 | 0.577 |
| Adjusted R² | 0.397 | 0.331 | 0.329 | 0.296 | 0.348 | 0.331 | 0.329 | 0.348 | 0.375 | 0.370 | 0.368 | 0.448 | 0.473 | 0.473 | 0.437 |

Notes: Each regression includes period fixed effects. Clustered standard errors in brackets. *p<0.1, **p<0.05, ***p<0.01.

# Supplementary Table 24. Emission Intensity Regressions: Hirschman-Herfindahl Index Robustness Check

| | *Dependent variable:* | | | | | | | | | | | | | |
|---|---|---|---|---|---|---|---|---|---|---|---|---|---|---|
| | GHG emissions per GDP (1996-99, 2000-03, 2004-07, 2008-11, 2012-15, 2016-19) | | | | | | | | | | | | | |
| | (1) | (2) | (3) | (4) | (5) | (6) | (7) | (8) | (9) | (10) | (11) | (12) | (13) | (14) |
| ECI (trade) | -5.007*** | | | | | | | | | | | | -4.264** | -0.749 |
| | (1.903) | | | | | | | | | | | | (2.044) | (0.630) |
| ECI (technology) | -4.059*** | | | | | | | | | | | | -4.182*** | -1.173** |
| | (1.370) | | | | | | | | | | | | (1.351) | (0.475) |
| ECI (research) | -5.928*** | | | | | | | | | | | | -6.043*** | -1.154* |
| | (1.987) | | | | | | | | | | | | (2.016) | (0.625) |
| ECI (trade) x ECI (technology) | 6.666*** | | | | | | | | | | | | 6.510*** | 2.112** |
| | (2.419) | | | | | | | | | | | | (2.446) | (0.834) |
| ECI (trade) x ECI (research) | 10.567*** | | | | | | | | | | | | 10.355*** | 1.388 |
| | (3.564) | | | | | | | | | | | | (3.654) | (1.102) |
| ECI (research) x ECI (technology) | 8.607*** | | | | | | | | | | | | 9.404*** | 2.314*** |
| | (2.662) | | | | | | | | | | | | (2.615) | (0.873) |
| ECI (trade) x ECI (research) x ECI (technology) | -15.268*** | | | | | | | | | | | | -15.678*** | -3.289** |
| | (4.293) | | | | | | | | | | | | (4.295) | (1.397) |
| HHI (trade) | | 0.890*** | | | 0.788** | 0.853** | | 0.759** | 0.588 | 0.928*** | | 0.495 | 0.582 | -0.095 |
| | | (0.327) | | | (0.332) | (0.332) | | (0.336) | (0.429) | (0.342) | | (0.447) | (0.400) | (0.197) |
| HHI (technology) | | | 0.504** | | 0.291 | | 0.472** | 0.275 | 0.157 | | 0.551** | 0.057 | | |
| | | | (0.207) | | (0.209) | | (0.206) | (0.208) | (0.241) | | (0.223) | (0.262) | | |
| HHI (research) | | | | 0.677 | | 0.474 | 0.574 | 0.436 | | 0.638 | 0.953** | 0.716 | | |
| | | | | (0.416) | | (0.413) | (0.401) | (0.411) | | (0.444) | (0.480) | (0.532) | | |
| HHI (trade) x HHI (technology) | | | | | | | | | 0.816 | | | 1.626* | | |
| | | | | | | | | | (0.882) | | | (0.891) | | |
| HHI (trade) x HHI (research) | | | | | | | | | | -1.374 | | -1.631 | | |
| | | | | | | | | | | (1.390) | | (7.619) | | |
| HHI (research) x HHI (technology) | | | | | | | | | | | -1.400 | -0.118 | | |
| | | | | | | | | | | | (1.275) | (2.585) | | |
| HHI (trade) x HHI (research) x HHI (technology) | | | | | | | | | | | | -2.111 | | |
| | | | | | | | | | | | | (9.653) | | |
| Log of population | 0.113*** | 0.036 | 0.067** | 0.055* | 0.051 | 0.046 | 0.077** | 0.060* | 0.049 | 0.047 | 0.080** | 0.057* | 0.091*** | 0.135 |
| | (0.033) | (0.031) | (0.031) | (0.031) | (0.031) | (0.031) | (0.031) | (0.032) | (0.031) | (0.031) | (0.031) | (0.032) | (0.034) | (0.093) |
| Log of human capital | 0.810*** | 0.587** | 0.566** | 0.462* | 0.646*** | 0.601*** | 0.582** | 0.655*** | 0.650*** | 0.594** | 0.573** | 0.645*** | 0.774*** | 0.918*** |
| | (0.271) | (0.231) | (0.244) | (0.242) | (0.232) | (0.229) | (0.242) | (0.231) | (0.230) | (0.231) | (0.246) | (0.229) | (0.267) | (0.211) |
| Log of natural resource exports per capita | 0.233*** | 0.164** | 0.237*** | 0.254*** | 0.168** | 0.172** | 0.243*** | 0.175*** | 0.167** | 0.174*** | 0.246*** | 0.176*** | 0.194*** | 0.040 |
| | (0.070) | (0.067) | (0.062) | (0.063) | (0.066) | (0.067) | (0.062) | (0.066) | (0.065) | (0.067) | (0.062) | (0.065) | (0.069) | (0.036) |
| Log of GDP per capita | -0.498*** | -0.594*** | -0.637*** | -0.684*** | -0.574*** | -0.594*** | -0.635*** | -0.575*** | -0.583*** | -0.593*** | -0.632*** | -0.588*** | -0.520*** | -0.478*** |
| | (0.119) | (0.103) | (0.105) | (0.107) | (0.103) | (0.102) | (0.104) | (0.102) | (0.103) | (0.102) | (0.104) | (0.103) | (0.120) | (0.076) |
| Fixed effects | No | No | No | No | No | No | No | No | No | No | No | No | No | Yes |
| Observations | 528 | 528 | 528 | 528 | 528 | 528 | 528 | 528 | 528 | 528 | 528 | 528 | 528 | 528 |
| $R^2$ | 0.415 | 0.346 | 0.320 | 0.310 | 0.352 | 0.350 | 0.326 | 0.355 | 0.354 | 0.352 | 0.328 | 0.363 | 0.425 | 0.476 |
| Adjusted $R^2$ | 0.397 | 0.334 | 0.307 | 0.297 | 0.338 | 0.336 | 0.312 | 0.340 | 0.339 | 0.337 | 0.312 | 0.343 | 0.406 | 0.314 |

Notes: Each regression includes period fixed effects. Clustered standard errors in brackets. *p<0.1, **p<0.05, ***p<0.01.

## Supplementary Table 25. Emission Intensity Regressions: Entropy Robustness Check

| | *Dependent variable:* | | | | | | | | | | | | | |
|---|---|---|---|---|---|---|---|---|---|---|---|---|---|---|
| | GHG emissions per GDP (1996-99, 2000-03, 2004-07, 2008-11, 2012-15, 2016-19) | | | | | | | | | | | | | |
| | (1) | (2) | (3) | (4) | (5) | (6) | (7) | (8) | (9) | (10) | (11) | (12) | (13) | (14) |
| ECI (trade) | -5.007*** | | | | | | | | | | | | -4.171** | -0.750 |
| | (1.903) | | | | | | | | | | | | (2.049) | (0.629) |
| ECI (technology) | -4.059*** | | | | | | | | | | | | -4.188*** | -1.173** |
| | (1.370) | | | | | | | | | | | | (1.350) | (0.475) |
| ECI (research) | -5.928*** | | | | | | | | | | | | -6.062*** | -1.157* |
| | (1.987) | | | | | | | | | | | | (2.015) | (0.624) |
| ECI (trade) x ECI (technology) | 6.666*** | | | | | | | | | | | | 6.489*** | 2.109** |
| | (2.419) | | | | | | | | | | | | (2.444) | (0.835) |
| ECI (trade) x ECI (research) | 10.567*** | | | | | | | | | | | | 10.344*** | 1.392 |
| | (3.564) | | | | | | | | | | | | (3.656) | (1.102) |
| ECI (research) x ECI (technology) | 8.607*** | | | | | | | | | | | | 9.494*** | 2.317*** |
| | (2.662) | | | | | | | | | | | | (2.616) | (0.872) |
| ECI (trade) x ECI (research) x ECI (technology) | -15.268*** | | | | | | | | | | | | -15.732*** | -3.291** |
| | (4.293) | | | | | | | | | | | | (4.294) | (1.398) |
| Entropy (trade) | | -0.850*** | | | -0.755** | -0.812*** | | -0.728** | -1.389** | -0.039 | | 0.659 | -0.592 | 0.093 |
| | | (0.293) | | | (0.299) | (0.298) | | (0.304) | (0.613) | (1.219) | | (2.795) | (0.372) | (0.191) |
| Entropy (technology) | | | -0.508** | | -0.274 | | -0.468** | -0.252 | -0.960 | | 0.820 | 1.188 | | |
| | | | (0.207) | | (0.209) | | (0.207) | (0.208) | (0.686) | | (1.079) | (7.427) | | |
| Entropy (research) | | | | -0.652 | | -0.441 | -0.535 | -0.400 | | 0.261 | 0.436 | 2.368 | | |
| | | | | (0.404) | | (0.396) | (0.390) | (0.393) | | (1.167) | (0.924) | (1.873) | | |
| Entropy (trade) x Entropy (technology) | | | | | | | | | 0.848 | | | -0.751 | | |
| | | | | | | | | | (0.798) | | | (7.941) | | |
| Entropy (trade) x Entropy (research) | | | | | | | | | | -0.825 | | -2.587 | | |
| | | | | | | | | | | (1.276) | | (2.982) | | |
| Entropy (research) x Entropy (technology) | | | | | | | | | | | -1.382 | -2.738 | | |
| | | | | | | | | | | | (1.157) | (7.581) | | |
| Entropy (trade) x Entropy (research) x Entropy (technology) | | | | | | | | | | | | 2.258 | | |
| | | | | | | | | | | | | (8.084) | | |
| Log of GDP per population | 0.113*** | 0.036 | 0.068** | 0.058* | 0.052 | 0.048 | 0.080** | 0.061* | 0.048 | 0.049 | 0.083*** | 0.059* | 0.089*** | 0.135 |
| | (0.033) | (0.031) | (0.031) | (0.031) | (0.032) | (0.031) | (0.031) | (0.032) | (0.032) | (0.032) | (0.032) | (0.033) | (0.034) | (0.093) |
| Log of human capital | 0.810*** | 0.593*** | 0.573** | 0.466* | 0.650*** | 0.608*** | 0.589*** | 0.660*** | 0.654*** | 0.604*** | 0.579*** | 0.646*** | 0.765*** | 0.918*** |
| | (0.271) | (0.228) | (0.243) | (0.241) | (0.229) | (0.226) | (0.242) | (0.228) | (0.227) | (0.228) | (0.245) | (0.227) | (0.265) | (0.212) |
| Log of natural resource exports per capita | 0.233*** | 0.159** | 0.237*** | 0.254*** | 0.163** | 0.167** | 0.243*** | 0.170** | 0.161** | 0.169** | 0.247*** | 0.173*** | 0.189*** | 0.041 |
| | (0.070) | (0.067) | (0.062) | (0.063) | (0.066) | (0.066) | (0.061) | (0.065) | (0.065) | (0.067) | (0.062) | (0.065) | (0.069) | (0.036) |
| Log of GDP per capita | -0.498*** | -0.585*** | -0.634*** | -0.681*** | -0.566*** | -0.584*** | -0.631*** | -0.567*** | -0.576*** | -0.583*** | -0.627*** | -0.579*** | -0.522*** | -0.478*** |
| | (0.119) | (0.103) | (0.105) | (0.107) | (0.103) | (0.102) | (0.104) | (0.102) | (0.103) | (0.102) | (0.104) | (0.103) | (0.120) | (0.076) |
| Fixed effects | No | No | No | No | No | No | No | No | No | No | No | No | No | Yes |
| Observations | 528 | 528 | 528 | 528 | 528 | 528 | 528 | 528 | 528 | 528 | 528 | 528 | 528 | 528 |
| $R^2$ | 0.415 | 0.350 | 0.321 | 0.311 | 0.355 | 0.354 | 0.327 | 0.358 | 0.358 | 0.355 | 0.329 | 0.365 | 0.427 | 0.476 |
| Adjusted $R^2$ | 0.397 | 0.337 | 0.308 | 0.298 | 0.341 | 0.340 | 0.313 | 0.343 | 0.343 | 0.340 | 0.314 | 0.346 | 0.408 | 0.314 |

Notes: Each regression includes period fixed effects. Clustered standard errors in brackets. *p<0.1, **p<0.05, ***p<0.01.

## Supplementary Table 26. Emission Intensity Regressions: Fitness Robustness Check

|  | *Dependent variable:* | | | | | | | | | | | | | |
|---|---|---|---|---|---|---|---|---|---|---|---|---|---|---|
|  | GHG emissions per GDP (1996-99, 2000-03, 2004-07, 2008-11, 2012-15, 2016-19) | | | | | | | | | | | | | |
|  | (1) | (2) | (3) | (4) | (5) | (6) | (7) | (8) | (9) | (10) | (11) | (12) | (13) | (14) |
| ECI (trade) | -5.007*** | | | | | | | | | | | | -3.317 | -0.782 |
|  | (1.903) | | | | | | | | | | | | (2.075) | (0.623) |
| ECI (technology) | -4.059*** | | | | | | | | | | | | -4.012*** | -1.188** |
|  | (1.370) | | | | | | | | | | | | (1.399) | (0.468) |
| ECI (research) | -5.928*** | | | | | | | | | | | | -6.055*** | -1.137* |
|  | (1.987) | | | | | | | | | | | | (2.039) | (0.631) |
| ECI (trade) x ECI (technology) | 6.666*** | | | | | | | | | | | | 6.020** | 2.130*** |
|  | (2.419) | | | | | | | | | | | | (2.516) | (0.822) |
| ECI (trade) x ECI (research) | 10.567*** | | | | | | | | | | | | 10.149*** | 1.364 |
|  | (3.564) | | | | | | | | | | | | (3.668) | (1.106) |
| ECI (research) x ECI (technology) | 8.607*** | | | | | | | | | | | | 9.643*** | 2.319*** |
|  | (2.662) | | | | | | | | | | | | (2.666) | (0.868) |
| ECI (trade) x ECI (research) x ECI (technology) | -15.268*** | | | | | | | | | | | | -15.557*** | -3.303** |
|  | (4.293) | | | | | | | | | | | | (4.378) | (1.389) |
| Log of fitness (trade) |  | -1.374*** | | | -1.394*** | -1.324*** | | -1.348 | -2.022*** | -1.440 | | -1.885 | -1.329*** | 0.176 |
|  |  | (0.362) | | | (0.364) | (0.370) | | (0.686) | (1.141) | | (1.495) | (0.451) | (0.302) | |
| Log of fitness (technology) |  |  | -0.251 | | 0.041 |  | -0.187 | 0.054 | -0.556 |  | 0.770 | -0.040 | | |
|  |  |  | (0.180) | | (0.158) |  | (0.183) | (0.183) | (0.651) |  | (0.622) | (2.494) | | |
| Log of fitness (research) |  |  |  | -0.627 |  | -0.188 | -0.567 | -0.198 |  | -0.299 | 0.448 | 0.759 | | |
|  |  |  |  | (0.417) |  | (0.378) | (0.418) | (0.418) |  | (1.092) | (0.663) | (1.845) | | |
| Log of fitness (trade) x Log of fitness (technology) |  |  |  |  |  |  |  |  | 0.874 |  |  | 0.966 | | |
|  |  |  |  |  |  |  |  |  | (0.887) |  |  | (3.394) | | |
| Log of fitness (trade) x Log of fitness (research) |  |  |  |  |  |  |  |  |  | 0.164 |  | -0.468 | | |
|  |  |  |  |  |  |  |  |  |  | (1.496) |  | (2.743) | | |
| Log of fitness (research) x Log of fitness (technology) |  |  |  |  |  |  |  |  |  |  | -1.520 | -1.238 | | |
|  |  |  |  |  |  |  |  |  |  |  | (0.967) | (3.586) | | |
| Log of fitness (trade) x Log of fitness (research) x Log of fitness (technology) |  |  |  |  |  |  |  |  |  |  |  | 0.458 | | |
|  |  |  |  |  |  |  |  |  |  |  |  | (4.753) | | |
| Log of population | 0.113*** | 0.083** | 0.059* | 0.074** | 0.080** | 0.091** | 0.084** | 0.088** | 0.072** | 0.090** | 0.096*** | 0.083** | 0.116*** | 0.144 |
|  | (0.033) | (0.033) | (0.032) | (0.034) | (0.033) | (0.036) | (0.035) | (0.035) | (0.034) | (0.037) | (0.036) | (0.037) | (0.032) | (0.093) |
| Log of human capital | 0.810*** | 0.711*** | 0.526** | 0.493** | 0.700*** | 0.719*** | 0.557** | 0.704*** | 0.708*** | 0.720*** | 0.552** | 0.700*** | 0.770*** | 0.916*** |
|  | (0.271) | (0.225) | (0.242) | (0.239) | (0.225) | (0.225) | (0.240) | (0.240) | (0.224) | (0.226) | (0.237) | (0.223) | (0.258) | (0.214) |
| Log of natural resource exports per capita | 0.233*** | 0.149** | 0.241*** | 0.250*** | 0.149** | 0.154** | 0.245*** | 0.153** | 0.141** | 0.153** | 0.255*** | 0.153** | 0.182*** | 0.039 |
|  | (0.070) | (0.066) | (0.064) | (0.063) | (0.065) | (0.064) | (0.063) | (0.063) | (0.065) | (0.065) | (0.065) | (0.064) | (0.063) | (0.036) |
| Log of GDP per capita | -0.498*** | -0.518*** | -0.653*** | -0.632*** | -0.522*** | -0.507*** | -0.610*** | -0.512*** | -0.527*** | -0.508*** | -0.593*** | -0.516*** | -0.531*** | -0.479*** |
|  | (0.119) | (0.103) | (0.112) | (0.116) | (0.107) | (0.109) | (0.118) | (0.118) | (0.107) | (0.108) | (0.121) | (0.111) | (0.114) | (0.075) |
| Fixed effects | No | No | No | No | No | No | No | No | No | No | No | No | No | Yes |
| Observations | 528 | 528 | 528 | 528 | 528 | 528 | 528 | 528 | 528 | 528 | 528 | 528 | 528 | 528 |
| R² | 0.415 | 0.380 | 0.308 | 0.315 | 0.380 | 0.381 | 0.318 | 0.381 | 0.383 | 0.381 | 0.327 | 0.387 | 0.446 | 0.477 |
| Adjusted R² | 0.397 | 0.368 | 0.295 | 0.302 | 0.367 | 0.368 | 0.304 | 0.367 | 0.369 | 0.367 | 0.312 | 0.368 | 0.427 | 0.314 |

Notes: Each regression includes period fixed effects. Clustered standard errors in brackets. *p<0.1, **p<0.05, ***p<0.01.

## Supplementary Table 27. Emission Intensity Regressions: i-ECI Robustness Check

|  | Dependent variable: | | | | | |
|---|---|---|---|---|---|---|
|  | GHG emissions per GDP (1996-99, 2000-03, 2004-07, 2008-11, 2012-15, 2016-19) | | | | | |
|  | (1) | (2) | (3) | (4) | (5) | (6) |
| ECI (trade) | -5.007*** |  |  |  | -5.181*** | -0.766 |
|  | (1.903) |  |  |  | (1.931) | (0.656) |
| ECI (technology) | -4.059*** |  |  |  | -3.698*** | -1.197*** |
|  | (1.370) |  |  |  | (1.366) | (0.464) |
| ECI (research) | -5.928*** |  | -0.561* | 0.120 | -5.633*** | -1.144* |
|  | (1.987) |  | (0.322) | (0.651) | (2.021) | (0.606) |
| ECI (trade) x ECI (technology) | 6.666*** |  |  |  | 5.821** | 2.105** |
|  | (2.419) |  |  |  | (2.485) | (0.821) |
| ECI (trade) x ECI (research) | 10.567*** |  |  |  | 9.894*** | 1.336 |
|  | (3.564) |  |  |  | (3.652) | (1.056) |
| ECI (research) x ECI (technology) | 8.607*** |  |  |  | 8.442*** | 2.387*** |
|  | (2.662) |  |  |  | (2.602) | (0.846) |
| ECI (trade) x ECI (research) x ECI (technology) | -15.268*** |  |  |  | -14.645*** | -3.334** |
|  | (4.293) |  |  |  | (4.320) | (1.375) |
| i-ECI |  | -0.776** | -0.692* | 0.286 | 1.004* | 0.113 |
|  |  | (0.387) | (0.381) | (1.039) | (0.585) | (0.224) |
| i-ECI x ECI (research) |  |  |  | -1.489 |  |  |
|  |  |  |  | (1.332) |  |  |
| Log of population | 0.113*** | 0.065** | 0.090*** | 0.086*** | 0.102*** | 0.130 |
|  | (0.033) | (0.032) | (0.033) | (0.033) | (0.035) | (0.093) |
| Log of human capital | 0.810*** | 0.653** | 0.701** | 0.634** | 0.720** | 0.911*** |
|  | (0.271) | (0.275) | (0.275) | (0.301) | (0.286) | (0.211) |
| Log of natural resource exports per capita | 0.233*** | 0.227*** | 0.230*** | 0.233*** | 0.219*** | 0.034 |
|  | (0.070) | (0.068) | (0.067) | (0.069) | (0.067) | (0.033) |
| Log of GDP per capita | -0.498*** | -0.588*** | -0.522*** | -0.536*** | -0.485*** | -0.480*** |
|  | (0.119) | (0.116) | (0.126) | (0.130) | (0.116) | (0.077) |
| Fixed effects | No | No | No | No | No | Yes |
| Observations | 528 | 528 | 528 | 528 | 528 | 528 |
| $R^2$ | 0.415 | 0.327 | 0.342 | 0.349 | 0.425 | 0.476 |
| Adjusted $R^2$ | 0.397 | 0.314 | 0.328 | 0.334 | 0.406 | 0.314 |

Notes: Each regression includes period fixed effects. Clustered standard errors in brackets. *p<0.1, **p<0.05, ***p<0.01.

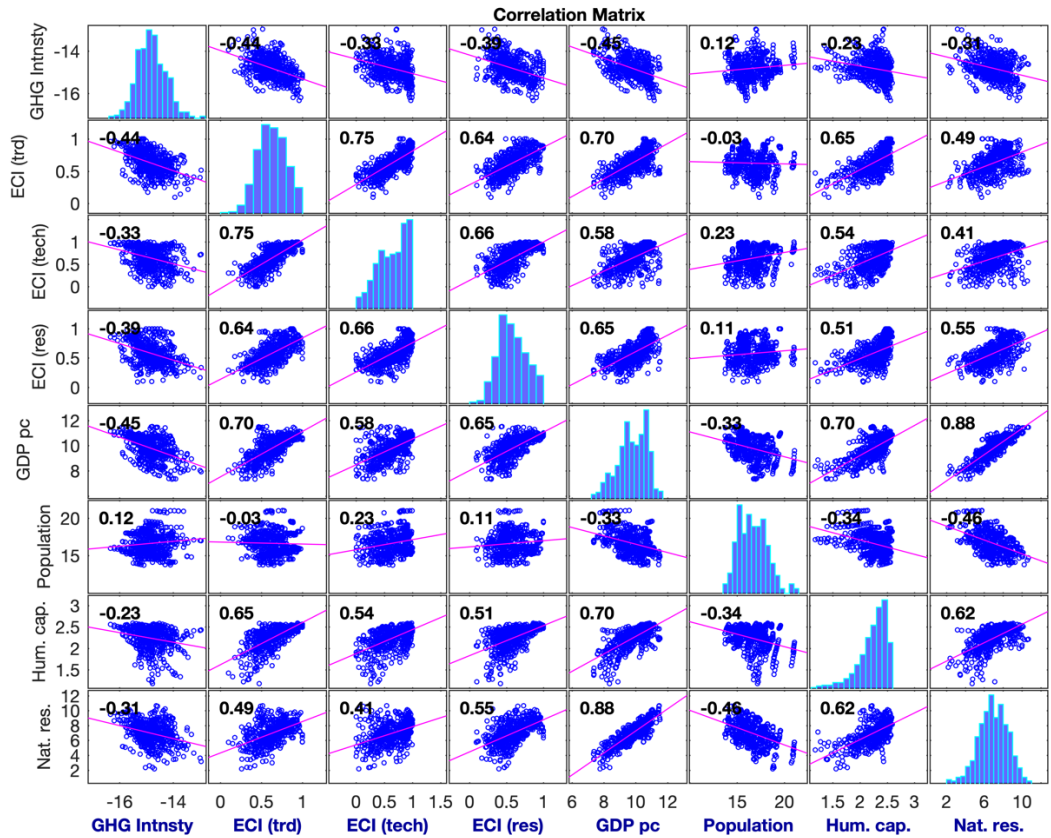

**Supplementary Figure 14. Correlations Between the Variables used in the Additional Explanatory Variables Emission Intensity Robustness Check Regression Analysis**

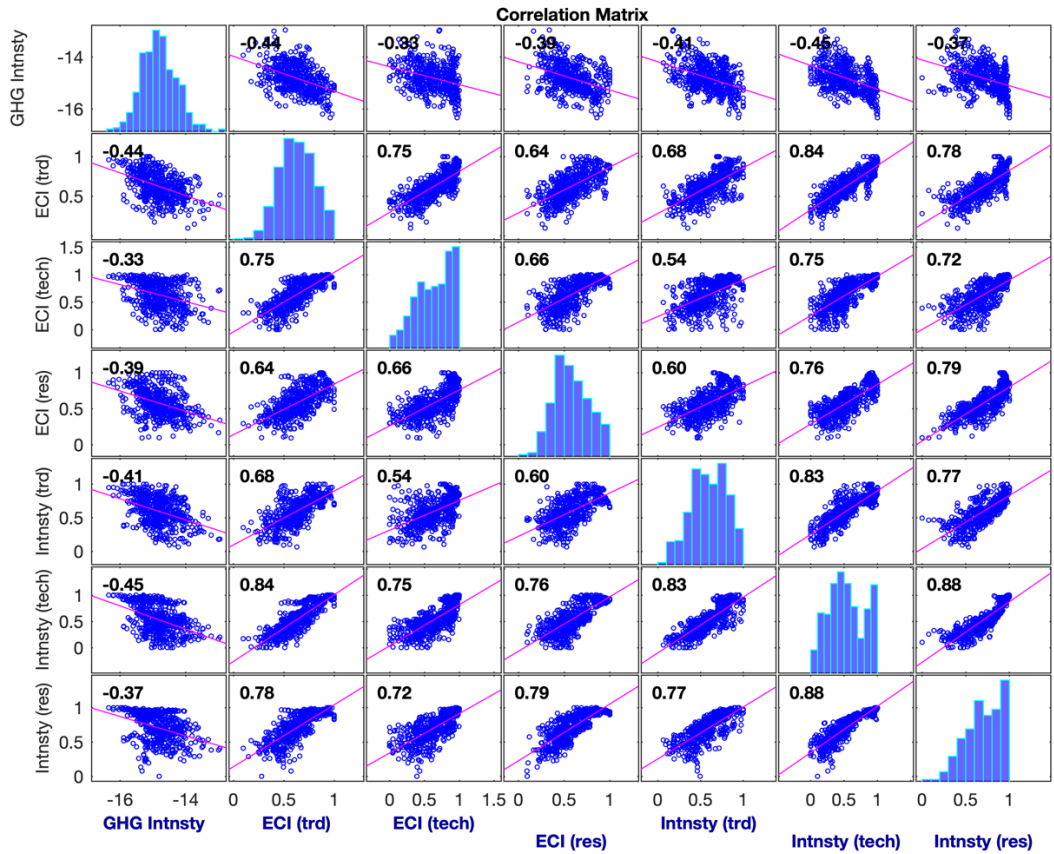

**Supplementary Figure 15. Correlations Between the Variables used in the Production Intensity Emission Intensity Robustness Check Regression Analysis**

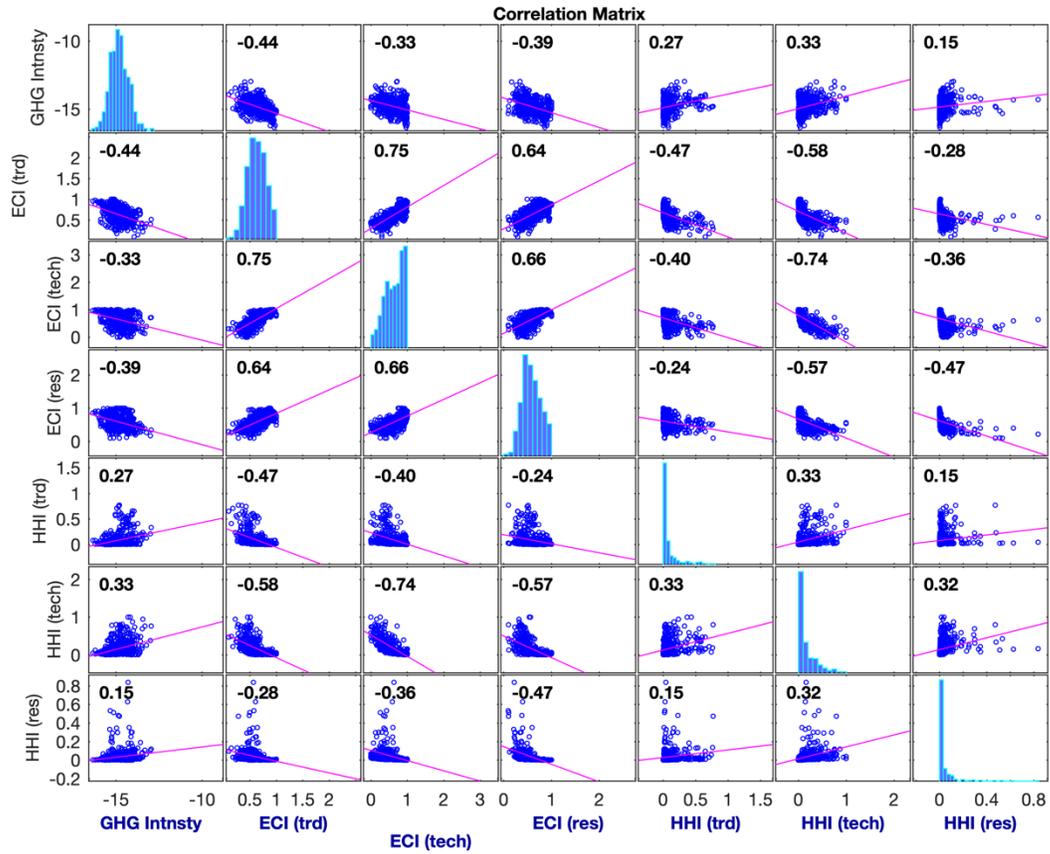

**Supplementary Figure 16. Correlations Between the Variables used in the HHI Emission Intensity Robustness Check Regression Analysis**

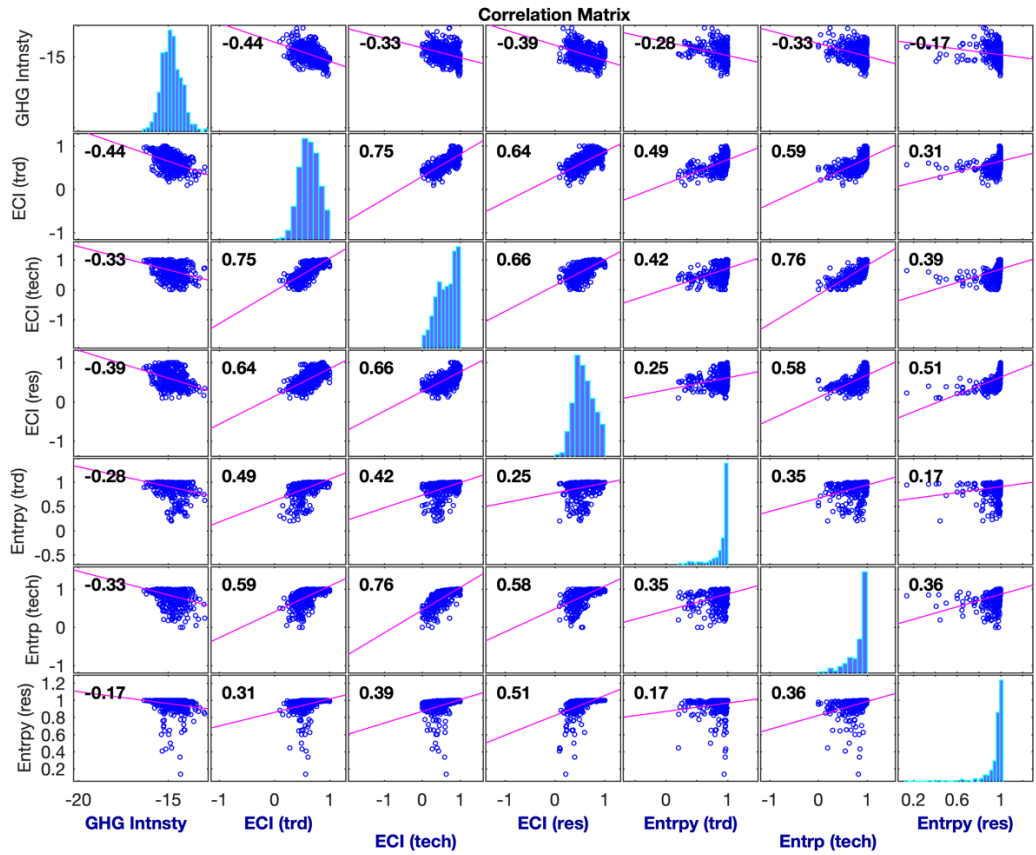

**Supplementary Figure 17. Correlations Between the Variables used in the Entropy Emission Intensity Robustness Check Regression Analysis**

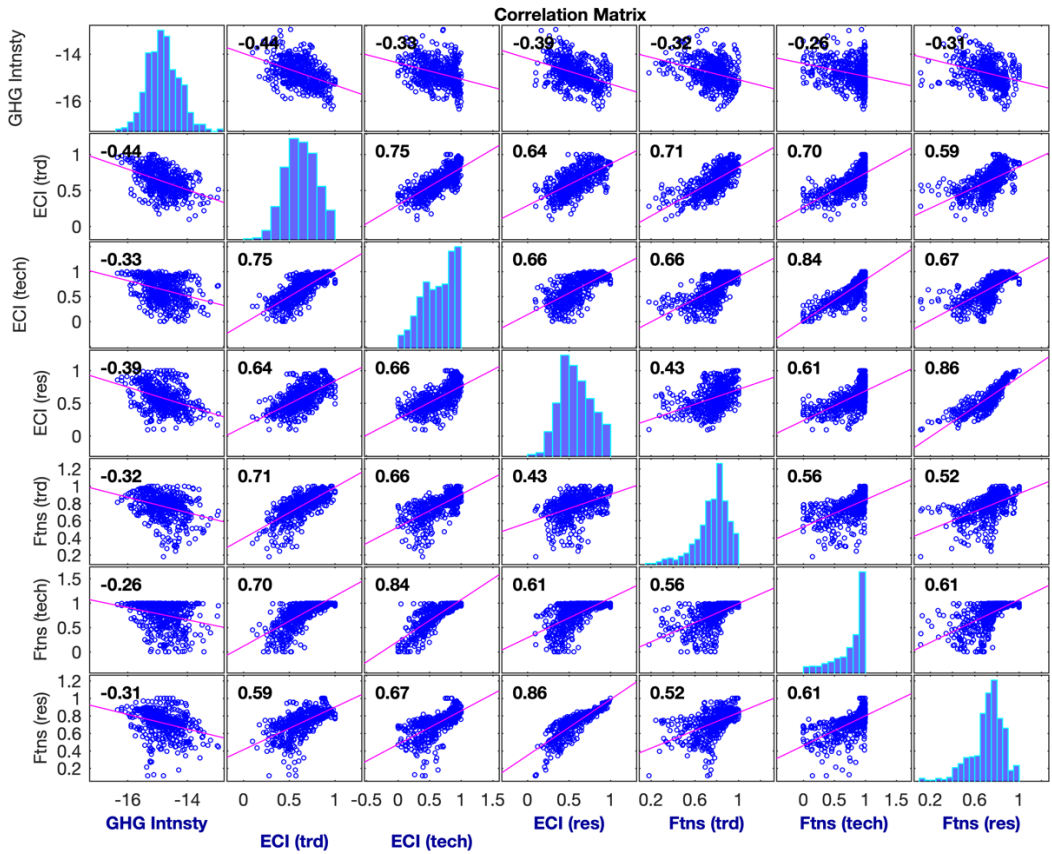

**Supplementary Figure 18. Correlations Between the Variables used in the Fitness Emission Intensity Robustness Check Regression Analysis**

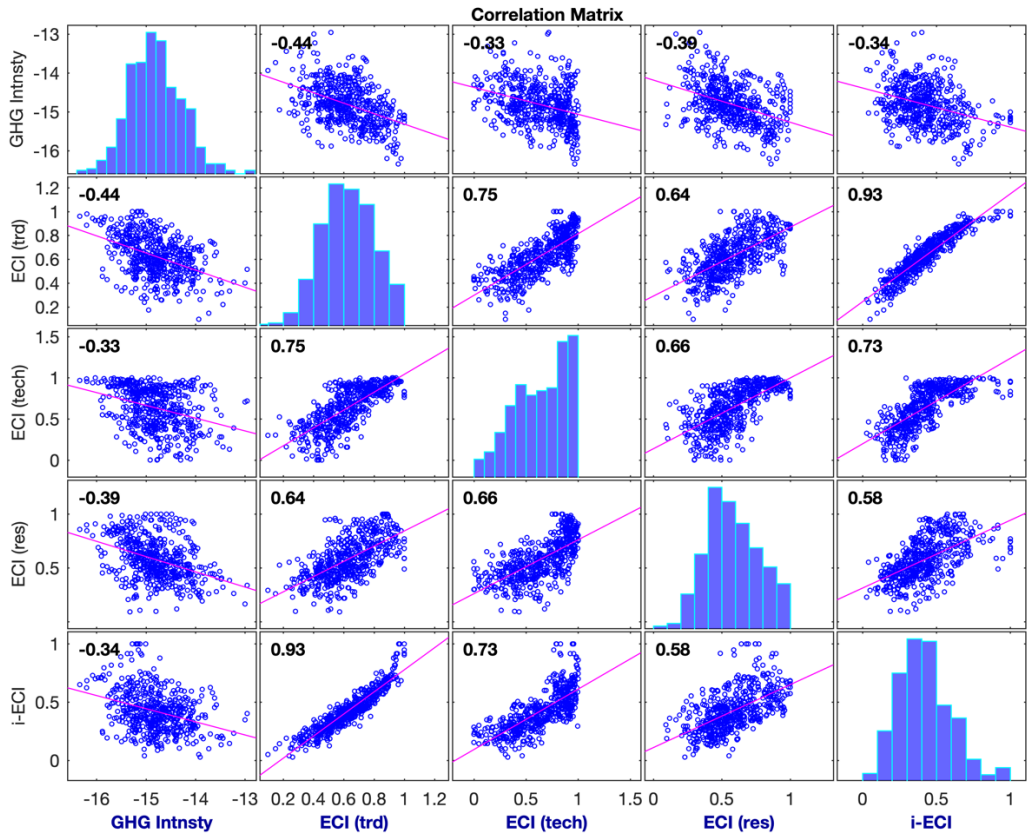

**Supplementary Figure 19. Correlations Between the Variables used in the i-ECI Emission Intensity Robustness Check Regression Analysis**

# Supplementary Note 7.     Economic complexity and embodied emissions

We also run an alternative model where instead of the emission intensities reported by countries, we use embodied emission intensities as the dependent. Embodied emissions add territorial and imported emissions and subtract exported emissions. We take the data for embodied emissions from the Global Carbon Project. They are available at:

www.icos-cp.eu/science-and-impact/global-carbon-budget/2021

We base our embodied emission intensity model on the same formulation as the model described in Supplementary Note 6. That is, in the models as controls we include as the log of the GDP per capita, log of population, log of human capital, and log of natural resources. The regression analysis is focused on the periods 1996-1999, 2000-2003, 2004-2007, 2008-2011, 2012-2015 and 2016-19. Because of the sparseness of the embodied greenhouse gas emissions dataset and slow temporal changes in the coefficients within a country we use average values for each panel.

In Supplementary Table 28 we estimate our embodied emission intensity ECI interaction models. We find that the multidimensional ECI model includes the trade ECI, technology ECI, and their interaction. Interestingly, it appears that both trade and technology ECI increase embodied emission intensity, and the interaction term suggests that they are weak substitutes. However, we also find that this result is not robust to alternate model specifications. Namely, Supplementary Tables 29-33 estimate the final production intensity, HHI, Entropy, Fitness, and i-ECI model. These results suggest that the multidimensional ECI embodied emissions model is not robust against the country-fixed effects and production intensity models.

## Supplementary Table 28. Embodied Emission Intensity Regressions

| | \multicolumn{13}{c}{Dependent variable:} |
|---|---|---|---|---|---|---|---|---|---|---|---|---|---|
| | \multicolumn{13}{c}{Embodied GHG per GDP (1996-99, 2000-03, 2004-07, 2008-11, 2012-15, 2016-19)} |
| | (1) | (2) | (3) | (4) | (5) | (6) | (7) | (8) | (9) | (10) | (11) | (12) | (13) |
| ECI (trade) | | -0.180 | | | 0.050 | -0.041 | | 0.030 | 2.223** | 1.359 | | -1.146 | 0.865 |
| | | (0.361) | | | (0.358) | (0.352) | | (0.359) | (0.926) | (0.894) | | (2.276) | (0.615) |
| ECI (technology) | | | -0.302 | | -0.317 | | -0.100 | -0.109 | 1.140** | | 0.746* | -1.315 | 0.791* |
| | | | (0.204) | | (0.201) | | (0.224) | (0.227) | (0.538) | | (0.402) | (1.364) | (0.430) |
| ECI (research) | | | | -0.779*** | | -0.774*** | -0.732** | -0.731** | | 0.748 | 0.536 | -2.933 | |
| | | | | (0.267) | | (0.264) | (0.285) | (0.286) | | (0.818) | (0.599) | (2.178) | |
| ECI (trade) x ECI (technology) | | | | | | | | | -2.725*** | | | 2.482 | -0.942 |
| | | | | | | | | | (0.945) | | | (2.558) | (0.681) |
| ECI (trade) x ECI (research) | | | | | | | | | | -2.236** | | 5.224 | |
| | | | | | | | | | | (1.125) | | (3.768) | |
| ECI (research) x ECI (technology) | | | | | | | | | | | -1.628** | 3.849 | |
| | | | | | | | | | | | (0.659) | (2.496) | |
| ECI (trade) x ECI (research) x ECI (technology) | | | | | | | | | | | | -7.796* | |
| | | | | | | | | | | | | (4.213) | |
| Log of population | 0.048 | 0.052 | 0.076** | 0.087** | 0.076** | 0.088** | 0.094*** | 0.094*** | 0.087*** | 0.081** | 0.087*** | 0.098*** | 0.383** |
| | (0.032) | (0.032) | (0.033) | (0.035) | (0.033) | (0.035) | (0.035) | (0.035) | (0.033) | (0.034) | (0.034) | (0.033) | (0.150) |
| Log of human capital | 0.597*** | 0.644*** | 0.714*** | 0.694*** | 0.707*** | 0.704*** | 0.727*** | 0.722*** | 0.607** | 0.640*** | 0.693*** | 0.647** | 0.879** |
| | (0.192) | (0.229) | (0.221) | (0.206) | (0.238) | (0.241) | (0.231) | (0.248) | (0.241) | (0.248) | (0.225) | (0.255) | (0.365) |
| Log of natural resource exports per capita | 0.174*** | 0.166*** | 0.178*** | 0.180*** | 0.180*** | 0.178*** | 0.181*** | 0.183*** | 0.213*** | 0.180*** | 0.188*** | 0.213*** | -0.052 |
| | (0.058) | (0.059) | (0.056) | (0.058) | (0.056) | (0.061) | (0.057) | (0.058) | (0.058) | (0.061) | (0.057) | (0.058) | (0.060) |
| Log of GDP per capita | -0.248*** | -0.219** | -0.210** | -0.147 | -0.216** | -0.142 | -0.141 | -0.145 | -0.270*** | -0.162 | -0.159 | -0.209** | -0.123 |
| | (0.096) | (0.101) | (0.096) | (0.098) | (0.100) | (0.105) | (0.099) | (0.104) | (0.097) | (0.108) | (0.099) | (0.101) | (0.121) |
| Fixed effects | No | No | No | No | No | No | No | No | No | No | No | No | Yes |
| Observations | 495 | 495 | 495 | 495 | 495 | 495 | 495 | 495 | 495 | 495 | 495 | 495 | 495 |
| $R^2$ | 0.163 | 0.165 | 0.175 | 0.209 | 0.175 | 0.209 | 0.210 | 0.210 | 0.225 | 0.231 | 0.230 | 0.258 | 0.254 |
| Adjusted $R^2$ | 0.148 | 0.148 | 0.158 | 0.193 | 0.156 | 0.191 | 0.192 | 0.191 | 0.206 | 0.211 | 0.211 | 0.233 | 0.037 |

Notes: Each regression includes period fixed effects. Clustered standard errors in brackets. *p<0.1, **p<0.05, ***p<0.01.

# Supplementary Table 29. Embodied Emission Intensity Regressions: Production Intensity Robustness Check

| | | | | | | | Dependent variable: | | | | | | | |
|---|---|---|---|---|---|---|---|---|---|---|---|---|---|---|
| | Embodied GHG emissions per GDP (1996-99, 2000-03, 2004-07, 2008-11, 2012-15, 2016-19) | | | | | | | | | | | | | |
| | (1) | (2) | (3) | (4) | (5) | (6) | (7) | (8) | (9) | (10) | (11) | (12) | (13) | (14) |
| ECI (trade) | 2.223** | | | | | | | | | | | | 1.044 | 0.858 |
| | (0.926) | | | | | | | | | | | | (0.954) | (0.607) |
| ECI (technology) | 1.140** | | | | | | | | | | | | 0.358 | 0.609 |
| | (0.538) | | | | | | | | | | | | (0.594) | (0.435) |
| ECI (trade) x ECI (technology) | -2.725*** | | | | | | | | | | | | -1.232 | -0.624 |
| | (0.945) | | | | | | | | | | | | (1.104) | (0.675) |
| Intensity (trade) | | 0.294 | | | 0.701 | 0.347 | | 0.691 | 2.427*** | 4.558*** | | 2.534 | 4.131*** | 2.136*** |
| | | (0.548) | | | (0.537) | (0.558) | | (0.531) | (0.764) | (0.985) | | (1.556) | (1.082) | (0.764) |
| Intensity (technology) | | | -0.401 | | -0.518* | | -0.584* | -0.694** | 1.607** | | 2.702*** | -0.255 | | |
| | | | (0.281) | | (0.286) | | (0.326) | (0.328) | (0.646) | | (0.904) | (1.941) | | |
| Intensity (research) | | | | -0.076 | | -0.120 | 0.350 | 0.342 | | 2.882*** | 1.515*** | 2.340* | 2.661*** | 1.057** |
| | | | | (0.330) | | (0.338) | (0.388) | (0.394) | | (0.631) | (0.414) | (1.263) | (0.711) | (0.528) |
| Intensity (trade) x Intensity (technology) | | | | | | | | | -2.810*** | | | 3.524 | | |
| | | | | | | | | | (0.742) | | | (3.163) | | |
| Intensity (trade) x Intensity (research) | | | | | | | | | | -4.842*** | | -2.638 | -4.150*** | -1.884** |
| | | | | | | | | | | (0.911) | | (2.339) | (1.138) | (0.866) |
| Intensity (research) x Intensity (technology) | | | | | | | | | | | -3.463*** | -0.134 | | |
| | | | | | | | | | | | (0.826) | (2.463) | | |
| Intensity (trade) x Intensity (research) x Intensity (technology) | | | | | | | | | | | | -3.354 | | |
| | | | | | | | | | | | | (3.446) | | |
| Log of GDP per population | 0.087*** | 0.048 | 0.054* | 0.049 | 0.056* | 0.050 | 0.051 | 0.053 | 0.062** | 0.048* | 0.065** | 0.056* | 0.076** | 0.353*** |
| | (0.033) | (0.032) | (0.033) | (0.032) | (0.033) | (0.032) | (0.033) | (0.032) | (0.029) | (0.028) | (0.029) | (0.029) | (0.032) | (0.135) |
| Log of human capital | 0.607** | 0.586*** | 0.739*** | 0.612*** | 0.753*** | 0.608*** | 0.735*** | 0.748*** | 0.503** | 0.435** | 0.519** | 0.454** | 0.466** | 0.624* |
| | (0.241) | (0.194) | (0.242) | (0.221) | (0.239) | (0.221) | (0.237) | (0.235) | (0.221) | (0.195) | (0.217) | (0.209) | (0.218) | (0.330) |
| Log of natural resource exports per capita | 0.213*** | 0.154** | 0.169*** | 0.173*** | 0.120* | 0.150** | 0.169*** | 0.121* | 0.119* | 0.118* | 0.204*** | 0.132** | 0.134* | -0.144** |
| | (0.058) | (0.072) | (0.058) | (0.058) | (0.072) | (0.073) | (0.057) | (0.071) | (0.070) | (0.067) | (0.061) | (0.066) | (0.074) | (0.072) |
| Log of GDP per capita | -0.270*** | -0.279*** | -0.166 | -0.236** | -0.217** | -0.265** | -0.185* | -0.235** | -0.350*** | -0.412*** | -0.285*** | -0.421*** | -0.439*** | -0.243* |
| | (0.097) | (0.103) | (0.103) | (0.103) | (0.108) | (0.108) | (0.105) | (0.110) | (0.119) | (0.109) | (0.107) | (0.115) | (0.105) | (0.145) |
| Fixed effects | No | No | No | No | No | No | No | No | No | No | No | No | No | Yes |
| Observations | 495 | 495 | 495 | 495 | 495 | 495 | 495 | 495 | 495 | 495 | 495 | 495 | 495 | 495 |
| R² | 0.225 | 0.165 | 0.178 | 0.164 | 0.186 | 0.166 | 0.183 | 0.191 | 0.273 | 0.304 | 0.278 | 0.316 | 0.318 | 0.286 |
| Adjusted R² | 0.206 | 0.148 | 0.161 | 0.146 | 0.167 | 0.147 | 0.164 | 0.170 | 0.255 | 0.287 | 0.260 | 0.293 | 0.296 | 0.072 |

Notes: Each regression includes period fixed effects. Clustered standard errors in brackets. *p<0.1, **p<0.05, ***p<0.01.

## Supplementary Table 30. Embodied Emission Intensity Regressions: HHI Robustness Check

| | *Dependent variable:* | | | | | | | | | | | | | |
|---|---|---|---|---|---|---|---|---|---|---|---|---|---|---|
| | Embodied GHG per GDP (1996-99, 2000-03, 2004-07, 2008-11, 2012-15, 2016-19) | | | | | | | | | | | | | |
| | (1) | (2) | (3) | (4) | (5) | (6) | (7) | (8) | (9) | (10) | (11) | (12) | (13) | (14) |
| ECI (trade) | 2.223** | | | | | | | | | | | | 2.612*** | 0.764 |
| | (0.926) | | | | | | | | | | | | (0.969) | (0.638) |
| ECI (technology) | 1.140** | | | | | | | | | | | | 1.405** | 0.767* |
| | (0.538) | | | | | | | | | | | | (0.570) | (0.435) |
| ECI (research) | -2.725*** | | | | | | | | | | | | -2.891*** | -0.932 |
| | (0.945) | | | | | | | | | | | | (0.964) | (0.680) |
| ECI (trade) x ECI (technology) | | 0.500* | | | 0.502 | 0.459 | | 0.474 | 0.294 | 0.295 | | 0.400 | 0.603* | -0.339 |
| | | (0.298) | | | (0.314) | (0.304) | | (0.319) | (0.429) | (0.395) | | (0.564) | (0.348) | (0.269) |
| HHI (trade) | | | 0.120 | | -0.005 | | 0.065 | -0.047 | -0.159 | | 0.049 | -0.120 | | |
| | | | (0.235) | | (0.251) | | (0.246) | (0.259) | (0.294) | | (0.281) | (0.333) | | |
| HHI (technology) | | | | 0.851 | | 0.722 | 0.822 | 0.740 | | 0.256 | 0.707 | 1.099 | | |
| | | | | (0.551) | | (0.524) | (0.557) | (0.532) | | (0.568) | (0.828) | (1.040) | | |
| HHI (research) | | | | | | | | | 0.966 | | | -0.109 | | |
| | | | | | | | | | (1.054) | | | (1.482) | | |
| HHI (trade) x HHI (technology) | | | | | | | | | | 4.437 | | -4.588 | | |
| | | | | | | | | | | (4.212) | | (8.169) | | |
| HHI (trade) x HHI (research) | | | | | | | | | | | 0.382 | -2.775 | | |
| | | | | | | | | | | | (2.264) | (2.945) | | |
| HHI (research) x HHI (technology) | | | | | | | | | | | | 27.381 | | |
| | | | | | | | | | | | | (19.280) | | |
| HHI (trade) x HHI (research) x HHI (technology) | 0.087*** | 0.046 | 0.054 | 0.063* | 0.045 | 0.059* | 0.066* | 0.057 | 0.042 | 0.059* | 0.066* | 0.054 | 0.063** | 0.384*** |
| | (0.033) | (0.032) | (0.033) | (0.034) | (0.034) | (0.035) | (0.035) | (0.036) | (0.034) | (0.034) | (0.036) | (0.036) | (0.032) | (0.148) |
| Log of population | 0.607** | 0.681*** | 0.629*** | 0.627*** | 0.680*** | 0.700*** | 0.644*** | 0.691*** | 0.675*** | 0.665*** | 0.643*** | 0.647*** | 0.567** | 0.859** |
| | (0.241) | (0.206) | (0.216) | (0.203) | (0.223) | (0.214) | (0.223) | (0.229) | (0.224) | (0.218) | (0.224) | (0.231) | (0.245) | (0.353) |
| Log of human capital | 0.213*** | 0.132** | 0.172*** | 0.183*** | 0.132** | 0.143** | 0.182*** | 0.143** | 0.131** | 0.144** | 0.181*** | 0.146** | 0.174*** | -0.036 |
| | (0.058) | (0.059) | (0.057) | (0.058) | (0.059) | (0.060) | (0.058) | (0.060) | (0.058) | (0.059) | (0.058) | (0.060) | (0.059) | (0.065) |
| Log of natural resource exports per capita | -0.270*** | -0.203** | -0.235** | -0.245*** | -0.204** | -0.205** | -0.239** | -0.208** | -0.214** | -0.202** | -0.239** | -0.219** | -0.282*** | -0.104 |
| | (0.097) | (0.094) | (0.094) | (0.094) | (0.093) | (0.093) | (0.093) | (0.092) | (0.094) | (0.092) | (0.093) | (0.094) | (0.096) | (0.115) |
| Fixed effects | No | No | No | No | No | No | No | No | No | No | No | No | No | Yes |
| Observations | 495 | 495 | 495 | 495 | 495 | 495 | 495 | 495 | 495 | 495 | 495 | 495 | 495 | 495 |
| $R^2$ | 0.225 | 0.183 | 0.165 | 0.175 | 0.183 | 0.191 | 0.175 | 0.191 | 0.186 | 0.195 | 0.175 | 0.204 | 0.242 | 0.257 |
| Adjusted $R^2$ | 0.206 | 0.166 | 0.148 | 0.158 | 0.164 | 0.172 | 0.157 | 0.171 | 0.166 | 0.175 | 0.155 | 0.177 | 0.222 | 0.040 |

Notes: Each regression includes period fixed effects. Clustered standard errors in brackets. *p<0.1, **p<0.05, ***p<0.01.

## Supplementary Table 31. Embodied Emission Entropy Regressions: Entropy Robustness Check

| | *Dependent variable:* | | | | | | | | | | | | | |
|---|---|---|---|---|---|---|---|---|---|---|---|---|---|---|
| | Embodied GHG emissions per GDP (1996-99, 2000-03, 2004-07, 2008-11, 2012-15, 2016-19) | | | | | | | | | | | | | |
| | (1) | (2) | (3) | (4) | (5) | (6) | (7) | (8) | (9) | (10) | (11) | (12) | (13) | (14) |
| ECI (trade) | 2.223** | | | | | | | | | | | | 2.635*** | 0.820 |
| | (0.926) | | | | | | | | | | | | (0.964) | (0.646) |
| ECI (technology) | 1.140** | | | | | | | | | | | | 1.471** | 0.837* |
| | (0.538) | | | | | | | | | | | | (0.573) | (0.437) |
| ECI (trade) x ECI (technology) | -2.725*** | | | | | | | | | | | | -2.904*** | -1.022 |
| | (0.945) | | | | | | | | | | | | (0.961) | (0.689) |
| Entropy (trade) | | -0.462* | | | -0.472 | -0.420 | | -0.444 | -1.148 | -4.345 | | -17.468* | -0.564* | 0.253 |
| | | (0.274) | | | (0.290) | (0.279) | | (0.294) | (0.771) | (3.025) | | (9.123) | (0.326) | (0.264) |
| Entropy (technology) | | | -0.110 | | 0.026 | | -0.047 | 0.075 | -0.685 | | -0.104 | -17.127 | | |
| | | | (0.236) | | (0.253) | | (0.248) | (0.262) | (0.863) | | (1.911) | (12.769) | | |
| Entropy (research) | | | | -0.747 | | -0.622 | -0.725 | -0.649 | | -4.213 | -0.766 | -15.476* | -0.593 | -0.538 |
| | | | | (0.491) | | (0.466) | (0.501) | (0.478) | | (2.977) | (1.508) | (8.650) | (0.474) | (0.361) |
| Entropy (trade) x Entropy (technology) | | | | | | | | | 0.875 | | | 20.123 | | |
| | | | | | | | | | (1.004) | | | (14.278) | | |
| Entropy (trade) x Entropy (research) | | | | | | | | | | 4.118 | | 17.317* | | |
| | | | | | | | | | | (3.256) | | (9.885) | | |
| Entropy (research) x Entropy (technology) | | | | | | | | | | | 0.060 | 17.438 | | |
| | | | | | | | | | | | (2.040) | (13.635) | | |
| Entropy (trade) x Entropy (research) x Entropy (technology) | | | | | | | | | | | | -20.324 | | |
| | | | | | | | | | | | | (15.198) | | |
| Log of GDP per population | 0.087*** | 0.046 | 0.054 | 0.064* | 0.045 | 0.060* | 0.067* | 0.057 | 0.041 | 0.060* | 0.067* | 0.055 | 0.072** | 0.428*** |
| | (0.033) | (0.032) | (0.034) | (0.035) | (0.035) | (0.035) | (0.036) | (0.037) | (0.035) | (0.035) | (0.036) | (0.037) | (0.033) | (0.138) |
| Log of human capital | 0.607** | 0.684*** | 0.628*** | 0.629*** | 0.678*** | 0.702*** | 0.642*** | 0.687*** | 0.671*** | 0.658*** | 0.641*** | 0.632*** | 0.570** | 0.831** |
| | (0.241) | (0.207) | (0.218) | (0.204) | (0.224) | (0.216) | (0.225) | (0.231) | (0.225) | (0.219) | (0.226) | (0.231) | (0.250) | (0.351) |
| Log of natural resource exports per capita | 0.213*** | 0.131** | 0.172*** | 0.183*** | 0.130** | 0.142** | 0.182*** | 0.141** | 0.129** | 0.142** | 0.182*** | 0.146** | 0.179*** | -0.031 |
| | (0.058) | (0.059) | (0.057) | (0.058) | (0.059) | (0.060) | (0.058) | (0.060) | (0.058) | (0.059) | (0.058) | (0.060) | (0.060) | (0.064) |
| Log of GDP per capita | -0.270*** | -0.201** | -0.236** | -0.243*** | -0.203** | -0.201** | -0.238** | -0.206** | -0.213** | -0.198** | -0.238** | -0.216** | -0.284*** | -0.095 |
| | (0.097) | (0.093) | (0.094) | (0.094) | (0.093) | (0.093) | (0.093) | (0.092) | (0.094) | (0.092) | (0.093) | (0.094) | (0.097) | (0.116) |
| Fixed effects | No | No | No | No | No | No | No | No | No | No | No | No | No | Yes |
| Observations | 495 | 495 | 495 | 495 | 495 | 495 | 495 | 495 | 495 | 495 | 495 | 495 | 495 | 495 |
| $R^2$ | 0.225 | 0.183 | 0.165 | 0.175 | 0.183 | 0.191 | 0.175 | 0.191 | 0.186 | 0.198 | 0.175 | 0.206 | 0.250 | 0.263 |
| Adjusted $R^2$ | 0.206 | 0.166 | 0.147 | 0.158 | 0.164 | 0.172 | 0.157 | 0.171 | 0.166 | 0.178 | 0.155 | 0.180 | 0.228 | 0.044 |

Notes: Each regression includes period fixed effects. Clustered standard errors in brackets. *p<0.1, **p<0.05, ***p<0.01.

## Supplementary Table 32. Embodied Emission Intensity Regressions: Fitness Robustness Check

| | *Dependent variable:* | | | | | | | | | | | | | |
|---|---|---|---|---|---|---|---|---|---|---|---|---|---|---|
| | Embodied GHG emissions per GDP (1996-99, 2000-03, 2004-07, 2008-11, 2012-15, 2016-19) | | | | | | | | | | | | | |
| | (1) | (2) | (3) | (4) | (5) | (6) | (7) | (8) | (9) | (10) | (11) | (12) | (13) | (14) |
| ECI (trade) | 2.223** | | | | | | | | | | | | 2.841*** | 0.892 |
| | (0.926) | | | | | | | | | | | | (0.992) | (0.732) |
| ECI (technology) | 1.140** | | | | | | | | | | | | 1.551** | 0.877* |
| | (0.538) | | | | | | | | | | | | (0.715) | (0.524) |
| ECI (trade) x ECI (technology) | -2.725*** | | | | | | | | | | | | -3.288*** | -1.145 |
| | (0.945) | | | | | | | | | | | | (1.082) | (0.840) |
| Log of fitness (technology) | | -0.444 | | | -0.577 | -0.273 | | -0.416 | -1.620 | -1.207 | | -12.309*** | -11.042** | -5.975*** |
| | | (0.348) | | | (0.351) | (0.359) | | (0.357) | (1.075) | (1.680) | | (4.540) | (4.318) | (2.252) |
| Log of fitness (research) | | | 0.119 | | 0.238 | | 0.213 | 0.287 | -0.687 | | 1.191** | -7.302** | -6.321** | -4.489** |
| | | | (0.180) | | (0.174) | | (0.192) | (0.183) | (0.900) | | (0.561) | (3.474) | (3.170) | (1.964) |
| Log of fitness (trade) x Log of fitness (technology) | | | | -0.711* | | -0.624 | -0.792** | -0.687* | | -1.534 | 0.312 | -11.126** | -9.154** | -5.959*** |
| | | | | (0.382) | | (0.390) | (0.400) | (0.402) | | (1.682) | (0.775) | (5.193) | (4.767) | (2.014) |
| Log of fitness (trade) x Log of fitness (research) | | | | | | | | | 1.376 | | | 12.909** | 10.339** | 6.308** |
| | | | | | | | | | (1.316) | | | (5.015) | (4.569) | (3.004) |
| Log of fitness (research) x Log of fitness (technology) | | | | | | | | | | 1.293 | | 17.397** | 13.654** | 8.592*** |
| | | | | | | | | | | (2.264) | | (7.491) | (6.920) | (3.120) |
| Log of fitness (trade) x Log of fitness (research) x Log of fitness (technology) | | | | | | | | | | | -1.564* | 10.411* | 7.730 | 6.302** |
| | | | | | | | | | | | (0.874) | (5.636) | (5.133) | (2.702) |
| Log of population | | | | | | | | | | | | -18.166** | -12.538* | -8.704** |
| | | | | | | | | | | | | (7.981) | (7.277) | (4.129) |
| Log of human capital | 0.087*** | 0.063* | 0.039 | 0.084** | 0.050 | 0.089** | 0.073* | 0.076** | 0.039 | 0.085** | 0.082** | 0.079** | 0.082** | 0.501*** |
| | (0.033) | (0.033) | (0.032) | (0.039) | (0.033) | (0.039) | (0.038) | (0.038) | (0.033) | (0.040) | (0.038) | (0.037) | (0.037) | (0.138) |
| Log of natural resource exports per capita | 0.607** | 0.688*** | 0.552*** | 0.664*** | 0.623*** | 0.712*** | 0.589*** | 0.636*** | 0.614*** | 0.700*** | 0.608*** | 0.566*** | 0.489** | 0.640* |
| | (0.241) | (0.214) | (0.210) | (0.215) | (0.221) | (0.227) | (0.228) | (0.233) | (0.224) | (0.227) | (0.219) | (0.217) | (0.242) | (0.381) |
| Log of GDP per capita | 0.213*** | 0.148** | 0.177*** | 0.180*** | 0.146** | 0.163*** | 0.186*** | 0.162*** | 0.137** | 0.160*** | 0.190*** | 0.166*** | 0.191*** | -0.041 |
| | (0.058) | (0.058) | (0.059) | (0.058) | (0.058) | (0.059) | (0.060) | (0.059) | (0.055) | (0.058) | (0.060) | (0.055) | (0.054) | (0.054) |
| log(gdp_start) | -0.270*** | -0.203** | -0.266*** | -0.189** | -0.225** | -0.169* | -0.215** | -0.193* | -0.235** | -0.174* | -0.199** | -0.171* | -0.272*** | -0.107 |
| | (0.097) | (0.093) | (0.101) | (0.096) | (0.098) | (0.095) | (0.100) | (0.099) | (0.100) | (0.095) | (0.101) | (0.096) | (0.094) | (0.112) |
| Fixed effects | No | No | No | No | No | No | No | No | No | No | No | No | No | Yes |
| Observations | 495 | 495 | 495 | 495 | 495 | 495 | 495 | 495 | 495 | 495 | 495 | 495 | 495 | 495 |
| R² | 0.225 | 0.174 | 0.165 | 0.187 | 0.181 | 0.190 | 0.193 | 0.200 | 0.189 | 0.192 | 0.205 | 0.249 | 0.300 | 0.300 |
| Adjusted R² | 0.206 | 0.157 | 0.148 | 0.170 | 0.162 | 0.172 | 0.174 | 0.181 | 0.169 | 0.172 | 0.186 | 0.224 | 0.272 | 0.080 |

Notes: Each regression includes period fixed effects. Clustered standard errors in brackets. *p<0.1, **p<0.05, ***p<0.01.

# Supplementary Table 33. Embodied Emission Intensity Regressions: i-ECI Robustness Check

| | Dependent variable: | | | | | |
|---|---|---|---|---|---|---|
| | Embodied GHG emissions per GDP (1996-99, 2000-03, 2004-07, 2008-11, 2012-15, 2016-19) | | | | | |
| | (1) | (2) | (3) | (4) | (5) | (6) |
| ECI (trade) | 2.223** | | | | 2.049** | 0.526 |
| | (0.926) | | | | (1.003) | (0.570) |
| ECI (technology) | 1.140** | | | | 1.226** | 0.853* |
| | (0.538) | | | | (0.560) | (0.442) |
| ECI (trade) x ECI (technology) | -2.725*** | | | | -2.862*** | -1.059 |
| | (0.945) | | | | (0.974) | (0.711) |
| i-ECI | | 0.134 | 0.305 | 2.323*** | 0.295 | 0.496 |
| | | (0.248) | (0.246) | (0.835) | (0.562) | (0.313) |
| ECI (research) | | | -0.513* | 0.927 | | |
| | | | (0.289) | (0.616) | | |
| i-ECI x ECI (research) | | | | -3.199*** | | |
| | | | | (1.151) | | |
| Log of population | 0.087*** | | | | 0.084** | 0.346** |
| | (0.033) | | | | (0.036) | (0.149) |
| Log of human capital | 0.607** | | | | 0.577** | 0.860** |
| | (0.241) | | | | (0.253) | (0.372) |
| Log of natural resource export per capita | 0.213*** | | | | 0.210*** | -0.059 |
| | (0.058) | | | | (0.058) | (0.059) |
| Log of GDP per capita | -0.270*** | 0.087 | 0.134** | 0.109* | -0.265*** | -0.121 |
| | (0.097) | (0.054) | (0.062) | (0.062) | (0.097) | (0.121) |
| Fixed effects | No | No | No | No | No | Yes |
| Observations | 495 | 495 | 495 | 495 | 495 | 495 |
| $R^2$ | 0.225 | 0.049 | 0.073 | 0.120 | 0.226 | 0.264 |
| Adjusted $R^2$ | 0.206 | 0.035 | 0.058 | 0.103 | 0.205 | 0.048 |

Notes: Each regression includes period fixed effects. Clustered standard errors in brackets. *p<0.1, **p<0.05, ***p<0.01.

# Supplementary Note 8. Instrumental variables robustness check

To account for the possible endogeneity between the various dimensions of ECI and economic growth, income inequality or emission intensity of a country, we re-estimate our results by using an instrumental variables (IV) approach. Endogeneity may arise because there might be omitted variables which affect both ECI and the macroeconomic outcome. The IV estimation approach corrects for possible endogeneity by using instruments: variables that do not belong in the explanatory equation but are correlated with ECI.

We use this approach to check the robustness of our final multidimensional models results to the inclusion of the additional variables, the production intensity, HHI, or entropy models to the analysis. We exclude the Fitness and i-ECI analyses from this robustness checks as they are endogenous to our complexity measure (both of them are alternate methods for calculating complexity and thus should be highly correlated with ECI).

**Defining the instrumental variable:** For each country $c$, we use the average complexity of the three economies that have the most similar specialization pattern as instrument with which the country does not share a land or maritime border. The rationale behind using this variable as an instrument is that countries with similar specialization patterns should also have similar complexity as the target country, but the complexity of the similar economies should not affect the macroeconomic outcome of the target country and it should not be related to other country specific features. We exclude neighboring countries as potential similar countries because it is known that local conditions usually diffuse geographically, and thus may bias our instrument.

We quantify the similarity of the specialization patterns between the target country $c$ and a paired country $c'$ by using the minimum conditional probability,

$$\phi_{cc'}^d = \frac{\sum_p M_{cp}^d M_{c'p}^d}{\max(M_c^d, M_{c'}^d)},$$

for the countries to be specialized in the same activity.

Supplementary Tables 34, 35, and 36 list, respectively, the countries and their 3 most similar economies in terms of trade, technology, and research specialization patterns in 2014. For example, in 2014 Japan was similar in terms of exports to Germany, South Korea, and Great Britain and Australia was closest in terms of technology to Great Britain, Spain, and Canada. This indeed suggests that the similarity-based metrics are not determined by the similarity in local conditions.

**Supplementary Table 34. List of Countries and their 3 Most Similar Economies in Terms of Trade Specialization Patterns in 2014**

| Country | Similar country 1 | Similar country 2 | Similar country 3 | Country | Similar country 1 | Similar country 2 | Similar country 3 | Country | Similar country 1 | Similar country 2 | Similar country 3 |
|---|---|---|---|---|---|---|---|---|---|---|---|
| **AGO** | GNQ | GAB | DZA | **GNQ** | AGO | LBY | COG | **NLD** | FRA | USA | POL |
| **ALB** | MAR | BIH | MDA | **GRC** | LBN | SRB | SYR | **NOR** | NAM | PRK | CAN |
| **ARE** | TGO | GEO | BHR | **GTM** | DOM | JOR | KEN | **NZL** | URY | ARG | CAN |
| **ARG** | NZL | PER | UKR | **HKG** | THA | PAN | VNM | **OMN** | TTO | BHR | MOZ |
| **ARM** | JAM | KGZ | MDA | **HND** | DOM | CRI | MUS | **PAK** | SYR | LKA | EGY |
| **AUS** | URY | IRL | RUS | **HRV** | LVA | BGR | EST | **PAN** | HKG | CHN | LKA |
| **AUT** | FRA | POL | USA | **HUN** | POL | CZE | BGR | **PER** | MAR | KEN | HND |
| **AZE** | YEM | SDN | DZA | **IDN** | LKA | PAK | TUN | **PHL** | LKA | THA | HKG |
| **BEL** | POL | USA | ESP | **IND** | TUR | EGY | VNM | **PNG** | GIN | CMR | YEM |
| **BEN** | CIV | PRY | CMR | **IRL** | URY | AUS | NZL | **POL** | PRT | AUT | BEL |
| **BFA** | SDN | ETH | GIN | **IRN** | UZB | SYR | KGZ | **PRK** | ALB | MMR | MAR |
| **BGD** | KHM | LSO | LKA | **IRQ** | SSD | DZA | LBY | **PRT** | TUR | POL | ITA |
| **BGR** | HRV | LTU | POL | **ISR** | CHE | USA | DOM | **PRY** | BEN | NIC | ECU |
| **BHR** | TTO | OMN | ARE | **ITA** | DEU | CHN | TUR | **QAT** | KWT | VEN | LBY |
| **BIH** | MKD | ALB | SVK | **JAM** | ARM | CRI | UGA | **ROU** | HRV | SVN | POL |
| **BLR** | HRV | BIH | SRB | **JOR** | GTM | DOM | LBN | **RUS** | AUS | CHL | CAN |
| **BOL** | BEN | LAO | MMR | **JPN** | DEU | GBR | CZE | **SAU** | TTO | NER | VEN |
| **BRA** | NZL | TZA | UKR | **KAZ** | OMN | ZMB | MOZ | **SDN** | BFA | MLI | NER |
| **BWA** | MNG | COD | GIN | **KEN** | GTM | JOR | DOM | **SEN** | UGA | TZA | KEN |
| **CAN** | FIN | SWE | NZL | **KGZ** | SYR | MKD | PAK | **SGP** | CHE | JPN | HKG |
| **CHE** | JPN | ISR | SWE | **KHM** | BGD | LSO | MMR | **SLE** | COD | NGA | PNG |
| **CHL** | MMR | NZL | URY | **KOR** | THA | HKG | FIN | **SLV** | DOM | MUS | KEN |
| **CHN** | ITA | PRT | ESP | **KWT** | QAT | VEN | LBY | **SRB** | LVA | POL | LTU |
| **CIV** | CMR | TGO | BEN | **LAO** | NIC | ETH | MDG | **SSD** | IRQ | AGO | DZA |
| **CMR** | CIV | GHA | BEN | **LBN** | EGY | GRC | JOR | **SVK** | HRV | ROU | BIH |
| **COD** | GIN | SLE | NGA | **LBR** | COG | NGA | GAB | **SVN** | CZE | POL | SWE |
| **COG** | LBR | NGA | GIN | **LBY** | VEN | QAT | KWT | **SWE** | AUT | CZE | SVN |
| **COL** | CIV | KEN | UGA | **LKA** | PAK | VNM | IDN | **SWZ** | SLV | MDG | MUS |
| **CRI** | HND | GTM | SLV | **LSO** | BGD | KHM | ETH | **SYR** | EGY | PAK | GTM |
| **CUB** | YEM | MOZ | ZMB | **LTU** | DNK | EST | BGR | **TCD** | MLI | COD | TKM |
| **CYP** | LVA | UGA | MDA | **LVA** | HRV | SRB | DNK | **TGO** | SEN | CIV | TZA |
| **CZE** | SVN | SWE | PRT | **MAR** | TUN | ALB | SYR | **THA** | PRT | CHN | HKG |
| **DEU** | USA | ITA | ESP | **MDA** | GTM | MKD | SYR | **TKM** | BFA | MLI | SDN |
| **DNK** | LTU | AUT | EST | **MDG** | LKA | MMR | KEN | **TTO** | BHR | OMN | QAT |
| **DOM** | GTM | SLV | JOR | **MEX** | TUN | HUN | DOM | **TUN** | MAR | DOM | PAK |
| **DZA** | QAT | IRQ | KWT | **MKD** | BIH | MDA | GTM | **TUR** | PRT | ITA | ESP |
| **ECU** | GHA | CIV | CMR | **MLI** | SDN | YEM | CMR | **TZA** | SEN | PAK | ETH |
| **EGY** | SYR | LBN | PAK | **MMR** | MDG | NIC | TZA | **UGA** | SEN | GTM | CRI |
| **ESP** | DEU | TUR | AUT | **MNG** | SDN | NGA | BOL | **UKR** | SRB | HRV | LVA |
| **EST** | LTU | DNK | POL | **MOZ** | CIV | TGO | BEN | **URY** | NZL | SEN | AUS |
| **ETH** | TZA | NIC | PAK | **MRT** | YEM | LBR | GIN | **USA** | DEU | FRA | GBR |
| **FIN** | AUT | SVN | DEU | **MUS** | SLV | HND | GTM | **UZB** | IRN | ETH | LAO |
| **FRA** | USA | AUT | NLD | **MWI** | UGA | ZWE | CIV | **VEN** | QAT | LBY | KWT |
| **GAB** | AGO | COD | LBR | **MYS** | HKG | KOR | EST | **VNM** | LKA | TUN | PAK |
| **GBR** | USA | JPN | SWE | **NAM** | NOR | AUS | NZL | **YEM** | SDN | NER | PNG |
| **GEO** | ALB | CHL | MMR | **NER** | SDN | YEM | TGO | **ZAF** | EGY | BGR | GRC |
| **GHA** | CMR | LAO | ECU | **NGA** | COD | GIN | SLE | **ZMB** | UGA | BEN | PRY |
| **GIN** | COD | PNG | NGA | **NIC** | ETH | LAO | MMR | **ZWE** | MWI | LAO | UGA |

**Supplementary Table 35. List of Countries and their 3 Most Similar Economies in Terms of Technology Specialization Patterns in 2014**

| Country | Similar country 1 | Similar country 2 | Similar country 3 | Country | Similar country 1 | Similar country 2 | Similar country 3 |
|---|---|---|---|---|---|---|---|
| **ARE** | ROU | BGR | MAR | **KAZ** | QAT | BLR | MAR |
| **ARG** | EGY | COL | MAR | **KEN** | MDA | SYR | CYP |
| **ARM** | CYP | TUN | PER | **KOR** | HKG | TUR | UKR |
| **AUS** | GBR | ESP | CAN | **KWT** | PAN | MUS | BOL |
| **AUT** | ESP | AUS | BRA | **LBN** | IRN | URY | BGD |
| **AZE** | QAT | VEN | DZA | **LKA** | LVA | IDN | SRB |
| **BEL** | PRT | CHE | DNK | **LTU** | EST | MAR | PHL |
| **BGD** | CUB | CRI | BHR | **LVA** | LKA | VNM | SRB |
| **BGR** | EGY | MAR | ARE | **MAR** | LTU | BGR | HRV |
| **BHR** | BGD | JOR | UZB | **MDA** | KEN | JOR | BLR |
| **BIH** | DZA | AZE | ECU | **MEX** | ZAF | ESP | POL |
| **BLR** | KAZ | IDN | QAT | **MUS** | KWT | OMN | TTO |
| **BOL** | GHA | TTO | UZB | **MYS** | IRL | FIN | NOR |
| **BRA** | AUS | GBR | ESP | **NGA** | JAM | UZB | ECU |
| **BWA** | CUB | GHA | JAM | **NLD** | NZL | DNK | AUS |
| **CAN** | AUS | ESP | RUS | **NOR** | AUS | NZL | CAN |
| **CHE** | ESP | BEL | GBR | **NZL** | ZAF | ESP | CAN |
| **CHL** | COL | PRT | THA | **OMN** | MUS | DZA | BHR |
| **CHN** | FIN | TUR | HUN | **PAK** | KAZ | QAT | PER |
| **COL** | SVN | CHL | THA | **PAN** | KWT | JOR | NGA |
| **CRI** | CUB | BGD | PAK | **PER** | PHL | KAZ | PAK |
| **CUB** | BGD | CRI | BWA | **PHL** | LTU | EST | HRV |
| **CYP** | ARM | TUN | VEN | **POL** | MEX | ZAF | ESP |
| **CZE** | RUS | ZAF | MEX | **PRK** | ARM | DZA | CRI |
| **DEU** | ITA | RUS | BRA | **PRT** | GRC | SVN | MEX |
| **DNK** | NZL | MEX | IRL | **QAT** | KAZ | PAK | MAR |
| **DOM** | DZA | PER | AZE | **ROU** | ARE | HRV | SVK |
| **DZA** | ECU | BIH | GEO | **RUS** | AUS | GBR | CAN |
| **ECU** | CUB | DZA | UZB | **SAU** | IND | MAR | LTU |
| **EGY** | BGR | ARG | MAR | **SGP** | PRT | IRL | USA |
| **ESP** | AUS | ZAF | CAN | **SLV** | BWA | UGA | BGD |
| **EST** | LTU | PHL | EGY | **SRB** | LTU | LVA | EGY |
| **FIN** | CAN | MYS | NZL | **SVK** | SVN | GRC | THA |
| **FRA** | RUS | AUT | ZAF | **SVN** | GRC | COL | PRT |
| **GBR** | AUS | BRA | RUS | **SWE** | NZL | AUS | ZAF |
| **GEO** | VEN | QAT | KAZ | **SYR** | URY | AZE | BIH |
| **GHA** | BOL | BWA | CUB | **THA** | SVN | COL | CHL |
| **GRC** | PRT | SVN | MEX | **TTO** | BOL | UGA | GHA |
| **HKG** | KOR | MEX | GRC | **TUN** | IRN | ARM | PAK |
| **HRV** | ROU | MAR | COL | **TUR** | ESP | MEX | ITA |
| **HUN** | CZE | PRT | MEX | **TZA** | KWT | MUS | SLV |
| **IDN** | MAR | LTU | HRV | **UGA** | TTO | BOL | PRK |
| **IND** | SGP | SAU | CAN | **UKR** | ZAF | GRC | KOR |
| **IRL** | DNK | NZL | MYS | **URY** | LBN | KAZ | CUB |
| **IRN** | VEN | TUN | KAZ | **USA** | ISR | GBR | IRL |
| **ISR** | USA | IRL | HKG | **UZB** | NGA | BGD | ECU |
| **ITA** | DEU | AUS | GBR | **VEN** | IRN | PER | KAZ |
| **JAM** | NGA | BWA | BGD | **VNM** | LTU | LVA | MAR |
| **JOR** | BGD | BHR | CRI | **ZAF** | AUS | ESP | NZL |
| **JPN** | DEU | UKR | SVK | | | | |

**Supplementary Table 36. List of Countries and their 3 Most Similar Economies in Terms of Research Specialization Patterns in 2014**

| Country | Similar country 1 | Similar country 2 | Similar country 3 | Country | Similar country 1 | Similar country 2 | Similar country 3 | Country | Similar country 1 | Similar country 2 | Similar country 3 |
|---|---|---|---|---|---|---|---|---|---|---|---|
| **AFG** | COD | TGO | AGO | **GNQ** | AGO | GIN | LBR | **NOR** | NZL | NLD | CAN |
| **AGO** | GIN | LBR | SLE | **GRC** | PRT | ESP | AUT | **NPL** | ETH | UGA | TZA |
| **ALB** | PNG | MUS | RWA | **GTM** | JAM | LAO | COG | **NZL** | CAN | NLD | GBR |
| **ARE** | MYS | CYP | SGP | **HKG** | SGP | PRT | FIN | **OMN** | IRQ | KWT | DZA |
| **ARG** | HUN | MEX | BEL | **HND** | DOM | COD | HTI | **PAK** | THA | SAU | SRB |
| **ARM** | MDA | BLR | UZB | **HRV** | NGA | JOR | THA | **PAN** | BFA | ETH | ECU |
| **AUS** | GBR | CAN | USA | **HTI** | SLV | COD | TGO | **PER** | TZA | KEN | GHA |
| **AUT** | DNK | ESP | BEL | **HUN** | ARG | CZE | CHL | **PHL** | KEN | GHA | TZA |
| **AZE** | MDA | BHR | UZB | **IDN** | NGA | COL | BGD | **PNG** | RWA | NIC | GAB |
| **BEL** | SWE | DNK | CAN | **IND** | SAU | KOR | EGY | **POL** | SVN | IRN | MEX |
| **BEN** | MDG | MLI | CIV | **IRL** | NZL | NLD | DNK | **PRT** | SVN | GRC | TUN |
| **BFA** | SDN | PAN | SEN | **IRN** | EGY | IND | TUN | **PRY** | DOM | NIC | PNG |
| **BGD** | ETH | KEN | GHA | **IRQ** | OMN | DZA | LTU | **PSE** | IRQ | YEM | SDN |
| **BGR** | SVK | VNM | MEX | **ISR** | NLD | IRL | ITA | **QAT** | KWT | DZA | OMN |
| **BHR** | AZE | MDA | YEM | **ITA** | DEU | ISR | BEL | **ROU** | CHN | DZA | SVK |
| **BIH** | MKD | YEM | KAZ | **JAM** | GTM | NER | TTO | **RUS** | DZA | CZE | SVK |
| **BLR** | MDA | ARM | KAZ | **JOR** | MYS | TUN | CYP | **RWA** | MDG | PNG | MOZ |
| **BOL** | PAN | BFA | CRI | **JPN** | IND | EGY | SGP | **SAU** | IND | CHN | SRB |
| **BRA** | TUR | MEX | NGA | **KAZ** | BLR | YEM | MNG | **SDN** | BFA | PAN | SEN |
| **BWA** | MLI | COG | MNG | **KEN** | GHA | PHL | PER | **SEN** | BFA | ZMB | MWI |
| **CAN** | AUS | GBR | NLD | **KGZ** | ALB | MNG | NAM | **SGP** | HKG | ARE | CHN |
| **CHE** | NLD | BEL | USA | **KHM** | MLI | CIV | MOZ | **SLE** | SWZ | MMR | AGO |
| **CHL** | MEX | ZAF | COL | **KOR** | IND | SAU | SGP | **SLV** | HTI | COD | DOM |
| **CHN** | DZA | SAU | SGP | **KWT** | OMN | QAT | JOR | **SRB** | SVN | SVK | SAU |
| **CIV** | KHM | MOZ | GAB | **LAO** | MDG | GAB | NIC | **SVK** | SRB | BGR | SVN |
| **CMR** | ETH | TZA | NPL | **LBN** | GRC | TUN | TUR | **SVN** | SRB | PRT | POL |
| **COD** | TGO | AFG | GIN | **LBR** | AGO | COD | SWZ | **SWE** | NLD | GBR | CAN |
| **COG** | MLI | ZWE | MDG | **LBY** | YEM | TTO | MDA | **SWZ** | SLE | MMR | AGO |
| **COL** | IDN | NGA | CHL | **LKA** | TZA | ETH | GHA | **SYR** | BFA | SDN | ZWE |
| **CRI** | TZA | ETH | UGA | **LTU** | DZA | IRQ | OMN | **TGO** | COD | AFG | AGO |
| **CUB** | BFA | CMR | UGA | **LVA** | IRQ | DZA | PSE | **THA** | NGA | PAK | BGD |
| **CYP** | ARE | JOR | HKG | **MAR** | PAK | BGR | VNM | **TJK** | NAM | ARM | KAZ |
| **CZE** | HUN | MEX | SVN | **MDA** | BLR | ARM | AZE | **TTO** | NER | RWA | JAM |
| **DEU** | ITA | ISR | ESP | **MDG** | GAB | BEN | MLI | **TUN** | PRT | VNM | IRN |
| **DNK** | NLD | BEL | CAN | **MEX** | CHL | ARG | SVN | **TUR** | POL | LBN | BRA |
| **DOM** | HND | PRY | COD | **MKD** | BIH | PSE | CUB | **TZA** | ETH | GHA | CMR |
| **DZA** | CHN | IRQ | MYS | **MLI** | KHM | COG | MDG | **UGA** | ETH | GHA | NPL |
| **ECU** | TZA | ETH | UGA | **MMR** | SLE | SWZ | AGO | **UKR** | DZA | LTU | CHN |
| **EGY** | IND | IRN | THA | **MNG** | BWA | BOL | KAZ | **URY** | ETH | CMR | KEN |
| **ESP** | BEL | AUT | NZL | **MOZ** | KHM | CIV | GAB | **USA** | GBR | AUS | NLD |
| **EST** | ARG | CHL | MEX | **MRT** | AGO | GIN | GNQ | **UZB** | MNG | AZE | ARM |
| **ETH** | TZA | GHA | CMR | **MUS** | JAM | NAM | LAO | **VEN** | PER | URY | TZA |
| **FIN** | NZL | NLD | CAN | **MWI** | ZWE | UGA | BFA | **VNM** | TUN | DZA | BGR |
| **FRA** | AUT | ISR | GRC | **MYS** | JOR | ARE | DZA | **YEM** | PSE | LBY | SDN |
| **GAB** | MDG | MOZ | CIV | **NAM** | PNG | LAO | NIC | **ZAF** | NZL | CHL | NOR |
| **GBR** | AUS | USA | CAN | **NER** | TTO | CIV | JAM | **ZMB** | SEN | MLI | BFA |
| **GEO** | PSE | ZMB | ECU | **NGA** | GHA | IDN | BGD | **ZWE** | MWI | COG | SEN |
| **GHA** | KEN | ETH | TZA | **NIC** | PNG | LAO | PRY | | | | |
| **GIN** | AGO | COD | SWZ | **NLD** | SWE | CAN | USA | | | | |

**Economic growth IV estimation First-Stage results:** The first-stage results for the IV estimation of the economic growth models are given in Supplementary Tables 37-38. Supplementary Table 37 presents the results for ECI (trade), whereas Supplementary Table 38 gives the results for ECI (technology). In each case, the F-statistic for the significance of the included instruments is above 10, suggesting that our instruments are strong.

**Economic growth IV estimation Second-Stage results:** The coefficients of the multidimensional ECI economic growth model remain significant, even when using the IV estimation approach (Supplementary Table 39, column (1)). Moreover, they remain robust even after including the additional explanatory variables (Supplementary Table 39, columns (2-6)), the multidimensional production intensity model (Supplementary Table 39, columns (7-8)), the multidimensional HHI model (Supplementary Table 39, columns (9-10)), and the multidimensional Entropy model (Supplementary Table 39, columns (11-12)).

# Supplementary Table 37. Economic Growth Regressions: Instrumental variables ECI (trade) First Stage results.

|  | \multicolumn{12}{c}{*Dependent variable:*} |
|---|---|---|---|---|---|---|---|---|---|---|---|---|
|  | \multicolumn{12}{c}{ECI (trade) (1999-09, 2009-19)} |
|  | (1) | (2) | (3) | (4) | (5) | (6) | (7) | (8) | (9) | (10) | (11) | (12) |
| ECI (trade), similarity | 0.769*** | 0.724*** | 0.666*** | 0.757*** | 0.612*** | 0.639*** | 0.802*** | 0.734*** | 0.566*** | 0.464*** | 0.569*** | 0.468*** |
|  | (0.110) | (0.111) | (0.107) | (0.109) | (0.106) | (0.105) | (0.107) | (0.095) | (0.135) | (0.153) | (0.136) | (0.154) |
| ECI (technology), similarity | 0.066 | 0.053 | 0.003 | 0.072 | 0.010 | 0.005 | 0.141 | 0.110 | -0.047 | -0.103 | -0.043 | -0.098 |
|  | (0.117) | (0.119) | (0.108) | (0.116) | (0.106) | (0.106) | (0.116) | (0.097) | (0.135) | (0.134) | (0.137) | (0.135) |
| ECI (trade), similarity x ECI (technology), similarity | 0.005 | 0.081 | 0.104 | -0.008 | 0.144 | 0.097 | -0.188 | -0.139 | 0.177 | 0.255 | 0.170 | 0.249 |
|  | (0.173) | (0.177) | (0.160) | (0.170) | (0.158) | (0.155) | (0.167) | (0.135) | (0.186) | (0.190) | (0.187) | (0.190) |
| Log of initial population |  | -0.010** |  |  | -0.010** |  |  |  |  |  |  |  |
|  |  | (0.004) |  |  | (0.005) |  |  |  |  |  |  |  |
| Log of initial human capital |  |  | 0.084*** |  | 0.073** | 0.093*** |  | 0.056** |  | 0.090*** |  | 0.091*** |
|  |  |  | (0.025) |  | (0.029) | (0.026) |  | (0.025) |  | (0.025) |  | (0.025) |
| Log of natural resource exports per capita |  |  |  | -0.013 | -0.023*** | -0.017** |  | -0.027*** |  | -0.013 |  | -0.013 |
|  |  |  |  | (0.009) | (0.008) | (0.009) |  | (0.010) |  | (0.009) |  | (0.009) |
| Intensity (trade) |  |  |  |  |  |  | 0.025 | 0.143 |  |  |  |  |
|  |  |  |  |  |  |  | (0.079) | (0.096) |  |  |  |  |
| Intensity (technology) |  |  |  |  |  |  | 0.173*** | 0.123** |  |  |  |  |
|  |  |  |  |  |  |  | (0.057) | (0.054) |  |  |  |  |
| HHI (trade) |  |  |  |  |  |  |  |  | -0.121** | -0.105** |  |  |
|  |  |  |  |  |  |  |  |  | (0.049) | (0.052) |  |  |
| HHI (technology) |  |  |  |  |  |  |  |  | 0.005 | -0.003 |  |  |
|  |  |  |  |  |  |  |  |  | (0.032) | (0.028) |  |  |
| Entropy (trade) |  |  |  |  |  |  |  |  |  |  | 0.112** | 0.096** |
|  |  |  |  |  |  |  |  |  |  |  | (0.046) | (0.048) |
| Entropy (technology) |  |  |  |  |  |  |  |  |  |  | -0.008 | -0.0001 |
|  |  |  |  |  |  |  |  |  |  |  | (0.033) | (0.029) |
| Log of initial GDP per capita | 0.036*** | 0.023** | 0.023*** | 0.056*** | 0.048*** | 0.049*** | 0.004 | 0.023 | 0.043*** | 0.049*** | 0.043*** | 0.049*** |
|  | (0.008) | (0.010) | (0.008) | (0.016) | (0.014) | (0.015) | (0.017) | (0.017) | (0.008) | (0.015) | (0.008) | (0.015) |
| Observations | 152 | 152 | 152 | 152 | 152 | 152 | 152 | 152 | 152 | 152 | 152 | 152 |
| $R^2$ | 0.863 | 0.869 | 0.873 | 0.866 | 0.882 | 0.877 | 0.878 | 0.888 | 0.870 | 0.882 | 0.870 | 0.882 |
| Adjusted $R^2$ | 0.859 | 0.863 | 0.868 | 0.860 | 0.875 | 0.871 | 0.872 | 0.881 | 0.864 | 0.875 | 0.864 | 0.875 |
| F Statistic for instrument significance | 346.50*** | 102.50*** | 115.70*** | 102.10*** | 73.00*** | 119.90*** | 87.00*** | 86.50*** | 123.50*** | 102.30*** | 122.10*** | 99.30*** |

Notes: Each regression includes period fixed effects. Clustered standard errors in brackets. *p<0.1, **p<0.05, ***p<0.01.

## Supplementary Table 38. Economic Growth Regressions: Instrumental variables ECI (technology) First Stage results.

| | Dependent variable: | | | | | | | | | | | |
|---|---|---|---|---|---|---|---|---|---|---|---|---|
| | ECI (technology) (1999-09, 2009-19) | | | | | | | | | | | |
| | (1) | (2) | (3) | (4) | (5) | (6) | (7) | (8) | (9) | (10) | (11) | (12) |
| ECI (technology), similarity | 0.966*** | 0.957*** | 0.880*** | 0.962*** | 0.880*** | 0.880*** | 1.037*** | 0.939*** | 1.039*** | 0.945*** | 1.031*** | 0.937*** |
| | (0.191) | (0.191) | (0.194) | (0.189) | (0.194) | (0.193) | (0.212) | (0.220) | (0.223) | (0.213) | (0.228) | (0.217) |
| ECI (trade), similarity | 0.127 | 0.092 | -0.015 | 0.134 | -0.013 | -0.011 | 0.158 | 0.033 | 0.259 | 0.113 | 0.250 | 0.102 |
| | (0.262) | (0.263) | (0.265) | (0.258) | (0.262) | (0.264) | (0.277) | (0.290) | (0.278) | (0.270) | (0.283) | (0.276) |
| ECI (trade), similarity x ECI (technology), similarity | -0.036 | 0.022 | 0.100 | -0.028 | 0.104 | 0.101 | -0.218 | -0.032 | -0.147 | -0.006 | -0.136 | 0.006 |
| | (0.308) | (0.308) | (0.311) | (0.305) | (0.305) | (0.310) | (0.357) | (0.374) | (0.325) | (0.314) | (0.330) | (0.320) |
| Log of initial population | | -0.007 | | | -0.001 | | | | | | | |
| | | (0.005) | | | (0.006) | | | | | | | |
| Log of initial human capital | | | 0.115*** | | 0.112*** | 0.114*** | | 0.090** | | 0.116*** | | 0.116*** |
| | | | (0.038) | | (0.039) | (0.040) | | (0.041) | | (0.040) | | (0.039) |
| Log of natural resource exports per capita | | | | 0.008 | 0.002 | 0.003 | | 0.002 | | -0.001 | | -0.001 |
| | | | | (0.012) | (0.013) | (0.012) | | (0.014) | | (0.012) | | (0.012) |
| Intensity (trade) | | | | | | | 0.026 | 0.004 | | | | |
| | | | | | | | (0.092) | (0.100) | | | | |
| Intensity (technology) | | | | | | | 0.162** | 0.096 | | | | |
| | | | | | | | (0.074) | (0.074) | | | | |
| HHI (trade) | | | | | | | | | 0.080 | 0.081 | | |
| | | | | | | | | | (0.084) | (0.080) | | |
| HHI (technology) | | | | | | | | | -0.005 | -0.012 | | |
| | | | | | | | | | (0.074) | (0.069) | | |
| Entropy (trade) | | | | | | | | | | | -0.069 | -0.069 |
| | | | | | | | | | | | (0.076) | (0.072) |
| Entropy (technology) | | | | | | | | | | | 0.007 | 0.013 |
| | | | | | | | | | | | (0.075) | (0.070) |
| Log of initial GDP per capita | 0.007 | -0.002 | -0.010 | -0.006 | -0.014 | -0.014 | -0.023 | -0.024 | 0.002 | -0.014 | 0.002 | -0.014 |
| | (0.013) | (0.015) | (0.014) | (0.021) | (0.020) | (0.020) | (0.023) | (0.024) | (0.014) | (0.020) | (0.015) | (0.020) |
| Observations | 152 | 152 | 152 | 152 | 152 | 152 | 152 | 152 | 152 | 152 | 152 | 152 |
| $R^2$ | 0.904 | 0.905 | 0.910 | 0.904 | 0.910 | 0.910 | 0.908 | 0.911 | 0.905 | 0.911 | 0.905 | 0.911 |
| Adjusted $R^2$ | 0.901 | 0.901 | 0.906 | 0.900 | 0.905 | 0.906 | 0.904 | 0.905 | 0.900 | 0.905 | 0.900 | 0.905 |
| F Statistic for instrument significance | 526.20*** | 122.20*** | 119.30*** | 122.10*** | 124.20*** | 123.10*** | 138.50*** | 133.70*** | 82.40*** | 93.00*** | 79.40*** | 89.70*** |

Notes: Each regression includes period fixed effects. Clustered standard errors in brackets. *p<0.1, **p<0.05, ***p<0.01.

## Supplementary Table 39. Economic Growth Regressions: Instrumental Variables Robustness Check Second Stage Results

| | *Dependent variable:* | | | | | | | | | | | |
|---|---|---|---|---|---|---|---|---|---|---|---|---|
| | Annualized GDP pc growth (1999-09, 2009-19) | | | | | | | | | | | |
| | (1) | (2) | (3) | (4) | (5) | (6) | (7) | (8) | (9) | (10) | (11) | (12) |
| ECI (trade), instrumented | 14.296*** | 13.032*** | 10.386*** | 15.421*** | 12.370*** | 12.200*** | 15.903*** | 12.230*** | 13.836*** | 9.232* | 13.981*** | 9.075* |
| | (3.790) | (3.684) | (3.632) | (3.354) | (3.690) | (3.527) | (3.442) | (3.459) | (5.345) | (5.156) | (5.383) | (5.200) |
| ECI (technology), instrumented | 9.652*** | 9.327*** | 7.631*** | 8.935*** | 7.405*** | 7.451*** | 10.606*** | 7.048** | 8.856*** | 5.359* | 8.920*** | 5.279* |
| | (2.773) | (2.687) | (2.683) | (2.594) | (2.540) | (2.533) | (2.705) | (2.774) | (3.196) | (3.076) | (3.228) | (3.114) |
| ECI (trade), instrumented x ECI (technology), instrumented | -14.020*** | -12.170*** | -10.649*** | -12.904*** | -10.683*** | -10.417*** | -16.574*** | -9.510** | -13.292*** | -7.725* | -13.397*** | -7.624* |
| | (4.316) | (4.186) | (4.060) | (3.780) | (3.789) | (3.676) | (4.321) | (4.340) | (5.000) | (4.490) | (5.004) | (4.511) |
| Log of initial population | | -0.219** | | | 0.057 | | | | | | | |
| | | (0.106) | | | (0.123) | | | | | | | |
| Log of initial human capital | | | 2.658*** | | 2.219*** | 2.083*** | | 2.245*** | | 2.273*** | | 2.283*** |
| | | | (0.750) | | (0.827) | (0.802) | | (0.841) | | (0.822) | | (0.824) |
| Log of natural resource exports per capita | | | | 0.907*** | 0.815*** | 0.781*** | | 0.732** | | 0.784*** | | 0.789*** |
| | | | | (0.248) | (0.271) | (0.244) | | (0.310) | | (0.230) | | (0.229) |
| Intensity (trade) | | | | | | | 5.216*** | 1.056 | | | | |
| | | | | | | | (1.623) | (2.209) | | | | |
| Intensity (technology) | | | | | | | 0.873 | -1.131 | | | | |
| | | | | | | | (1.724) | (1.698) | | | | |
| HHI (trade) | | | | | | | | | 0.018 | -1.064 | | |
| | | | | | | | | | (1.393) | (1.300) | | |
| HHI (technology) | | | | | | | | | -0.538 | -0.656 | | |
| | | | | | | | | | (1.054) | (0.972) | | |
| Entropy (trade) | | | | | | | | | | | -0.082 | 1.056 |
| | | | | | | | | | | | (1.341) | (1.240) |
| Entropy (technology) | | | | | | | | | | | 0.522 | 0.642 |
| | | | | | | | | | | | (1.072) | (0.983) |
| Log of initial GDP per capita | -1.892*** | -2.172*** | -2.210*** | -3.385*** | -3.426*** | -3.426*** | -3.099*** | -3.434*** | -1.904*** | -3.357*** | -1.911*** | -3.355*** |
| | (0.252) | (0.253) | (0.257) | (0.503) | (0.458) | (0.460) | (0.402) | (0.450) | (0.309) | (0.487) | (0.314) | (0.487) |
| Observations | 152 | 152 | 152 | 152 | 152 | 152 | 152 | 152 | 152 | 152 | 152 | 152 |
| $R^2$ | 0.423 | 0.447 | 0.485 | 0.497 | 0.534 | 0.534 | 0.459 | 0.535 | 0.422 | 0.539 | 0.422 | 0.539 |
| Adjusted $R^2$ | 0.404 | 0.424 | 0.464 | 0.476 | 0.508 | 0.511 | 0.433 | 0.506 | 0.394 | 0.510 | 0.394 | 0.510 |

Notes: Each regression includes period fixed effects. Clustered standard errors in brackets. *p<0.1, **p<0.05, ***p<0.01.

**Income inequality IV estimation First-Stage results:** The first-stage results for the IV estimation of the income inequality models are given in Tables S40-S41. Supplementary Table 40 presents the results for ECI (trade), whereas Supplementary Table 41 gives the results for ECI (technology). In each case, the F-statistic for the significance of the included instruments is above 10, suggesting that our instruments are strong.

**Income inequality IV estimation Second-Stage results:** The coefficient of ECI (technology) of the multidimensional ECI income inequality model loses significance when we exclude the controls of the final model, and we use the IV estimation approach (Supplementary Table 42, columns (1-4,7,9,11)). But all of the multidimensional ECI coefficients regain significance after including additional explanatory variables (Supplementary Table 42, columns (5-6)), and final production intensity (Supplementary Table 42, column (8)), HHI (Supplementary Table 42, column (10)), and Entropy models (Supplementary Table 42, column (12)). This signifies the robustness of the multidimensional ECI income inequality model even when using the IV approach.

# Supplementary Table 40. Income Inequality Regressions: Instrumental variables ECI (trade) First Stage results.

| | *Dependent variable:* | | | | | | | | | | | |
|---|---|---|---|---|---|---|---|---|---|---|---|---|
| | ECI (trade) (1996-99, 2000-03, 2004-07, 2008-11, 2012-15) | | | | | | | | | | | |
| | (1) | (2) | (3) | (4) | (5) | (6) | (7) | (8) | (9) | (10) | (11) | (12) |
| ECI (trade), similarity | 0.530*** | 0.532*** | 0.510*** | 0.516*** | 0.486*** | 0.515*** | 0.491*** | 0.481*** | 0.508*** | 0.497*** | 0.507*** | 0.496*** |
| | (0.056) | (0.057) | (0.061) | (0.057) | (0.063) | (0.061) | (0.064) | (0.071) | (0.068) | (0.074) | (0.069) | (0.075) |
| ECI (technology), similarity | 0.057* | 0.079** | 0.060* | 0.056* | 0.086** | 0.074** | 0.028 | 0.014 | 0.058* | 0.077* | 0.059* | 0.078* |
| | (0.030) | (0.037) | (0.031) | (0.029) | (0.035) | (0.036) | (0.028) | (0.041) | (0.035) | (0.039) | (0.035) | (0.040) |
| Log of population | | -0.007 | | | -0.009 | -0.004 | | 0.003 | | -0.003 | | -0.003 |
| | | (0.005) | | | (0.005) | (0.005) | | (0.007) | | (0.006) | | (0.006) |
| Log of human capital | | | 0.058 | | 0.055 | 0.046 | | 0.007 | | 0.050 | | 0.050 |
| | | | (0.035) | | (0.040) | (0.039) | | (0.041) | | (0.041) | | (0.041) |
| Log of natural resource exports per capita | | | | -0.012 | -0.022* | | | | | | | |
| | | | | (0.011) | (0.012) | | | | | | | |
| Intensity (trade) | | | | | | | 0.091 | 0.104 | | | | |
| | | | | | | | (0.073) | (0.072) | | | | |
| Intensity (technology) | | | | | | | 0.164*** | 0.173** | | | | |
| | | | | | | | (0.060) | (0.074) | | | | |
| Intensity (research) | | | | | | | -0.069 | -0.057 | | | | |
| | | | | | | | (0.069) | (0.073) | | | | |
| HHI (trade) | | | | | | | | | -0.055 | -0.038 | | |
| | | | | | | | | | (0.042) | (0.044) | | |
| HHI (technology) | | | | | | | | | 0.013 | 0.021 | | |
| | | | | | | | | | (0.033) | (0.033) | | |
| Entropy (trade) | | | | | | | | | | | 0.049 | 0.034 |
| | | | | | | | | | | | (0.040) | (0.043) |
| Entropy (technology) | | | | | | | | | | | -0.014 | -0.022 |
| | | | | | | | | | | | (0.034) | (0.033) |
| Log of GDP per capita | 0.096 | 0.051 | 0.010 | 0.102 | -0.031 | -0.003 | 0.148 | 0.162 | 0.097 | 0.010 | 0.096 | 0.009 |
| | (0.110) | (0.119) | (0.132) | (0.110) | (0.130) | (0.133) | (0.104) | (0.135) | (0.114) | (0.133) | (0.115) | (0.134) |
| Log of GDP per capita, squared | -0.003 | -0.001 | 0.001 | -0.002 | 0.005 | 0.002 | -0.007 | -0.008 | -0.003 | 0.001 | -0.002 | 0.001 |
| | (0.006) | (0.006) | (0.007) | (0.006) | (0.007) | (0.007) | (0.005) | (0.007) | (0.006) | (0.007) | (0.006) | (0.007) |
| Observations | 332 | 332 | 332 | 332 | 332 | 332 | 332 | 332 | 332 | 332 | 332 | 332 |
| $R^2$ | 0.860 | 0.862 | 0.863 | 0.861 | 0.869 | 0.864 | 0.873 | 0.873 | 0.861 | 0.865 | 0.861 | 0.865 |
| Adjusted $R^2$ | 0.856 | 0.858 | 0.859 | 0.858 | 0.865 | 0.860 | 0.868 | 0.868 | 0.857 | 0.859 | 0.857 | 0.859 |
| F Statistic | 324.60*** | 318.40*** | 293.00*** | 285.00*** | 183.60*** | 203.20*** | 96.60*** | 49.10*** | 133.20*** | 92.60*** | 125.70*** | 89.50*** |

Notes: Each regression includes period fixed effects. Clustered standard errors in brackets. *p<0.1, **p<0.05, ***p<0.01.

## Supplementary Table 41. Income Inequality Regressions: Instrumental variables ECI (technology) First Stage results.

|  | \multicolumn{12}{c}{Dependent variable:} |
| --- | --- | --- | --- | --- | --- | --- | --- | --- | --- | --- | --- | --- |
|  | ECI (technology) (1996-99, 2000-03, 2004-07, 2008-11, 2012-15) | | | | | | | | | | | |
|  | (1) | (2) | (3) | (4) | (5) | (6) | (7) | (8) | (9) | (10) | (11) | (12) |
| ECI (technology), similarity | 0.755*** | 0.793*** | 0.758*** | 0.757*** | 0.785*** | 0.788*** | 0.733*** | 0.793*** | 0.686*** | 0.723*** | 0.685*** | 0.721*** |
|  | (0.028) | (0.030) | (0.027) | (0.027) | (0.031) | (0.030) | (0.028) | (0.040) | (0.025) | (0.028) | (0.025) | (0.028) |
| ECI (trade), similarity | 0.031 | 0.033 | 0.007 | 0.048 | 0.024 | 0.018 | 0.009 | 0.038 | 0.007 | 0.010 | 0.005 | 0.009 |
|  | (0.034) | (0.029) | (0.032) | (0.031) | (0.029) | (0.028) | (0.040) | (0.035) | (0.031) | (0.033) | (0.032) | (0.033) |
| Log of population |  | -0.011*** |  |  | -0.008* | -0.009** |  | -0.013** |  | -0.008** |  | -0.008** |
|  |  | (0.004) |  |  | (0.004) | (0.004) |  | (0.006) |  | (0.003) |  | (0.004) |
| Log of human capital |  |  | 0.067** |  | 0.041 | 0.043 |  | 0.016 |  | 0.036 |  | 0.037 |
|  |  |  | (0.029) |  | (0.033) | (0.033) |  | (0.036) |  | (0.024) |  | (0.024) |
| Log of natural resource exports per capita |  |  |  | 0.014* | 0.005 |  |  |  |  |  |  |  |
|  |  |  |  | (0.008) | (0.009) |  |  |  |  |  |  |  |
| Intensity (trade) |  |  |  |  |  |  | -0.016 | -0.072 |  |  |  |  |
|  |  |  |  |  |  |  | (0.061) | (0.062) |  |  |  |  |
| Intensity (technology) |  |  |  |  |  |  | 0.180*** | 0.112* |  |  |  |  |
|  |  |  |  |  |  |  | (0.056) | (0.062) |  |  |  |  |
| Intensity (research) |  |  |  |  |  |  | -0.158** | -0.190*** |  |  |  |  |
|  |  |  |  |  |  |  | (0.064) | (0.061) |  |  |  |  |
| HHI (trade) |  |  |  |  |  |  |  |  | -0.075** | -0.039 |  |  |
|  |  |  |  |  |  |  |  |  | (0.035) | (0.037) |  |  |
| HHI (technology) |  |  |  |  |  |  |  |  | -0.121*** | -0.113*** |  |  |
|  |  |  |  |  |  |  |  |  | (0.032) | (0.032) |  |  |
| Entropy (trade) |  |  |  |  |  |  |  |  |  |  | 0.069** | 0.036 |
|  |  |  |  |  |  |  |  |  |  |  | (0.032) | (0.034) |
| Entropy (technology) |  |  |  |  |  |  |  |  |  |  | 0.119*** | 0.112*** |
|  |  |  |  |  |  |  |  |  |  |  | (0.033) | (0.032) |
| Log of GDP per capita | 0.194 | 0.118 | 0.094 | 0.187 | 0.075 | 0.068 | 0.260* | 0.136 | 0.106 | 0.007 | 0.099 | 0.0004 |
|  | (0.162) | (0.168) | (0.184) | (0.164) | (0.183) | (0.184) | (0.158) | (0.191) | (0.089) | (0.093) | (0.090) | (0.094) |
| Log of GDP per capita, squared | -0.010 | -0.007 | -0.006 | -0.011 | -0.005 | -0.005 | -0.014* | -0.007 | -0.005 | -0.001 | -0.005 | -0.001 |
|  | (0.008) | (0.008) | (0.009) | (0.008) | (0.009) | (0.009) | (0.008) | (0.010) | (0.005) | (0.005) | (0.005) | (0.005) |
| Observations | 332 | 332 | 332 | 332 | 332 | 332 | 332 | 332 | 332 | 332 | 332 | 332 |
| $R^2$ | 0.929 | 0.932 | 0.931 | 0.930 | 0.933 | 0.932 | 0.933 | 0.936 | 0.933 | 0.935 | 0.933 | 0.935 |
| Adjusted $R^2$ | 0.927 | 0.930 | 0.929 | 0.928 | 0.930 | 0.930 | 0.931 | 0.933 | 0.931 | 0.933 | 0.930 | 0.933 |
| F Statistic | 880.00*** | 713.00*** | 949.00*** | 896.80*** | 775.20*** | 769.70*** | 724.80*** | 397.80*** | 427.80*** | 382.10*** | 402.80*** | 365.00*** |

Notes: Each regression includes period fixed effects. Clustered standard errors in brackets. *p<0.1, **p<0.05, ***p<0.01.

## Supplementary Table 42. Income Inequality Regressions: Instrumental Variables Robustness Check Second Stage Results

|  | *Dependent variable:* | | | | | | | | | | | |
|---|---|---|---|---|---|---|---|---|---|---|---|---|
|  | Gini coefficient (1996-99, 2000-03, 2004-07, 2008-11, 2012-15) | | | | | | | | | | | |
|  | (1) | (2) | (3) | (4) | (5) | (6) | (7) | (8) | (9) | (10) | (11) | (12) |
| ECI (trade), instrumented | -21.690*** | -21.830*** | -15.665*** | -25.930*** | -19.530*** | -18.456*** | -14.492** | -17.798*** | -15.907** | -16.940*** | -15.143** | -16.294*** |
|  | (6.483) | (4.924) | (5.642) | (5.555) | (4.466) | (4.533) | (6.027) | (5.112) | (7.215) | (5.971) | (7.270) | (6.145) |
| ECI (technology), instrumented | -2.909 | -9.569*** | -3.906 | -2.826 | -8.791*** | -9.176*** | -0.521 | -8.341*** | -0.157 | -8.100** | -0.076 | -7.865** |
|  | (3.056) | (2.839) | (2.502) | (2.880) | (2.735) | (2.722) | (2.785) | (3.083) | (4.285) | (3.863) | (4.353) | (3.911) |
| Log of population |  | 1.535*** |  |  | 1.237*** | 1.317*** |  | 1.059*** |  | 1.265*** |  | 1.248*** |
|  |  | (0.284) |  |  | (0.311) | (0.288) |  | (0.341) |  | (0.281) |  | (0.279) |
| Log of human capital |  |  | -8.891*** |  | -4.754** | -4.951*** |  | -5.379*** |  | -5.094** |  | -5.176*** |
|  |  |  | (2.399) |  | (2.014) | (1.922) |  | (1.908) |  | (1.994) |  | (1.990) |
| Log of natural resource exports |  |  |  | -1.866** | -0.418 |  |  |  |  |  |  |  |
|  |  |  |  | (0.770) | (0.676) |  |  |  |  |  |  |  |
| Intensity (trade) |  |  |  |  |  |  | -15.628*** | -9.851** |  |  |  |  |
|  |  |  |  |  |  |  | (4.989) | (4.471) |  |  |  |  |
| Intensity (technology) |  |  |  |  |  |  | -7.603 | 3.200 |  |  |  |  |
|  |  |  |  |  |  |  | (5.160) | (4.852) |  |  |  |  |
| Intensity (research) |  |  |  |  |  |  | -4.683 | -4.796 |  |  |  |  |
|  |  |  |  |  |  |  | (3.909) | (3.830) |  |  |  |  |
| HHI (trade) |  |  |  |  |  |  |  |  | 8.177* | 2.048 |  |  |
|  |  |  |  |  |  |  |  |  | (4.289) | (4.330) |  |  |
| HHI (technology) |  |  |  |  |  |  |  |  | 3.146 | 1.049 |  |  |
|  |  |  |  |  |  |  |  |  | (3.347) | (3.120) |  |  |
| Entropy (trade) |  |  |  |  |  |  |  |  |  |  | -8.089** | -2.513 |
|  |  |  |  |  |  |  |  |  |  |  | (3.978) | (4.112) |
| Entropy (technology) |  |  |  |  |  |  |  |  |  |  | -3.113 | -1.193 |
|  |  |  |  |  |  |  |  |  |  |  | (3.386) | (3.181) |
| Log of GDP per capita | -6.608 | 4.915 | 6.223 | -5.216 | 9.852 | 10.419 | -9.390 | 10.629 | -4.588 | 10.882 | -4.103 | 11.140 |
|  | (10.606) | (8.416) | (9.498) | (10.034) | (8.033) | (8.163) | (8.815) | (8.692) | (10.680) | (8.538) | (10.695) | (8.653) |
| Log of GDP per capita, squared | 0.297 | -0.186 | -0.312 | 0.398 | -0.395 | -0.456 | 0.707 | -0.362 | 0.145 | -0.494 | 0.115 | -0.513 |
|  | (0.555) | (0.437) | (0.490) | (0.519) | (0.415) | (0.419) | (0.459) | (0.457) | (0.563) | (0.447) | (0.563) | (0.454) |
| Observations | 332 | 332 | 332 | 332 | 332 | 332 | 332 | 332 | 332 | 332 | 332 | 332 |
| $R^2$ | 0.568 | 0.668 | 0.628 | 0.596 | 0.688 | 0.688 | 0.649 | 0.704 | 0.585 | 0.691 | 0.587 | 0.693 |
| Adjusted $R^2$ | 0.557 | 0.659 | 0.618 | 0.585 | 0.677 | 0.678 | 0.637 | 0.692 | 0.572 | 0.680 | 0.574 | 0.681 |

Notes: Each regression includes period fixed effects. Clustered standard errors in brackets. *p<0.1, **p<0.05, ***p<0.01.

**Emission intensity IV estimation First-Stage results:** The first-stage results for the IV estimation of the emission intensity models are given in Supplementary Tables 43-45. Supplementary Table 43 presents the results for ECI (trade), Supplementary Table 44 gives the results for ECI (technology), and Supplementary Table 45 shows the results for ECI (research). In each case, the F-statistic for the significance of the included instruments is above 10, suggesting that our instruments are strong.

**Emission intensity IV estimation Second-Stage results:** The main variables of the multidimensional ECI emission intensity model remain significant (the three-way term), even when using the IV estimation approach (Supplementary Table 46, column (1)). Moreover, the model is robust even after accounting for country-fixed effects (Supplementary Table 46, column (2)), the multidimensional HHI model (Supplementary Table 46, columns (5-6)), and the multidimensional Entropy model (Supplementary Table 46, columns (7-8)). The multidimensional ECI model loses significance when we add the variables included in the multidimensional production intensity model (Supplementary Table 46, column (3)), but regain significance when we also add the country fixed effects specification (Supplementary Table 46, column (4)). This is the more restrictive case, and therefore we use it as a sign of the robustness of the multidimensional ECI emission intensity model even when using the IV approach.

## Supplementary Table 43. Emission Intensity Regressions: Instrumental variables ECI (trade) First Stage results.

|  | *Dependent variable:* | | | | | | | |
|---|---|---|---|---|---|---|---|---|
|  | ECI (trade) (1996-99, 2000-03, 2004-07, 2008-11, 2012-15, 2016-19) | | | | | | | |
|  | (1) | (2) | (3) | (4) | (5) | (6) | (7) | (8) |
| ECI (trade), similarity | 0.898*** | 0.299* | 0.900*** | 0.378** | 0.784*** | 0.314* | 0.849*** | 0.301* |
|  | (0.268) | (0.172) | (0.294) | (0.181) | (0.280) | (0.168) | (0.274) | (0.169) |
| ECI (technology), similarity | 0.204 | -0.252 | 0.396 | -0.141 | 0.181 | -0.199 | 0.288 | -0.216 |
|  | (0.293) | (0.154) | (0.262) | (0.158) | (0.296) | (0.152) | (0.302) | (0.169) |
| ECI (research), similarity | 0.307 | -0.491*** | 0.341 | -0.383** | 0.295 | -0.437** | 0.375 | -0.458*** |
|  | (0.289) | (0.170) | (0.326) | (0.164) | (0.293) | (0.175) | (0.288) | (0.174) |
| ECI (trade), similarity x ECI (technology), similarity | -0.109 | 0.388 | -0.360 | 0.173 | -0.033 | 0.317 | -0.139 | 0.330 |
|  | (0.425) | (0.247) | (0.419) | (0.269) | (0.431) | (0.244) | (0.427) | (0.248) |
| ECI (trade), similarity x ECI (research), similarity | -0.628 | 0.817*** | -0.353 | 0.721** | -0.553 | 0.741** | -0.660 | 0.767** |
|  | (0.510) | (0.300) | (0.575) | (0.293) | (0.515) | (0.308) | (0.498) | (0.305) |
| ECI (technology), similarity x ECI (research), similarity | -0.383 | 0.810*** | -0.521 | 0.588** | -0.415 | 0.716*** | -0.533 | 0.743*** |
|  | (0.539) | (0.240) | (0.503) | (0.246) | (0.545) | (0.239) | (0.544) | (0.243) |
| ECI (trade), similarity x ECI (technology), similarity x ECI (research), similarity | 0.608 | -1.253*** | 0.304 | -1.046*** | 0.566 | -1.125*** | 0.712 | -1.156*** |
|  | (0.762) | (0.380) | (0.778) | (0.395) | (0.768) | (0.382) | (0.754) | (0.379) |
| Intensity (trade) |  |  | 0.422** | -0.101 |  |  |  |  |
|  |  |  | (0.174) | (0.193) |  |  |  |  |
| Intensity (technology) |  |  | 0.235 | 0.185 |  |  |  |  |
|  |  |  | (0.147) | (0.119) |  |  |  |  |
| Intensity (research) |  |  | 0.262 | 0.165 |  |  |  |  |
|  |  |  | (0.294) | (0.197) |  |  |  |  |
| Intensity (trade) x Intensity (technology) |  |  | 0.199 | 0.433 |  |  |  |  |
|  |  |  | (0.472) | (0.357) |  |  |  |  |
| Intensity (trade) x Intensity (research) |  |  | -0.572* | -0.194 |  |  |  |  |
|  |  |  | (0.300) | (0.242) |  |  |  |  |
| Intensity (technology) x Intensity (research) |  |  | -0.328 | -0.366 |  |  |  |  |
|  |  |  | (0.390) | (0.317) |  |  |  |  |
| Intensity (trade) x Intensity (technology) x Intensity (research) |  |  | 0.387 | 0.166 |  |  |  |  |
|  |  |  | (0.524) | (0.465) |  |  |  |  |
| HHI (trade) |  |  |  |  | -0.059 | -0.115* |  |  |
|  |  |  |  |  | (0.045) | (0.059) |  |  |
| Entropy (trade) |  |  |  |  |  |  | 0.062 | 0.112* |
|  |  |  |  |  |  |  | (0.043) | (0.057) |
| Entropy (technology) |  |  |  |  |  |  | -0.045 | 0.005 |
|  |  |  |  |  |  |  | (0.028) | (0.024) |
| Log of population | -0.008 | 0.056** | 0.005 | 0.050** | -0.006 | 0.048* | -0.005 | 0.047* |
|  | (0.005) | (0.027) | (0.005) | (0.022) | (0.005) | (0.026) | (0.005) | (0.026) |
| Log of human capital | 0.053** | 0.031 | 0.030 | 0.043 | 0.056** | 0.024 | 0.059** | 0.025 |
|  | (0.025) | (0.059) | (0.024) | (0.057) | (0.026) | (0.058) | (0.026) | (0.058) |
| Log of natural resource exports per capita | -0.022*** | -0.005 | -0.038*** | -0.004 | -0.018** | 0.002 | -0.018** | 0.002 |
|  | (0.009) | (0.008) | (0.012) | (0.014) | (0.009) | (0.009) | (0.009) | (0.009) |
| Log of GDP per capita | 0.067*** | 0.076*** | 0.029** | 0.060*** | 0.067*** | 0.076*** | 0.068*** | 0.075*** |
|  | (0.015) | (0.023) | (0.014) | (0.016) | (0.016) | (0.022) | (0.016) | (0.022) |
| Fixed effects | No | Yes | No | Yes | No | Yes | No | Yes |
| Observations | 528 | 528 | 528 | 528 | 528 | 528 | 528 | 528 |
| $R^2$ | 0.889 | 0.528 | 0.917 | 0.592 | 0.890 | 0.539 | 0.890 | 0.539 |
| Adjusted $R^2$ | 0.885 | 0.383 | 0.913 | 0.457 | 0.886 | 0.396 | 0.887 | 0.395 |
| F Statistic | 313.80*** | 359.20*** | 177.60*** | 212.00*** | 202.00*** | 290.20*** | 216.80*** | 36.117*** |

Notes: Each regression includes period fixed effects. Clustered standard errors in brackets. *p<0.1, **p<0.05, ***p<0.01.

# Supplementary Table 44. Emission Intensity Regressions: Instrumental variables ECI (technology) First Stage results.

| | *Dependent variable:* | | | | | | | |
|---|---|---|---|---|---|---|---|---|
| | ECI (technology) (1996-99, 2000-03, 2004-07, 2008-11, 2012-15, 2016-19) | | | | | | | |
| | (1) | (2) | (3) | (4) | (5) | (6) | (7) | (8) |
| ECI (trade), similarity | 0.796 | 0.473 | 0.901* | 0.578 | 0.823 | 0.481 | 0.575 | 0.451 |
| | (0.516) | (0.556) | (0.515) | (0.472) | (0.536) | (0.550) | (0.500) | (0.553) |
| ECI (technology), similarity | 1.385*** | 1.304*** | 1.451*** | 1.476*** | 1.390*** | 1.331*** | 1.041*** | 1.270*** |
| | (0.348) | (0.369) | (0.346) | (0.331) | (0.351) | (0.365) | (0.355) | (0.377) |
| ECI (research), similarity | 0.249 | -0.119 | 0.445 | 0.128 | 0.252 | -0.091 | -0.008 | -0.143 |
| | (0.523) | (0.497) | (0.523) | (0.462) | (0.522) | (0.502) | (0.478) | (0.485) |
| ECI (trade), similarity x ECI (technology), similarity | -1.118* | -0.923 | -1.138* | -1.219** | -1.136* | -0.960 | -0.769 | -0.915 |
| | (0.626) | (0.672) | (0.612) | (0.576) | (0.636) | (0.665) | (0.607) | (0.669) |
| ECI (trade), similarity x ECI (research), similarity | -1.019 | -0.078 | -1.232 | -0.328 | -1.037 | -0.118 | -0.671 | -0.067 |
| | (0.966) | (0.904) | (0.969) | (0.810) | (0.976) | (0.907) | (0.900) | (0.898) |
| ECI (technology), similarity x ECI (research), similarity | -0.374 | -0.113 | -0.546 | -0.505 | -0.366 | -0.161 | 0.008 | -0.083 |
| | (0.638) | (0.600) | (0.624) | (0.552) | (0.628) | (0.599) | (0.585) | (0.581) |
| ECI (trade), similarity x ECI (technology), similarity x ECI (research), similarity | 1.417 | 0.524 | 1.495 | 0.926 | 1.427 | 0.590 | 0.936 | 0.517 |
| | (1.127) | (1.050) | (1.105) | (0.923) | (1.130) | (1.047) | (1.049) | (1.034) |
| Intensity (trade) | | | 0.045 | -0.559 | | | | |
| | | | (0.274) | (0.405) | | | | |
| Intensity (technology) | | | -0.149 | -0.210 | | | | |
| | | | (0.190) | (0.256) | | | | |
| Intensity (research) | | | 0.236 | 0.894* | | | | |
| | | | (0.327) | (0.494) | | | | |
| Intensity (trade) x Intensity (technology) | | | -0.812 | -0.966 | | | | |
| | | | (0.548) | (0.801) | | | | |
| Intensity (trade) x Intensity (research) | | | -0.020 | 0.442 | | | | |
| | | | (0.382) | (0.485) | | | | |
| Intensity (technology) x Intensity (research) | | | -0.009 | -0.735 | | | | |
| | | | (0.351) | (0.637) | | | | |
| Intensity (trade) x Intensity (technology) x Intensity (research) | | | 0.634 | 1.017 | | | | |
| | | | (0.560) | (0.894) | | | | |
| HHI (trade) | | | | | 0.014 | -0.060 | | |
| | | | | | (0.054) | (0.109) | | |
| Entropy (trade) | | | | | | | -0.016 | 0.050 |
| | | | | | | | (0.050) | (0.107) |
| Entropy (technology) | | | | | | | 0.138*** | 0.023 |
| | | | | | | | (0.049) | (0.058) |
| Log of population | 0.002 | 0.032 | 0.002 | 0.006 | 0.001 | 0.028 | 0.001 | 0.026 |
| | (0.005) | (0.048) | (0.006) | (0.044) | (0.005) | (0.049) | (0.005) | (0.050) |
| Log of human capital | 0.059** | 0.122 | 0.052** | 0.118 | 0.058*** | 0.118 | 0.051** | 0.118 |
| | (0.023) | (0.096) | (0.023) | (0.104) | (0.022) | (0.098) | (0.022) | (0.099) |
| Log of natural resource exports per capita | 0.010 | 0.001 | 0.018** | 0.042 | 0.009 | 0.004 | 0.011 | 0.004 |
| | (0.007) | (0.017) | (0.009) | (0.028) | (0.008) | (0.017) | (0.008) | (0.017) |
| Log of GDP per capita | -0.027* | 0.004 | -0.014 | -0.002 | -0.027* | 0.004 | -0.029** | 0.003 |
| | (0.014) | (0.030) | (0.016) | (0.031) | (0.014) | (0.030) | (0.014) | (0.030) |
| Fixed effects | No | Yes | No | Yes | No | Yes | No | Yes |
| Observations | 528 | 528 | 528 | 528 | 528 | 528 | 528 | 528 |
| $R^2$ | 0.921 | 0.629 | 0.923 | 0.651 | 0.921 | 0.629 | 0.925 | 0.630 |
| Adjusted $R^2$ | 0.919 | 0.514 | 0.920 | 0.536 | 0.919 | 0.514 | 0.922 | 0.513 |
| F Statistic | 1108.60*** | 281.30*** | 479.40*** | 240.00*** | 600.60*** | 193.60*** | 526.40*** | 165.70*** |

Notes: Each regression includes period fixed effects. Clustered standard errors in brackets. *p<0.1, **p<0.05, ***p<0.01.

# Supplementary Table 45. Emission Intensity Regressions: Instrumental variables ECI (research) First Stage results.

| | *Dependent variable:* | | | | | | | |
|---|---|---|---|---|---|---|---|---|
| | ECI (research) (1996-99, 2000-03, 2004-07, 2008-11, 2012-15, 2016-19) | | | | | | | |
| | (1) | (2) | (3) | (4) | (5) | (6) | (7) | (8) |
| ECI (trade), similarity | -0.023 | 0.302 | 0.080 | 0.365* | -0.093 | 0.309 | -0.121 | 0.303 |
| | (0.210) | (0.226) | (0.224) | (0.220) | (0.219) | (0.222) | (0.220) | (0.224) |
| ECI (technology), similarity | -0.00004 | 0.109 | 0.097 | 0.175 | -0.014 | 0.136 | -0.060 | 0.127 |
| | (0.149) | (0.190) | (0.178) | (0.195) | (0.151) | (0.188) | (0.162) | (0.194) |
| ECI (research), similarity | 0.754*** | 1.006*** | 0.828*** | 1.017*** | 0.747*** | 1.034*** | 0.712*** | 1.022*** |
| | (0.221) | (0.218) | (0.218) | (0.223) | (0.223) | (0.218) | (0.223) | (0.222) |
| ECI (trade), similarity x ECI (technology), similarity | -0.006 | -0.238 | -0.216 | -0.380 | 0.040 | -0.274 | 0.086 | -0.265 |
| | (0.261) | (0.324) | (0.308) | (0.334) | (0.266) | (0.320) | (0.270) | (0.322) |
| ECI (trade), similarity x ECI (research), similarity | 0.232 | -0.551 | 0.061 | -0.589 | 0.277 | -0.589 | 0.323 | -0.574 |
| | (0.422) | (0.433) | (0.417) | (0.425) | (0.430) | (0.431) | (0.427) | (0.434) |
| ECI (technology), similarity x ECI (research), similarity | 0.149 | -0.055 | -0.019 | -0.136 | 0.130 | -0.103 | 0.183 | -0.087 |
| | (0.279) | (0.316) | (0.304) | (0.329) | (0.277) | (0.315) | (0.280) | (0.321) |
| ECI (trade), similarity x ECI (technology), similarity x ECI (research), similarity | -0.212 | 0.400 | 0.041 | 0.522 | -0.238 | 0.466 | -0.303 | 0.445 |
| | (0.485) | (0.549) | (0.514) | (0.549) | (0.491) | (0.546) | (0.490) | (0.551) |
| Intensity (trade) | | | -0.099 | -0.302 | | | | |
| | | | (0.142) | (0.187) | | | | |
| Intensity (technology) | | | -0.013 | -0.095 | | | | |
| | | | (0.109) | (0.130) | | | | |
| Intensity (research) | | | -0.133 | -0.202 | | | | |
| | | | (0.207) | (0.199) | | | | |
| Intensity (trade) x Intensity (technology) | | | 0.395 | 0.624 | | | | |
| | | | (0.378) | (0.410) | | | | |
| Intensity (trade) x Intensity (research) | | | 0.062 | 0.368 | | | | |
| | | | (0.234) | (0.306) | | | | |
| Intensity (technology) x Intensity (research) | | | 0.295 | 0.236 | | | | |
| | | | (0.279) | (0.295) | | | | |
| Intensity (trade) x Intensity (technology) x Intensity (research) | | | -0.476 | -0.673 | | | | |
| | | | (0.427) | (0.513) | | | | |
| HHI (trade) | | | | | -0.036 | -0.059 | | |
| | | | | | (0.025) | (0.038) | | |
| Entropy (trade) | | | | | | | 0.030 | 0.051 |
| | | | | | | | (0.024) | (0.038) |
| Entropy (technology) | | | | | | | 0.018 | 0.002 |
| | | | | | | | (0.022) | (0.019) |
| Log of population | 0.001 | 0.008 | 0.005 | 0.010 | 0.003 | 0.004 | 0.002 | 0.004 |
| | (0.003) | (0.031) | (0.004) | (0.028) | (0.004) | (0.031) | (0.004) | (0.031) |
| Log of human capital | 0.012 | 0.019 | -0.002 | 0.022 | 0.013 | 0.016 | 0.013 | 0.017 |
| | (0.016) | (0.050) | (0.015) | (0.050) | (0.015) | (0.050) | (0.015) | (0.050) |
| Log of natural resource exports per capita | -0.013** | 0.024*** | -0.008 | 0.025** | -0.011* | 0.027*** | -0.011* | 0.027*** |
| | (0.006) | (0.006) | (0.006) | (0.011) | (0.006) | (0.007) | (0.006) | (0.007) |
| Log of GDP per capita | 0.038*** | 0.006 | 0.028** | 0.002 | 0.038*** | 0.005 | 0.038*** | 0.005 |
| | (0.011) | (0.019) | (0.012) | (0.020) | (0.011) | (0.019) | (0.011) | (0.019) |
| Fixed effects | No | Yes | No | Yes | No | Yes | No | Yes |
| Observations | 528 | 528 | 528 | 528 | 528 | 528 | 528 | 528 |
| $R^2$ | 0.951 | 0.671 | 0.953 | 0.677 | 0.951 | 0.672 | 0.951 | 0.672 |
| Adjusted $R^2$ | 0.949 | 0.569 | 0.951 | 0.570 | 0.949 | 0.570 | 0.949 | 0.569 |
| F Statistic | 2041.20*** | 618.20*** | 1194.40*** | 636.40*** | 1059.90*** | 419.90*** | 487.30*** | 272.40*** |

Notes: Each regression includes period fixed effects. Clustered standard errors in brackets. *p<0.1, **p<0.05, ***p<0.01.

## Supplementary Table 46. Emission Intensity Regressions: Instrumental Variables Robustness Check

| | Dependent variable: | | | | | | | |
|---|---|---|---|---|---|---|---|---|
| | GHG emissions per GDP (1996-99, 2000-03, 2004-07, 2008-11, 2012-15, 2016-19) | | | | | | | |
| | (1) | (2) | (3) | (4) | (5) | (6) | (7) | (8) |
| ECI (trade), instrumented | -3.698 | -0.969 | -2.124 | -0.118 | -2.095 | -0.992 | -1.855 | -0.999 |
| | (2.573) | (0.862) | (2.700) | (0.948) | (2.962) | (0.886) | (2.964) | (0.890) |
| ECI (technology), instrumented | -3.211 | -1.493** | -1.483 | -1.219* | -3.241 | -1.450** | -3.187 | -1.452** |
| | (2.018) | (0.646) | (2.285) | (0.684) | (2.023) | (0.664) | (2.034) | (0.666) |
| ECI (research), instrumented | -5.159** | -2.042** | -4.355* | -1.890** | -5.308** | -1.998** | -5.299** | -2.011** |
| | (2.509) | (0.941) | (2.513) | (0.852) | (2.581) | (0.964) | (2.582) | (0.962) |
| ECI (trade), instrumented x ECI (technology), instrumented | 4.824 | 2.520** | 0.938 | 1.362 | 3.995 | 2.470** | 3.831 | 2.470** |
| | (3.318) | (1.133) | (3.813) | (1.333) | (3.412) | (1.159) | (3.415) | (1.164) |
| ECI (trade), instrumented x ECI (research), instrumented | 8.298* | 2.493 | 5.227 | 1.969 | 7.688* | 2.443 | 7.603* | 2.461 |
| | (4.378) | (1.528) | (4.126) | (1.478) | (4.600) | (1.562) | (4.588) | (1.563) |
| ECI (research), instrumented x ECI (technology), instrumented | 7.709** | 3.946*** | 4.722 | 3.634*** | 8.757** | 3.866*** | 8.806** | 3.881*** |
| | (3.781) | (1.389) | (3.891) | (1.197) | (3.882) | (1.417) | (3.916) | (1.412) |
| ECI (trade), instrumented x ECI (technology), instrumented x ECI (research), instrumented | -12.677** | -5.367*** | -6.250 | -4.453** | -12.770** | -5.273** | -12.700** | -5.293** |
| | (5.511) | (2.047) | (5.685) | (1.858) | (5.592) | (2.081) | (5.605) | (2.080) |
| Intensity (trade) | | | -5.105** | 0.976 | | | | |
| | | | (2.028) | (0.629) | | | | |
| Intensity (technology) | | | -5.073** | -0.897 | | | | |
| | | | (2.178) | (0.718) | | | | |
| Intensity (research) | | | -1.638 | 1.404*** | | | | |
| | | | (1.697) | (0.517) | | | | |
| Intensity (trade) x Intensity (technology) | | | 13.694*** | 3.097*** | | | | |
| | | | (3.605) | (1.174) | | | | |
| Intensity (trade) x Intensity (research) | | | 5.637* | -2.036* | | | | |
| | | | (3.299) | (1.078) | | | | |
| Intensity (research) x Intensity (technology) | | | 7.108** | 1.213 | | | | |
| | | | (2.799) | (1.176) | | | | |
| Intensity (trade) x Intensity (technology) x Intensity (research) | | | -16.516*** | -2.559 | | | | |
| | | | (4.239) | (1.779) | | | | |
| HHI (trade) | | | | | 0.712 | -0.109 | | |
| | | | | | (0.464) | (0.197) | | |
| Entropy (trade) | | | | | | | -0.739* | 0.111 |
| | | | | | | | (0.434) | (0.194) |
| Log of population | 0.113*** | 0.117 | 0.135*** | 0.127 | 0.090** | 0.113 | 0.087** | 0.113 |
| | (0.036) | (0.093) | (0.038) | (0.081) | (0.037) | (0.096) | (0.037) | (0.096) |
| Log of human capital | 0.799*** | 0.861*** | 0.692** | 0.608*** | 0.719** | 0.856*** | 0.700** | 0.857*** |
| | (0.285) | (0.218) | (0.291) | (0.187) | (0.291) | (0.217) | (0.291) | (0.218) |
| Log of natural resource exports per capital | 0.223*** | 0.041 | 0.320*** | 0.021 | 0.190*** | 0.047 | 0.185*** | 0.047 |
| | (0.077) | (0.037) | (0.107) | (0.042) | (0.073) | (0.038) | (0.072) | (0.039) |
| Log of initial GDP per capita | -0.468*** | -0.475*** | -0.581*** | -0.588*** | -0.529*** | -0.472*** | -0.537*** | -0.472*** |
| | (0.131) | (0.084) | (0.129) | (0.065) | (0.137) | (0.085) | (0.137) | (0.086) |
| Fixed effects | No | Yes | No | Yes | No | Yes | No | Yes |
| Observations | 528 | 528 | 528 | 528 | 528 | 528 | 528 | 528 |
| R² | 0.411 | 0.458 | 0.486 | 0.557 | 0.419 | 0.458 | 0.420 | 0.458 |
| Adjusted R² | 0.392 | 0.291 | 0.463 | 0.410 | 0.400 | 0.290 | 0.401 | 0.289 |

Notes: Each regression includes period fixed effects. Clustered standard errors in brackets. *p<0.1, **p<0.05, ***p<0.01.

# Supplementary Note 9. What about services?

Services are an integral part of an economy, and the service sector is becoming a rising share of international trade and within-country employment. Due to this, there have been several studies which produced attempts to incorporate the service dimension into the economic complexity framework. Nevertheless, the studies have found that the service dimension only moderately improves the complexity rankings and does not lead to significant improvement in macroeconomic outcomes regressions[14].

To explore whether the service dimension is significant in explaining international variations in economic growth, income inequality and emission intensity, we collect service trade data from OECD's Input Output Database. This database distinguishes 20 service sectors across 61 countries. We use these data to estimate the service dimension ECI, and then use it as an independent variable in individual regression analyses which are defined in the same way as the ones described in Supplementary Note 3.

The results are given in Supplementary Tables 47-49, respectively for the economic growth, income inequality, and emission intensity regressions. In each case, the service ECI is not a significant or robust explanatory variable. Thus, we assert that services data still does not provide enough information to be included in the multidimensional economic complexity analysis.

## Supplementary Table 47. ECI (services) Economic Growth Regressions

| | Dependent variable: | | | | | | | | | | | |
|---|---|---|---|---|---|---|---|---|---|---|---|---|
| | Annualized GDP pc growth (1999-09, 2009-19) | | | | | | | | | | | |
| | (1) | (2) | (3) | (4) | (5) | (6) | (7) | (8) | (9) | (10) | (11) | (12) |
| ECI (services) | | 0.486 | 0.994 | 0.320 | 0.531 | 0.400 | 0.726 | 0.426 | 0.430 | 2.956** | 0.087 | 0.495 |
| | | (0.681) | (0.624) | (0.615) | (0.657) | (0.615) | (0.656) | (0.662) | (0.662) | (1.195) | (0.693) | (0.672) |
| Log of initial population | | | -0.234* | | | -0.016 | | | | | | |
| | | | (0.129) | | | (0.130) | | | | | | |
| Log of natural resource exports per capita | | | | 3.990*** | | 3.583*** | | | | | | |
| | | | | (0.706) | | (0.660) | | | | | | |
| Log of initial human capital | | | | | 0.673*** | 0.437* | | | | | | |
| | | | | | (0.221) | (0.233) | | | | | | |
| Intensity (services) | | | | | | | -2.649 | | | | | |
| | | | | | | | (3.140) | | | | | |
| HHI (services) | | | | | | | | -0.755 | | | | |
| | | | | | | | | (1.167) | | | | |
| Entropy (services) | | | | | | | | | 0.665 | | | |
| | | | | | | | | | (1.166) | | | |
| Fitness (services) | | | | | | | | | | -3.134*** | | |
| | | | | | | | | | | (1.004) | | |
| i-ECI | | | | | | | | | | | 1.921 | |
| | | | | | | | | | | | (1.170) | |
| Log of initial GDP per capita | -1.484*** | -1.579*** | -1.906*** | -2.419*** | -2.491*** | -2.948*** | -0.874 | -1.641*** | -1.634*** | -1.693*** | -1.786*** | -1.581*** |
| | (0.262) | (0.341) | (0.342) | (0.379) | (0.413) | (0.405) | (0.994) | (0.365) | (0.366) | (0.313) | (0.360) | (0.342) |
| Observations | 113 | 113 | 113 | 113 | 113 | 113 | 113 | 113 | 113 | 113 | 113 | 113 |
| $R^2$ | 0.378 | 0.382 | 0.412 | 0.519 | 0.437 | 0.543 | 0.388 | 0.385 | 0.384 | 0.453 | 0.398 | 0.382 |
| Adjusted $R^2$ | 0.367 | 0.365 | 0.390 | 0.501 | 0.416 | 0.517 | 0.366 | 0.362 | 0.361 | 0.433 | 0.376 | 0.365 |

Notes: Each regression includes period fixed effects. Clustered standard errors in brackets. *p<0.1, **p<0.05, ***p<0.01.
The last column includes the instrumental variables estimate (see Supplementary Note 8).

## Supplementary Table 48. ECI (services) Income Inequality Regressions

|  | | | | | | *Dependent variable:* | | | | | | |
|---|---|---|---|---|---|---|---|---|---|---|---|---|
|  | | | | | | Gini coefficient (1996-99, 2000-03, 2004-07, 2008-11, 2012-15) | | | | | | |
|  | (1) | (2) | (3) | (4) | (5) | (6) | (7) | (8) | (9) | (10) | (11) | (12) |
| ECI (services) |  | -1.035 | -2.583 | 0.367 | -1.062 | -0.826 | -0.233 | -0.135 | -0.164 | -1.879 | 0.970 | -1.610 |
|  |  | (2.144) | (2.038) | (1.898) | (2.031) | (1.892) | (1.758) | (2.029) | (2.026) | (2.468) | (2.457) | (2.524) |
| Log of population |  |  | 0.907*** |  |  | 0.587** |  |  |  |  |  |  |
|  |  |  | (0.286) |  |  | (0.268) |  |  |  |  |  |  |
| Log of human capital |  |  |  | -10.535*** |  | -9.147*** |  |  |  |  |  |  |
|  |  |  |  | (2.547) |  | (2.292) |  |  |  |  |  |  |
| Log of natural resource exports |  |  |  |  | -1.362** | -0.252 |  |  |  |  |  |  |
|  |  |  |  |  | (0.666) | (0.683) |  |  |  |  |  |  |
| Intensity (services) |  |  |  |  |  |  | -21.851*** |  |  |  |  |  |
|  |  |  |  |  |  |  | (4.788) |  |  |  |  |  |
| HHI (services) |  |  |  |  |  |  |  | 16.783*** |  |  |  |  |
|  |  |  |  |  |  |  |  | (3.176) |  |  |  |  |
| Entropy (services) |  |  |  |  |  |  |  |  | -15.011*** |  |  |  |
|  |  |  |  |  |  |  |  |  | (3.139) |  |  |  |
| Fitness (services) |  |  |  |  |  |  |  |  |  | 1.029 |  |  |
|  |  |  |  |  |  |  |  |  |  | (2.519) |  |  |
| i-ECI |  |  |  |  |  |  |  |  |  |  | -9.490 |  |
|  |  |  |  |  |  |  |  |  |  |  | (6.438) |  |
| Log of GDP per capita | 4.707 | 1.770 | 7.096 | 20.899* | 3.184 | 22.091** | 1.890 | 5.691 | 5.810 | 1.085 | 5.222 | 0.138 |
|  | (12.127) | (13.429) | (12.781) | (11.664) | (13.109) | (11.063) | (12.228) | (12.911) | (12.949) | (13.823) | (14.201) | (13.676) |
| Log of GDP per capita, squared | -0.494 | -0.333 | -0.544 | -1.209** | -0.304 | -1.225** | -0.102 | -0.544 | -0.549 | -0.295 | -0.458 | -0.244 |
|  | (0.615) | (0.692) | (0.656) | (0.592) | (0.674) | (0.572) | (0.623) | (0.660) | (0.662) | (0.712) | (0.724) | (0.708) |
| Observations | 264 | 264 | 264 | 264 | 264 | 264 | 264 | 264 | 264 | 264 | 264 | 264 |
| $R^2$ | 0.462 | 0.463 | 0.511 | 0.553 | 0.487 | 0.576 | 0.566 | 0.522 | 0.523 | 0.464 | 0.500 | 0.463 |
| Adjusted $R^2$ | 0.450 | 0.449 | 0.496 | 0.539 | 0.471 | 0.559 | 0.552 | 0.507 | 0.508 | 0.447 | 0.485 | 0.448 |

Notes: Each regression includes period fixed effects. Clustered standard errors in brackets. *p<0.1, **p<0.05, ***p<0.01.
The last column includes the instrumental variables estimate (see Supplementary Note 8).

## Supplementary Table 49. ECI (services) Emission intensity Regressions

|  | *Dependent variable:* | | | | | | | | | | | |
|---|---|---|---|---|---|---|---|---|---|---|---|---|
|  | GHG emissions per GDP (1996-99, 2000-03, 2004-07, 2008-11, 2012-15, 2016-19) | | | | | | | | | | | |
|  | (1) | (2) | (3) | (4) | (5) | (6) | (7) | (8) | (9) | (10) | (11) | (12) |
| ECI (services) |  | -0.212 | -0.131 | -0.082 | -0.353* | -0.220 | -0.386 | -0.277 | -0.353 | -0.268 | -0.134* | -0.437** |
|  |  | (0.227) | (0.201) | (0.207) | (0.209) | (0.221) | (0.248) | (0.199) | (0.251) | (0.217) | (0.075) | (0.220) |
| Log of population |  |  |  |  |  | -1.013 |  |  |  |  |  |  |
|  |  |  |  |  |  | (0.748) |  |  |  |  |  |  |
| Log of human capital |  |  |  |  |  |  | -0.114 |  |  |  |  |  |
|  |  |  |  |  |  |  | (0.363) |  |  |  |  |  |
| Intensity (services) |  |  |  |  |  |  |  | -0.818* |  |  |  |  |
|  |  |  |  |  |  |  |  | (0.461) |  |  |  |  |
| HHI (services) |  |  |  |  |  |  |  |  | -0.00005 |  |  |  |
|  |  |  |  |  |  |  |  |  | (0.188) |  |  |  |
| Entropy (services) |  |  |  |  |  |  |  |  |  | -0.464 |  |  |
|  |  |  |  |  |  |  |  |  |  | (0.377) |  |  |
| Fitness (services) |  | 0.054 |  |  | 0.118*** | 0.112*** | 0.122*** | 0.106*** | 0.118*** | 0.126*** | 0.605*** | 0.123*** |
|  |  | (0.037) |  |  | (0.038) | (0.038) | (0.041) | (0.036) | (0.038) | (0.039) | (0.177) | (0.039) |
| i-ECI |  |  | 0.819*** |  | 0.919*** | 0.906*** | 0.924*** | 0.995*** | 0.919*** | 1.040*** | 0.838*** | 0.931*** |
|  |  |  | (0.257) |  | (0.251) | (0.252) | (0.253) | (0.264) | (0.251) | (0.303) | (0.157) | (0.253) |
| Log of initial GDP per capita |  |  |  | 0.086 | 0.187** | 0.177** | 0.191** | 0.134 | 0.187** | 0.169** | 0.099** | 0.190** |
|  |  |  |  | (0.082) | (0.083) | (0.080) | (0.085) | (0.084) | (0.083) | (0.080) | (0.048) | (0.083) |
| log(gdp_start) | -0.323*** | -0.230** | -0.486*** | -0.429*** | -0.623*** | -0.341 | -0.635*** | -0.594*** | -0.623*** | -0.563*** | -0.397*** | -0.609*** |
|  | (0.066) | (0.092) | (0.111) | (0.128) | (0.131) | (0.257) | (0.142) | (0.121) | (0.130) | (0.124) | (0.083) | (0.133) |
| Fixed effects | No | No | No | No | No | No | No | No | No | No | Yes | No |
| Observations | 342 | 342 | 342 | 342 | 342 | 342 | 342 | 342 | 342 | 342 | 342 | 342 |
| R² | 0.262 | 0.288 | 0.336 | 0.276 | 0.424 | 0.438 | 0.425 | 0.448 | 0.424 | 0.434 | 0.421 | 0.423 |
| Adjusted R² | 0.249 | 0.271 | 0.320 | 0.259 | 0.407 | 0.419 | 0.406 | 0.430 | 0.405 | 0.415 | 0.276 | 0.406 |

Notes: Each regression includes period fixed effects. Clustered standard errors in brackets. *p<0.1, **p<0.05, ***p<0.01.
The last column includes the instrumental variables estimate (see Supplementary Note 8).

# Supplementary References